\documentclass[utf8]{aa} 
\usepackage{multirow,longtable,threeparttable}
\usepackage{xcolor}
\def\mbh{$M_\mathrm{BH}$\/}
\def\lledd{$\lambda_\mathrm{E}$\/}

\def\rfe{$R_{\rm FeII}$}

\def\feiiq{\rm Fe{\sc ii}$\lambda$4570\/}
\def\msol{M$_\odot$\/}

\def\civ{{\sc{Civ}}$\lambda$1549\/}

\def\cm3{cm$^{-3}$\/}
\def\hb{{\sc{H}}$\beta$\/}

\def\hbbc{{\sc{H}}$\beta_{\rm BC}$\/}
\def\hbvbc{{\sc{H}}$\beta_{\rm VBC}$\/}

\def\oiiiopt{{\sc{[Oiii]}}$\lambda\lambda$4959,5007\/}

\def\feii{{Fe\sc{ii}}\/}

\def\fe{{\sc{Fe}}\/}

\def\fe76087{{\sc [Fe vii]}$\lambda$6087\/}
\def\oiii{{\sc [Oiii]}$\lambda$5007}

\def\kms{km~s$^{-1}$}

\def\rk{$R_{\rm K'}$\/}
\def\ergss{erg s$^{-1}$\/}

\def\aap{A\&Ap\/}

\def\apjs{ApJS}

\def\apjl{ApJL}

 \date{ }
\begin{document}
\titlerunning{Radio-loudness along the quasar main sequence}
\title{Radio-loudness along the quasar main sequence} 
\author{ V. Ganci\inst{1} \and P. Marziani\inst{2}  \and M. D'Onofrio\inst{1} \and A. del Olmo\inst{3} \and 
E. Bon\inst{4} \and N. Bon\inst{4}  \and C.A. Negrete\inst{5}}
\institute{ 
{Dipartimento di Fisica \& Astronomia ``Galileo Galilei'', Universit\`a di Padova, Padova,  Italy}
\and{National Institute for Astrophysics (INAF), Padua Astronomical Observatory, IT 35122, Padova, Italy}
\and {Instituto de Astrofis\'{\i}ca de Andaluc\'{\i}a, IAA-CSIC, E-18008 Granada, Spain}
 \and {Belgrade Astronomical Observatory, 11060, Belgrade, Serbia}
 \and {Instituto de Astronom\'{\i}a, UNAM, DF 04510, Mexico}
}

\abstract
{When can an active galactic nucleus (AGN) be considered radio-loud? Following the established view of the AGNs inner workings, an AGN is radio-loud if associated with  relativistic  ejections emitting a radio synchrotron spectrum (i.e., it is  a ``jetted'' AGN). In this paper we exploit  the AGN main  sequence that offers a powerful tool to contextualize radio properties.}
{If large samples of optically-selected quasars are considered,  AGNs are identified as radio-loud if their Kellermann's radio loudness ratio $R_\mathrm{K} > 10$.   Our aims are   to characterize the optical properties of different classes based on radio-loudness within the main sequence and to test whether the condition $R_\mathrm{K} > 10$ is sufficient for the identification of RL AGNs, since the origin of relatively strong radio emission may not be necessarily due to relativistic ejection.}
{A sample of  355  quasars was selected by cross-correlating the FIRST survey with the SDSS DR14 quasar catalog. We classified the optical spectra according to their spectral types along the main sequence of quasars. For each spectral type, we distinguished compact and extended morphology (providing a FIRST-based atlas of radio maps in the latter case), and three classes of radio-loudness: detected ( specific flux ratio in the $g$ band and at 1.4GHz,  $R_\mathrm{K'}<$10), intermediate (10 $\le R_\mathrm{K'} <$ 70), and radio loud ($R_\mathrm{K'} \geq $70).}
{The analysis revealed systematic differences between radio-detected (i.e., radio-quiet), radio-intermediate, and radio-loud in each spectral type along the main sequence.  We show that spectral bins that contain the extreme Population A sources have radio power compatible with emission by mechanisms ultimately due to star formation processes.  Radio-loud sources of Population B are characteristically jetted. Their broad \hb\ profiles  can be interpreted as due to a binary broad-line region. We suggest that RL Population B sources should be preferential targets for the search of black hole binaries, and present a sample of binary black hole AGN candidates. }
{The validity of the Kellermann's criterion may be dependent on the source location along the quasar main sequence. The consideration of the MS trends allowed to distinguish between sources whose radio emission mechanisms is ``jetted'' from the ones where the mechanism is likely to be fundamentally different. 

}

\maketitle


\section{Introduction}
\label{intro}

The main sequence (MS) of quasars has proved to be a powerful tool to contextualize the observational properties of type-1 active galactic nuclei (AGNs; see \citealt{sulenticmarziani15} for a recent review). The MS concept stemmed from an application of the Principal Component Analysis to quasar spectra  \citep{borosongreen92} that yielded a first fundamental correlation vector (called the Eigenvector-1, E1) of type-1 AGNs. More precisely, the E1 is associated with anti-correlations between the strength of FeII$\lambda$4570 and the \oiii\ peak intensity, and between strength of FeII$\lambda$4570 and  FWHM of H$\beta$ \citep{borosongreen92}. The FeII$\lambda$4570 strength is usually represented by the parameter $R_\mathrm{FeII} =$ I(FeII$\lambda$4570)/I(H$\beta$), the ratio between the integrated flux of the FeII$\lambda$4570 blend of multiplets  between 4434 \AA\ and 4684  \AA, and that of the H$\beta$ \ full broad component\footnote{The full broad component refers to the full profile without the narrow component, that is, as in previous works   \citep[e.g.,][]{brothertonetal94a,corbinboroson96,bonetal06,huetal08,zietal15,adhikarietal16}, subdivided into a broad (\hbbc) and a very-broad component (\hbvbc).}. The distribution of data points in the E1 optical plane, or in other words FWHM(H$\beta$) vs. $R_\mathrm{FeII}$\, defines the  quasar main sequence, in  analogy to the stellar main sequence on the Hertzsprung-Russell diagram \citep{sulenticetal01,sulenticetal01a,sulenticetal08,shenho14}.  

\begin{figure}[htbp!]
\begin{center}
\includegraphics[width=.45\textwidth]{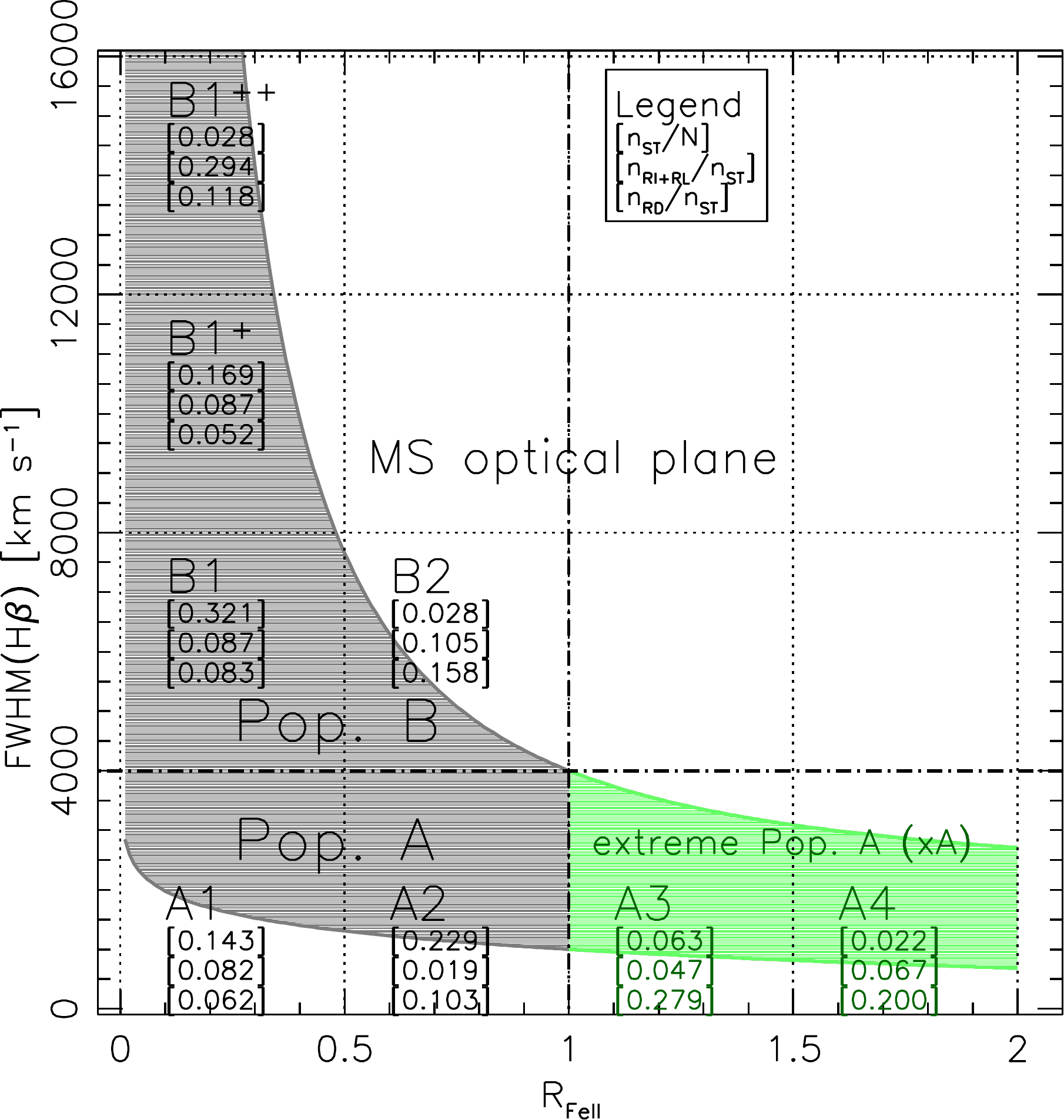}
 \quad
\caption{Sketch illustrating the definition of the spectral types along the MS, as a function of \rfe\ and FWHM(\hb). The numbers in square brackets yield from top to bottom, the prevalences of each spectral bin ($n_\mathrm{ST}$) in an optically selected sample  \citep{marzianietal13a}, of the  RI and RL {sources}  ($n_\mathrm{RI+RL}$) and of RD ($n_\mathrm{RD}$) in each spectral bin. The total number of sources is N=680. The gray and pale green ares trace the source occupation in the plane; the green area separates extreme Population A from the rest of the MS.} 
\label{fig:e1r}
\end{center}
\end{figure}

The shape of the MS for  quasars of luminosity $L \lesssim 10^{47}$ erg s$^{-1}$, and $z < 0.7$ allows for the subdivision of the optical E1 plane into two Populations, and a grid of bins of FWHM(H$\beta$) and FeII emission strength ($R_\mathrm{FeII}$) that defines a sequence of spectral types (STs; see the sketch in Fig. \ref{fig:e1r}). Population A (FWHM(H$\beta) \le 4000$ kms$^{-1}$) and B  (FWHM(H$\beta)  > 4000$ \kms) separate different behaviors along the E1 sequence, possibly associated with differences in accretion modes \citep[e.g.,][and references therein]{wangetal14,duetal16}.   Population A spectral types are defined in terms of increasing $R_\mathrm{FeII}$ with bin size $\Delta R_\mathrm{FeII} = 0.5$. Spectral types from A1, with $R_\mathrm{FeII} \le 0.5$, to A4, with $1.5 < R_\mathrm{FeII} \le 2.0$\ account for almost all Population A sources, save for a few FeII-strong ``outliers'' \citep{liparietal93,grahametal96,marzianietal13a}. Population B STs are defined in terms of increasing FWHM H$\beta$ with $\Delta$FWHM(H$\beta) = 4000$ kms$^{-1}$. Spectral types from B1 ($4000 < FWHM$(H$\beta$) $\le 8000$ \kms),  to B1$^{++}$   ($12000 < FWHM$(H$\beta$) $\le 16000$ \kms) account for the wide majority of Population B sources. Spectral types are intended to isolate sources with similar broad line physics and/or viewing angle.   From here onwards, we will consider STs A3 and A4 as extreme population A (xA;  the area shaded green in Fig. \ref{fig:e1r}), in accordance with the recent analysis by \citet{negreteetal18}.  

There is no doubt that several physical parameters affect the E1 MS. Recent works \citep[e.g., ][]{sulenticetal11,fraix-burnetetal17} present a list of them and of their relation to observed parameters. The physical interpretation of the MS is still being debated \citep{shenho14,pandaetal18}, although  there is a growing consensus that the main factors shaping the MS occupation in the optical plane \rfe-FWHM(\hb) (at least for a low-z sample) are the Eddington ratio (\lledd) and the viewing angle defined as the angle between the line-of-sight and the symmetry axis of the active nucleus.   $R_\mathrm{FeII}$ correlates with metallicity, ionization conditions, density and column density, and ultimately with Eddington ratio   \citep[][and references therein]{grupeetal99,kuraszkiewiczetal09,aietal10,dongetal11,duetal16,pandaetal19}. FWHM(H$\beta$) correlates with the velocity dispersion in the low-ionization lines emitting region of the broad-line region (BLR)  and provides a measurement of the virial broadening associated with ionized gas motion around the central massive black hole  \citep[][and references therein]{shen13}. The FWHM is affected by physical parameters such as black hole mass (\mbh) and  \lledd, but also by the viewing angle \citep{pandaetal19}. Growing evidence indicates that the low-ionization part of the broad line region is highly flattened \citep[][and references therein]{mejia-restrepoetal18a}.  It is therefore straightforward to think about a physically-motivated distinction between Population A and Population B. In a word, Population A sources are fast-accreting objects with relatively small black hole masses (at least at low-$z$) and population B are the ones with high black hole mass and low $\lambda_\mathrm{E}$\ \citep[e.g.,][]{marconietal09,fraix-burnetetal17}.  

The MS as drawn in Fig. \ref{fig:e1r} has been built for an optically selected sample at low-$z$\ ($\lesssim 1$; \citealt{marzianietal13a}), and includes both radio-quiet (RQ) and radio-loud (RL) AGNs. The behavior of the RL sources in the MS is still poorly known, since  RL quasars are a minority,  only $\sim 10 \%$ of all quasars  \citep[a fact that became known a few years after the discovery of quasars,][]{sandage65}. 
The physical definition of RL sources involves the presence of a strong relativistic jet (i.e., sources are ``jetted,'' \citealt{padovani17}),  whereas in RQ sources the jet is expected to be non-relativistic or intrinsically  weaker than in RL \citep[e.g.,][]{middelbergetal04,ulvestadetal05,gallimoreetal06}, and other physical mechanisms due to nuclear activity could be important   components. RQ sources do have a radio emission and their radio powers can be even a few orders of magnitude lower than those of their RL counterparts for the same optical power \citep[e.g.,][]{padovani16}.  

The presence of   relativistic ejections gives rise to a host of phenomenologies over the full frequency range of the electromagnetic spectrum. Apart from the enhancement in radio power with respect to RQ quasars, differences can be seen between the hard X-ray and the $\gamma$-ray wavelength ranges. RL sources emit up to GeV (2.4 $\times$ 10$^{23}$ Hz, \citealt[][Fig. 1]{padovani17}) energies or in some cases to TeV whereas RQ sources show a sharp cut-off at energies $\sim$ 1 MeV \citep[e.g.,][]{maliziaetal14}. No RQ AGN has been detected in $\gamma$-rays \citep{ackermannetal12}. These properties let us think about two, at least in part, intrinsically different types of sources:  through the accretion process, RLs emit a large fraction of their energy non-thermally  over the whole electromagnetic spectrum, whereas RQ quasars emit most of their power thermally from viscous dissipation in the torus and the accretion disk.  

The customary selection of RL sources is based on the ratio $R_\mathrm{K} = f_\mathrm{\nu , radio}/f_\mathrm{\nu,opt}, $\ where the numerator is the radio flux density at 5GHz and the denominator is the optical flux density  in the $B$ \ band \citep{kellermannetal89}. RL sources are defined as the ones having $R_\mathrm{K} > 10$. The classification is made also employing a surrogate Kellermann's parameter  $R_\mathrm{K'}$ with the $g$\ band and the specific radio flux at 20 cm (Sect. \ref{rk} and  \ref{datanrad}). 

Applying the Kellermann's criterion, the prevalence of RL sources is not constant along the main sequence. The data of Fig. \ref{fig:e1r} are from a 680-strong quasar  SDSS sample \citep{marzianietal13a}: in that sample, 30\%\ of extreme Pop. B ({i.e., spectral type B1$^{++}$) are RIs or RLs}; ST B1 and B1$^{+}$, which include half of the sample, have a prevalence of $\sim 9 \%$ each consistent with previous studies \citep{kellermannetal89,urrypadovani95,zamfiretal08}.  A minimum prevalence of just 2\%\ is reached in ST A2, whereas higher \rfe\ ST A3 and A4  show a surprising increase.  Previous works have shown that ``jetted'' sources are more frequent among Pop. B sources or, equivalently at low Eddington ratio \citep{sikoraetal07,zamfiretal08}. Pop. A and Pop. B sources are known to have a different spectral energy distributions (SEDs), flatter in the case of Pop. B.  \citet{zamfiretal08} suggest a more restrictive criterion:  $R_\mathrm{K} >70$\ as a sufficient condition to identify ``jetted'' sources. This criterion may indeed be sufficient but  may lead to the loss of a significant number of intrinsically jetted sources.  \citet{zamfiretal08} observed that ``intermediate'' sources with $10 \lesssim R_\mathrm{K'} \lesssim 70$\ are distributed across the whole main sequence. This did not explain their physical nature but did not exclude that a fraction of them could be ``jetted''. On the other hand,  ST A3 and A4 are associated with high Eddington ratio and concomitant high star-formation rate ($SFR$; e.g., \citealt{sanietal10}). Star forming processes are   believed to be associated with accretion in RQ quasars \citep[e.g.,][]{sandersetal09}. They become most evident in the FIR domain of their spectral energy distribution \citep[e.g.,][]{sandersetal88,haasetal03,sanietal10}. High $SFR$\  leads to correspondingly high radio-power \citep[e.g.,][and references therein]{condon92,mirabelsanders96}. Therefore, 
the trends along the MS concerning the prevalence and the nature of radio emission call into question the validity of the  criterion $R_\mathrm{K} > 10$  as a necessary condition of  ``jetted''   sources.

Optically selected samples even if $\sim 1000$ in size contain relatively few sources in each spectral type to make a reliable assessment of their properties along the MS.   This work tries to find new clues on the selection of truly jetted sources through a systematic study of a large radio-detected AGNs sample (Section \ref{sample}). Sample sources were subdivided  on the basis of radio-morphology (core-dominated, Fanaroff-Riley II), radio-loudness range (radio-detected, radio-intermediate, radio-loud, Sect. \ref{sample} and Sect. \ref{datanrad}). Following the MS approach, the sample was subdivided  in optical STs (Sect. \ref{datanopt}). For most STs, there is a sufficient number of sources in each radio loudness and morphology class. Results are reported for the radio classes along the main sequence (Sect. \ref{res}). The most relevant aspects are the possibility of ``thermal'' radio emission (i.e., due to emission mechanisms associated with stars in their late evolutionary stages) in extreme Population A (Sect. \ref{radfir}) and  the relatively high prevalence of \hb\ profiles that can be expected from a binary broad line region (BLR) in jetted sources (Sect. \ref{bbh}). Other MS   trends are briefly discussed in terms of physical phenomena that may affect both the radio and the optical properties (Sect. \ref{disc}).    Throughout this paper, we use $H_\mathrm{0} = 70$ km s$^{-1}$ Mpc$^{-1}$, $\Omega_\mathrm{M} =$ 0.3 and $\Omega_\mathrm{\Lambda}=$ 0.7.

\section{Sample selection}
\label{sample}

The AGNs studied in this work were selected from the twelfth release of the Sloan Digital Sky Survey (SDSS) Quasar Catalog  published in 2017 \citep{parisetal17}. The selected objects have an $i$-band magnitude $m_\mathrm{i}$ value below 19.5 and a Baryon Oscillation Spectroscopic Survey \citep[BOSS,][]{dawsonetal13} pipeline redshift value $z \lesssim$  1.0. Sources with a higher magnitude with respect to this limit have low S/N individual spectra that make it difficult to estimate the FWHM(H$\beta$), the \feii\ flux and henceforth to assign a proper E1 ST. 

\subsection{Selection on the basis of radio morphology}

The BOSS sample was then cross-matched with the Very Large Array (VLA) Faint Images of the Radio Sky at Twenty-Centimeters (FIRST) survey \citep{beckeretal95}. Through the cross-match we selected only the SDSS objects that have one or more radio sources that are no more distant than 120" with respect to the SDSS object position, given by their right ascension (J2000) and declination (J2000). All the objects with only one radio source within an annulus of 2.5'' $<$ r $\leq$ 120'' were rejected. SDSS objects with one radio source within 120" were classified as core dominated (CD) AGNs if that radio source is within 2.5" from the optical coordinates of the quasar. Objects with two radio sources within 120" were classified as CD AGNs if one of the radio sources is within 2.5"; in this case the other radio source is not considered part of the system. We considered the peak radio position and the peak radio flux $F_\mathrm{peak}$, even if a fraction of radio sources show some resolved structures \citep{ivezicetal02}. A significant fraction of radio sources  in the FIRST catalog shows an integrated flux $F_\mathrm{int} \lesssim F_\mathrm{peak}$, probably because of over-resolution. Sources with $F_\mathrm{int} \gtrsim 1.3 F_\mathrm{peak}$\ are $\lesssim$10\%\ and for about one half of low S/N. Some remaining cases with a possible significant extended  emission are identified in the Appendix \ref{CDphys}. 

The remaining objects are SDSS objects with two or more radio components within 120". These are FRII AGNs candidates.\footnote{The term FRII is used here to signify sources with strong extended emission in the FIRST; they corresponds to core-lobe, lobe-core-lobe, lobe-lobe radio morphologies; CD corresponds to core, core-jet according to the radio morphology classification of \citet{kimballetal11}, with the possibility of   compact-steep spectrum (CSS) or FRI sources. \textcolor{red}{}However, only three sources listed in Table \ref{tab:FRIIphys} are somewhat below (at most by $0.2$ dex) the formal limit of 10$^{31}$ erg/s/Hz at 5GHz (which correspond to $\sim 3 \cdot 10^{31}$ erg/s/Hz at 1.5GHz). We therefore use the term FRII for all sources in the sample.} 

In order to select FRII AGNs systems with a high probability of being physical, we adopted the statistical procedure suggested by \citet{devriesetal06}. The procedure  considers  the radio sources around a SDSS quasar two-by-two, with  each pair being a possible set of radio lobes. Pairs were ranked by the value of the following parameter: 

\begin{equation} 
w_\mathrm{i,j}=\dfrac{\Psi / 50^{\circ}}{(r_\mathrm{i}+r_\mathrm{j})^{2}}, 
\end{equation} 

where $\Psi$ is the opening angles (in degrees) as seen from the quasar position and $r_\mathrm{i,j}$ are the distance rank numbers of the components under consideration, equal to zero for the closest component to the quasar, equal to one for the next closest and so on. In this way the  radio components closest to the quasar optical position have more weight. The opening angle is divided by 50$^{\circ}$ to weight against pairs of sources unrelated to the quasar that tend to have small opening angles. Then a higher value of $w_\mathrm{i,j}$, means radio components closer to the quasar with an opening angle closer to 180$^{\circ}$, and therefore a higher probability of the configuration to be a FRII AGN. The configuration with the highest $w_\mathrm{i,j}$ value for each SDSS objects is kept and considered as a real FRII AGN if $130^{\circ} \leq \Psi \leq 180^{\circ}$ in the case the two radio components are within 60" or if $150^{\circ} \leq \Psi \leq 180^{\circ}$ in the case one or both are more distant than 60" but closer than 120". Finally we checked whether each radio source corresponding to a radio lobe has an optical counterpart (within 5") through the NASA/IPAC Extragalactic Database (NED) \footnote{https://ned.ipac.caltech.edu/}; the FRII classification is rejected if one or both lobes have an optical counterpart. This sample (not yet final) has 482 SDSS objects, 407 CD and 75 FRII. 

\subsection{Selection on the basis of the radio-loudness parameter}
\label{rk}

We  classified the sample sources in radio-detected (RD), radio-intermediate (RI) and radio-loud (RL) classes on the basis of the Kellermann's parameter.

Sources with log$R_\mathrm{K} < 1.0$ are classified as RD, sources with $1.0 \le$ log$R_\mathrm{K} < 1.8$   as RI and sources with log$R_\mathrm{K} \geq 1.8$ as RL. These limits are defined on the 1.4 GHz  and $g$\ magnitude estimates of $R_\mathrm{K}$ that is about 1.4 the value as originally defined by \citet{kellermannetal89}  for a power-law spectral index $a = 0.3$. The 1.4 GHz/$g$  ratio  has been used in  recent works   \citep[][see also \citealt{gurkanetal15}]{zamfiretal08} and will be used also in the present one  keeping the notation $R_\mathrm{K}$. We did not update the limit also because of  systematic underestimates of the FIRST radio fluxes by at least about 30\%\ for unresolved sources (Table 1 of \citealt{beckeretal95}). {Previous analyzes considered the condition $\log R_\mathrm{K} > 1.0$\ as sufficient to identify RL sources \citep{kellermannetal89}. More recent work has questioned the universal validity of the criterion, on the basis of two considerations: (1) $R_\mathrm{K}$\ is sensitive to intrinsic continuum shape differences and by  internal extinction of the host  that depresses the $g$ flux; (2) large radio power can be associated with processes related to the last evolutionary stages of stars \citep[e.g.,][and references therein]{dubnergiacani15}. 


\citet{padovani17} stressed the need to distinguish between truly “jetted” and “non-jetted” sources, where for “jetted” sources only highly-relativistic jets should be considered. From the observational point of view, the inadequacy of the condition $R_\mathrm{K} \gtrsim 10$ \ to identify truly “jetted” sources was illustrated by \citet{zamfiretal08}: classical radio sources are segregated within Pop. B, whereas sources with $10 \lesssim R_\mathrm{K} \lesssim 70$ show the uniform distribution of RQ quasars along the MS. On the other hand, the nature of the $10 \lesssim R_\mathrm{K} \lesssim 70$ “intermediate” sources was not clarified by \citet{zamfiretal08}. The condition  $R_\mathrm{K} \gtrsim 70$\ is very restrictive and, albeit sufficient, may not be necessary to identify truly jetted sources (we may miss jetted sources).  

\subsection{Optical spectral type classification}

Once we had grouped the sources in radio-loudness classes, the objects were classified on the basis of their optical spectroscopic characteristics in terms of the E1. We took the spectrum of each object from the SDSS Data Release 14 (DR14) that includes the complete dataset of optical spectroscopy collected by the BOSS. First of all, we rejected objects that show a type II AGN spectrum, and spectra in which the host galaxy contamination is therefore strong to make   impossible   a reliable measure of the FWHM(H$\beta$). In addition, we excluded data with instrumental problems:  spectra spoiled by noise and spectra with poor reduction  due to strong atmosphere contamination in the NIR, whenever H$\beta$ was falling in that range. After screening, the sample consisted of 355 objects, of which 289 CD and 66 FRII. 

Figure \ref{fig:Zsample} shows the redshift and the luminosity at 5100 \textup{\AA} distributions of the sample. Regarding $L_\mathrm{5100}$, we see no substantial differences between the different   radio morphology and radio power classes.  Kolmogorov-Smirnov tests indicate that  the distributions of CD RD on the one hand and RI and RL (both CD and FRII) are significantly different at a $\sim$ 2 - 2.5$\sigma$\ confidence level. No significant differences are found between RI and RL (both CD and FRII).

\begin{figure*}[htbp!]
\begin{center}
\includegraphics[width=.65\textwidth]{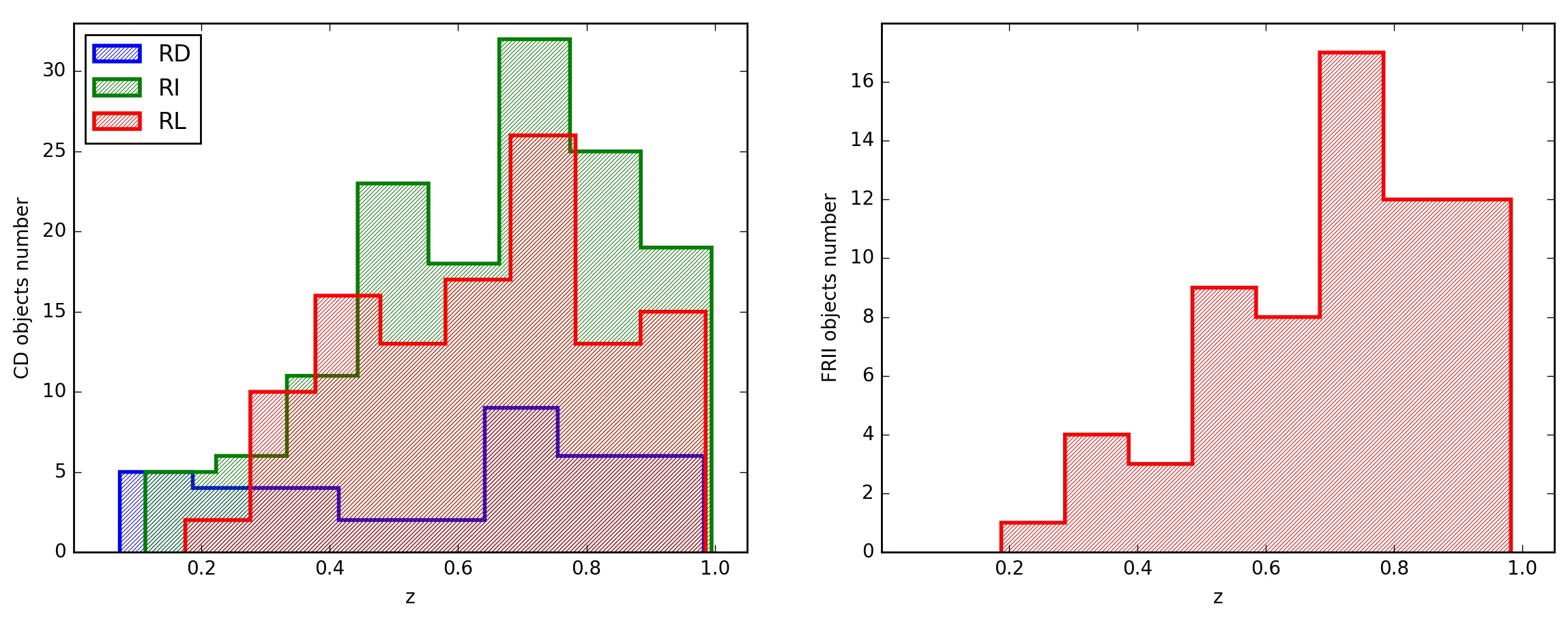}\\[1.em]
\includegraphics[width=.65\textwidth]{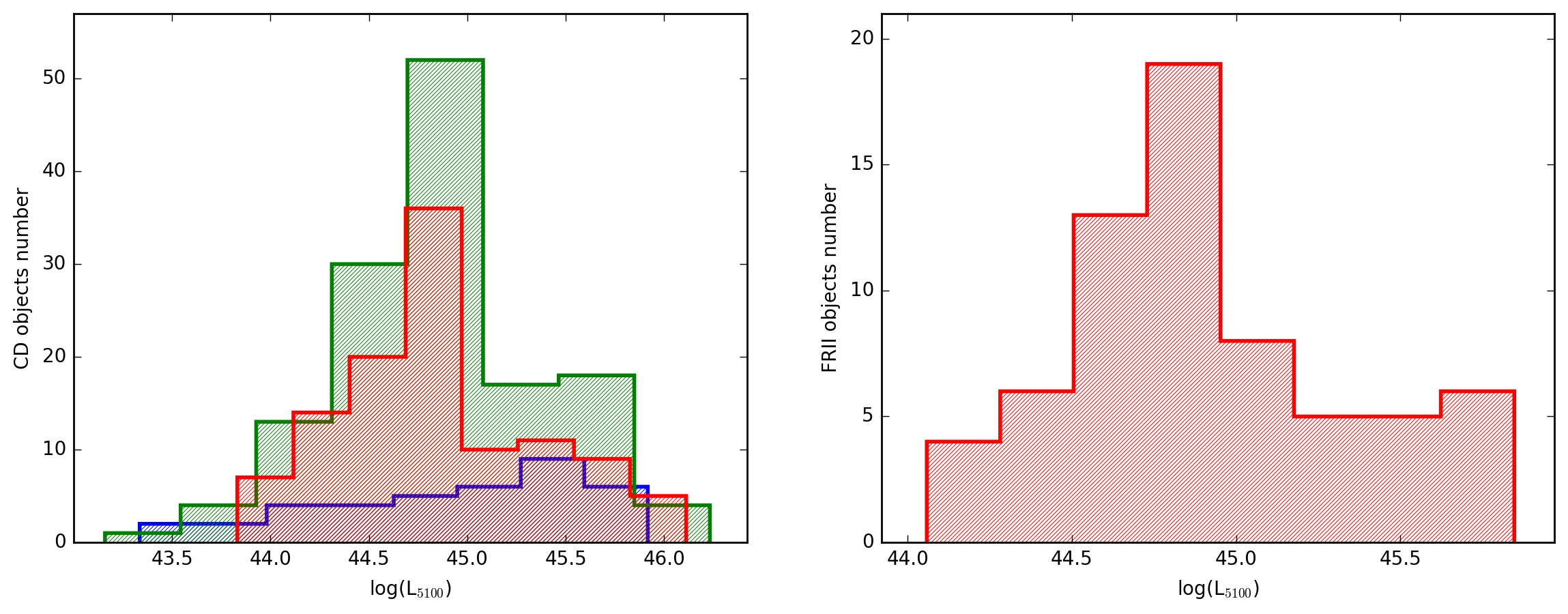}\\
\caption{Redshift (top) and luminosity at 5100 \textup{\AA} (bottom) distributions of the sample. The left plots refer to CD sources, the right ones to FRII ones. Blue refers to the RD class, green to the RI one and red to the RL one.
}
\label{fig:Zsample}
\end{center}
\end{figure*}

\section{Data Analysis}
\label{datan}

\subsection{Radio and optical K-corrections}
\label{datanrad}

To get the optical flux density $f_\mathrm{\nu , opt}$, we started from the $g$-band magnitudes $m_\mathrm{g}$ retrieved from the SDSS quasar catalog. First, we corrected them for the galactic extinction using the absorption parameter $A_\mathrm{b}$ values for each object, acquired from NED. We computed the flux density in mJy from the magnitude through the Pogson law. The last step before the calculation of $R_\mathrm{K}$ is the $K$\ correction. From: 

\begin{equation} f_\mathrm{\nu_\mathrm{o}}d \nu_\mathrm{o} = \dfrac{L_\mathrm{\nu_\mathrm{e}} d\nu_\mathrm{e}}{4 \pi d^{2}_\mathrm{L}},\end{equation} 

where $d_\mathrm{L}$ is the luminosity distance, the subscript \emph{o} refers to quantities  in the observer's frame, and the subscript \emph{e} refers to quantities in the quasar rest-frame,  we can write the flux density as: 

\begin{equation}
f_\mathrm{\nu_\mathrm{o}} = \frac{L_\mathrm{\nu_\mathrm{o,e}}}{4 \pi d_\mathrm{L}^2 }\left[ \frac{f_\mathrm{\nu_\mathrm{e}}}{f_\mathrm{\nu_\mathrm{o,e}}} (1+z) \right]
\label{eq:kcorr}
\end{equation}

where the subscript {\it o,e} means that the observed and emission frequency are the same.
Across limited wavelength domains the AGNs spectra can be represented by  power laws: $L_\mathrm{\nu} \propto \nu^{-\alpha}$, where $\alpha$ is the spectral index at that frequency, and $L_\mathrm{\nu}$ is the luminosity. In this case, Eq. \ref{eq:kcorr} can be written as:

\begin{equation} f_\mathrm{\nu_\mathrm{o}}=\dfrac{L_\mathrm{\nu_\mathrm{o,e}}(1+z)^{-\alpha}(1+z)}{4 \pi d^{2}_\mathrm{L}}=f_\mathrm{\nu_\mathrm{o,e}}(1+z)^{1-\alpha}. \end{equation} 

The $f_{\nu_\mathrm{o,e}}$\ becomes
\begin{equation}
f_\mathrm{\nu_\mathrm{o,e}} = f_\mathrm{\nu,o}[(1+z)^{\alpha-1}]
 \end{equation} 

where the factor within square brackets is the K-correction. For the SDSS sources and the FIRST objects corresponding to an SDSS source we used $\alpha = 0.3$, for the radio lobes we used $\alpha = 1$  \citep[0.3 is the most frequent value measured for the sources of our sample for which radio spectra are available in the \citet{vollmer09} catalog][and references therein]{urrypadovani95}. Finally, we calculated $R_\mathrm{K}$ using the peak flux density at 1.4 GHz $f_\mathrm{1.4 GHz, P}$ for CD sources and the integrated flux density at 1.4 GHz $f_\mathrm{1.4 GHz, I}$ for FRII sources  acquired from the FIRST catalog. Henceforth, we drop the \textit{P} and \textit{I} subscripts for $f_\mathrm{1.4 GHz}$ keeping in mind that we use the peak flux density for the CD objects and the integrated fluxes for the FRII ones. It should be noted that in the case of FRII AGNs, the radio flux density is the flux density sum of all the radio components, core (if detected) and lobes. 

\subsection{Radio power and pseudo $SFR$}

From the flux density at 1.4 GHz we calculated a formal estimate of the star formation rate (hereafter pseudo-$SFR$, $pSFR$) of each object following \citet{yunetal01}: 

\begin{equation} p{SFR}_\mathrm{radio}=5.8 \times 10^{-22} P_\mathrm{1.4 GHz}, \end{equation} 

with $P_\mathrm{1.4 GHz}$ being the radio-power at 1.4 GHz: 

\begin{equation} \mathrm{log}(P_\mathrm{1.4 GHz})=20.08 + 2 \, \mathrm{log}d_\mathrm{L} + \mathrm{log}f_\mathrm{1.4 GHz}, \end{equation} 

 $d_\mathrm{L}$ the luminosity distance in Mpc, and $f_\mathrm{1.4 GHz}$ the flux density at 1.4 GHz in W/Hz. 

The physical basis of the connection between radio emission and $SFR$\ is the synchrotron emission associated with supernova remnants. As the supernova remnants cooling times are relatively short, radio emission can be considered as a measure of instantaneous $SFR$. Computing a $SFR$\ from the radio power is certainly incorrect in the case of ``jetted''  RL sources.  However, there are cases in which the radio  power is relatively high, and in which the dominant contribution is not associated with the relativistic jet but, indirectly, to star formation processes \citep[e.g., ][who found that 75\%\ of RL NLSy1s have WISE color indices  consistent with star formation]{caccianigaetal15}. We therefore compute the $pSFR$\ for all radio classes, being aware that this $pSFR$\ is   not meaningful in the case of ``jetted'' RL sources. A term of comparison is offered by the $SFR$\ computed from the FIR luminosity \citep[e.g., ][]{calzetti13a}.

\subsection{FIR power and K-correction}

To obtain an estimate of the $SFR$\ from the IR luminosity we used the local calibration derived by \citet{lietal10}:

\begin{equation}
SFR_\mathrm{70\mu m}  \approx 9.4  \cdot 10^{-44} L_\mathrm{70\mu m},
\end{equation}

with $SFR_\mathrm{70\mu m}$ in M$_\odot$/yr and $L_\mathrm{70\mu m}$ in erg/s, mostly derived  from the Herschel flux density.  

The observed flux needs to be K-corrected to obtain the rest-frame flux at   $70\mu m$. 
The spectral energy distribution in the FIR can be represented by a single, modified black body:

\begin{equation}
f_\nu \propto \frac{\nu^{3+\beta}}{\exp(h\nu/kT)-1},
\end{equation}

where $h$ is the Planck constant, $k$ the Boltzmann constant and $T$ the dust temperature \citep{smithetal14}. We assume $T \approx 60$ K because this is the temperature of the HII regions gas that correlates to the $SFR$\ in contrast with the galactic cirrus whose gas is at $T \approx  20$ K and  not connected to the $SFR$. The $\beta$ term, referred to as emissivity index, modifies the classic Planck function with the assumption that the dust emissivity varies as a power law of frequency. We assume a constant $\beta = 1.82$ following \citet{smithetal14}. Our goal is to get an  estimate of the $SFR$ for comparison with the $pSFR$\ derived from the radio power. The results can be found in Section 4.4. Even if a more consistent approach would have been the fitting of the IR SED with the various components emitting in the FIR spectral ranges, our estimates should be anyway fairly reliable because we expect that the main contributor to the flux at 70$\mu$ m, the dusty torus, has a steeply declining emission from a maximum at around 10-20$\mu$ m \citep[][and references therein]{durasetal17}.\footnote{\citet{durasetal17} point out that the only model that predicts 70$\mu$m emission from the torus at a level close to the maximum of the torus spectral energy distribution  is probably not correct { (see Fig.  Four torus emission models are in overall agreement. }}  The $K$-correction (Eq. \ref{eq:kcorr}) factor becomes: 
\begin{equation}
K_\mathrm{FIR} = \left[ \frac{f_\mathrm{\nu_\mathrm{e}}}{f_\mathrm{\nu_\mathrm{o,e}}} (1+z) \right] =   \frac{\nu_\mathrm{e}^{3+\beta}}{\nu_\mathrm{70\mu m}^{3+\beta}} \frac{\exp(h\nu_\mathrm{70\mu m}/kT)-1}{\exp(h\nu_\mathrm{e}/kT)-1} (1+z)
\end{equation}
where $\nu_\mathrm{e} = \nu_\mathrm{o}(1+z)$.  The $K$\ correction in the FIR is of order unity for the seven sources listed in Table \ref{tab:bonz}; all the observed fluxes are affected by   a factor in the range $\sim 0.8 - 2.4$.  

\subsection{Optical spectra analysis and derivation of physical parameters}
\label{datanopt}

The optical spectral study was done with the Image Reduction and Analysis Facility (IRAF) \footnote{http://ast.noao.edu/data/software}. The first steps involved the conversion to rest-frame wavelength and the flux scale of the spectra, using the $z$ values retrieved from the SDSS quasars catalog. 
Some sources show a different redshift with respect to that listed by
the SDSS. In this case, a better estimate was calculated from the position of the [OII]$\lambda \lambda$3726-3729 emission line doublet that provides a redshift identification that can minimize confusion with other single emission lines \citep[][Bon et al. in preparation]{comparatetal13}, or, if this doublet is not detected, from the H$\beta$ narrow component position. An accurate rest frame definition is relevant to estimate a correct physical interpretation of internal broad and narrow line shifts, which are an important part of the results presented in this paper.  After the $z$-correction, we normalized each spectrum to their continuum value at 5100 \textup{\AA}, that is, the mean over a region between 5050 and 5110 \textup{\AA}. Then, we measured the FWHM(H$\beta$) and its flux through a Gaussian fit of the emission line, and the \feii\ blend flux by marking two continuum points at 4440 and 4680 \textup{\AA} and summing the pixels in this range, considering partial pixels at the ends. In this way we obtained an estimate of \rfe.  

We computed the black hole (BH) mass of each source through the equation of \citet{shenliu12} using the scaling parameters of \citet{assefetal11}: 

\begin{eqnarray} 
\mathrm{log}(M_\mathrm{BH}/M_\mathrm{\odot}) & =& a + b \, \mathrm{log}(L_\mathrm{5100}/10^{44} \mathrm{erg \, s}^{-1} ) \\ \nonumber
& & + c \, \mathrm{log}(\mathrm{FWHM(H}\beta)/ \mathrm{km \, s}^{-1}), \end{eqnarray} 

where $a = 0.895$, $b = 0.520$, $c = 2.0$, and $L_\mathrm{5100}$ is the luminosity at 5100 \textup{\AA} in erg s$^{-1}$, $L_\mathrm{5100}=4 \pi \, d^{2}_\mathrm{c} \, (\lambda f_\mathrm{\lambda})$,  with $d_\mathrm{c}$ being the radial comoving distance, $\lambda=5100$ \textup{\AA} and $f_\mathrm{\lambda}$ the continuum flux density at the quasar rest-frame. The Eddington ratio follows from: 
 
 \begin{equation} \lambda_\mathrm{E} = \dfrac{L_\mathrm{}}{L_\mathrm{E}} \approx \dfrac{10 \, L_\mathrm{5100}}{1.5\times 10^{38}\, (M_\mathrm{BH}/M_\mathrm{\odot})}, \end{equation} 
 
where $L_\mathrm{} \approx 10 \, L_\mathrm{5100}$ is the bolometric luminosity and $L_\mathrm{E}$ is the Eddington luminosity, the maximum luminosity allowed for objects powered by  steady-state accretion \citep[e.g.,][]{netzer13}. From the FWHM(H$\beta$) and \rfe\ values, we classified the objects in the E1 spectral types. Higher S/N composite spectra (compared  to individual spectra)  were computed as median composite spectra.
The next step was the creation of a model, through a non-linear $\chi^{2}$\ minimization procedure (in IRAF with the package {\tt specfit}, \citealt{kriss94}), fitting the composite spectrum of each class in the H$\beta$ wavelength range from 4430 to 5450 {\AA}. The models for Pop. A and Pop. B AGNs involve a similar number of components with the following constraints:

\begin{itemize} 
\item Continuum was fit with a power-law for both populations. 
\item Balmer lines were fit in different ways based on the population of the source. For Pop. A objects, the H$\beta$ line was fit with a Lorentzian profile, to account for the broad-component (BC), with a Gaussian profile, to account for the narrow-component (NC), and with a second Gaussian profile, called blueshifted component (BLUE), if needed, to account for a strong blue excess especially seen in sources with a high $R_\mathrm{FeII}$. For Pop. B objects, the line was fit with three Gaussian profiles, to account for the broad, narrow, and very-broad (VBC) components. The broader VBC may be associated with a distinct emitting region referred as very-broad line region (VBLR; e.g., \citealt{sulenticetal00c,sneddengaskell07,wangli11}). 
\item\, \oiiiopt\ doublet -- For Pop. B sources, two Gaussian profiles were used to fit each doublet line, one to account for the blue excess of the line. For Pop. A objects, each doublet line was fit with three Gaussian profiles. 
\item The \feiiq\ emission was fit through a template based on I Zw 1 \citep{borosongreen92,marzianietal03a}.
\end{itemize}
Finally, each E1 class composite spectrum was analyzed measuring the equivalent width ($W$), FWHM, and centroid shifts, $c(x)$, of H$\beta$ and \oiii\ emission lines.  The centroid shifts were defined as proposed by \citet{zamfiretal10}: 

\begin{equation} c(x) = \dfrac{v_\mathrm{r,R}(x) + v_\mathrm{r,B}(x)}{2} \end{equation} 

where $x$ is a specific height of the profile (we considered 1/4, 1/2 and 9/10), $v_\mathrm{r,R}$ and $v_\mathrm{r,B}$ refer respectively to the velocity shift on the red and blue wing in the rest-frame of the quasar. The interpretation of centroid shifts is mainly based on the Doppler effect due to gas motion with respect to the observer, along with selective obscuration \citep[e.g.,][]{marzianietal16,negreteetal18}.

\section{Results}
\label{res}

\begin{figure*}[htbp!]
\centering
\includegraphics[width=.95\textwidth]{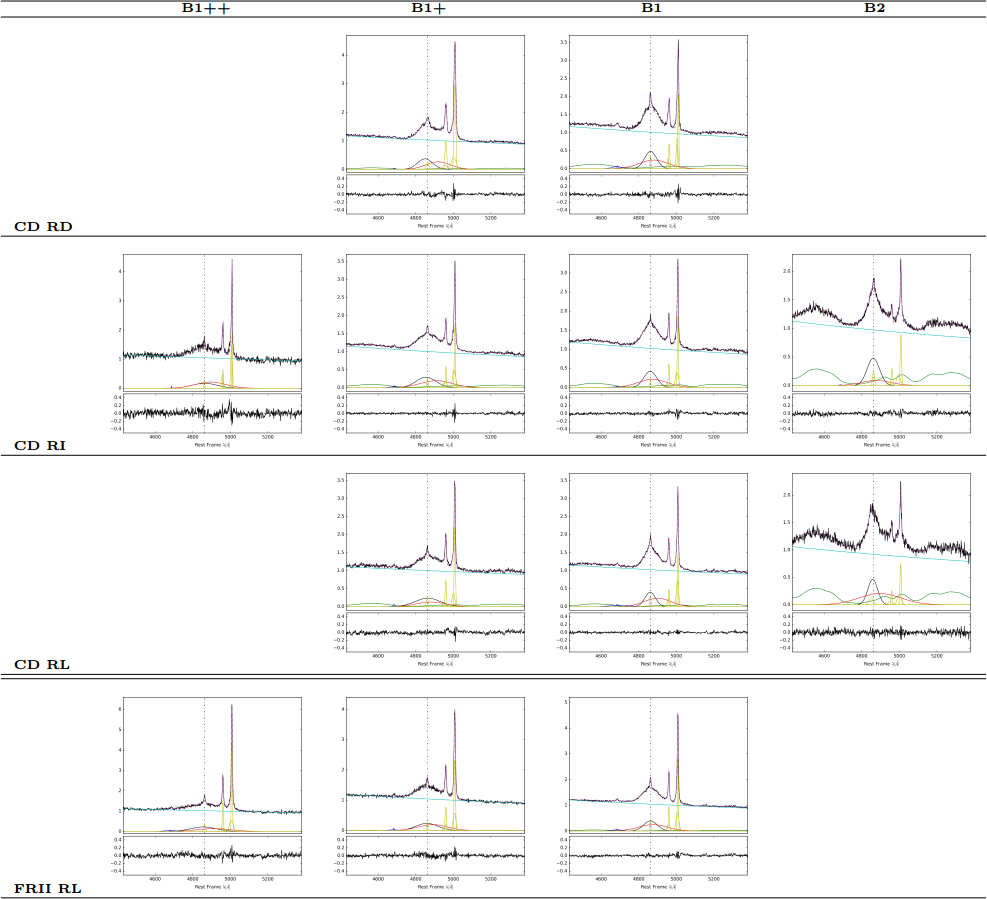}\\[1.em]
\caption{Composite of spectral types along the MS, ordered from B1$^{++}$ to B2. Each panel shows the original composite, along with the modeled continuum (cyan line). The lower part of the panels shows the result of the {\tt specfit} analysis on the emission lines. \hbbc: black line, \hbvbc: red line; \feii: green line; yellow lines: \hb\ narrow components and \oiiiopt.} 
\label{fig:sp}
\end{figure*}

\begin{figure*}[htbp!]
\centering
\includegraphics[width=.95\textwidth]{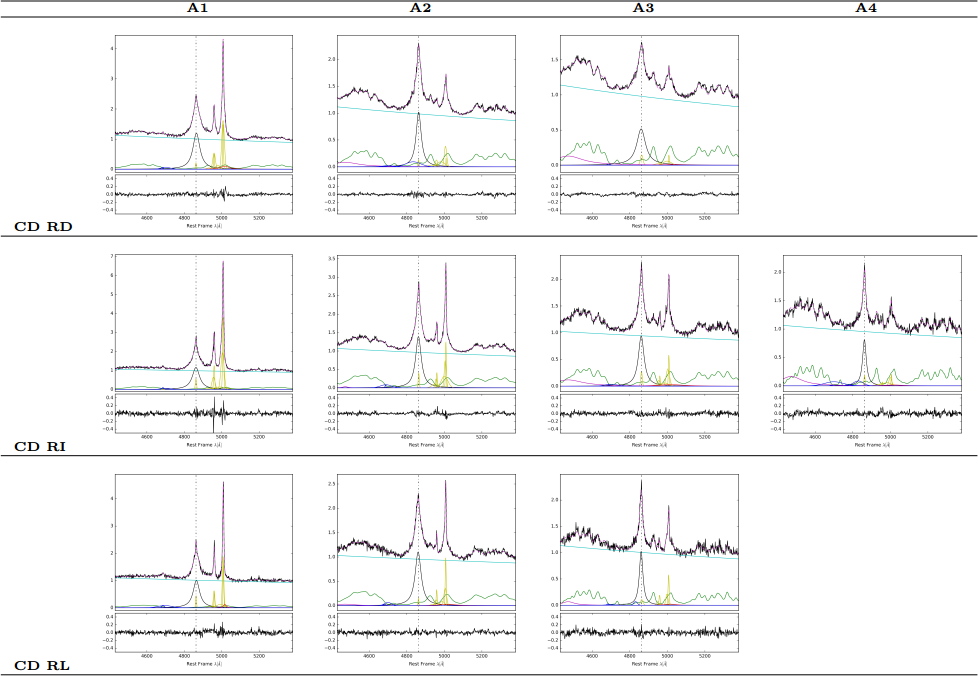}\\[1.em]
\caption{Composite of spectral types along the MS, ordered from A1 to A4. Color scheme same as in Fig. \ref{fig:sp}.} 
\label{fig:spa}
\end{figure*}

Core-dominated sources are identified in Table \ref{tab:CDphys} of Appendix \ref{CDphys}.  Table \ref{tab:CDphys} reports SDSS identification,   redshift, $g$\ band specific flux in mJy, peak flux density at 1.4 GHz in mJy, integrated flux density at 1.4 GHz in mJy if $f_\mathrm{1.4 GHz,I} \geq f_\mathrm{1.4 GHz,P}$, the decimal logarithm of the Kellermann's ratio $\log$\rk, the decimal logarithm of rest-frame continuum luminosity at 5100 \AA, the decimal logarithm of radio power at 1.4GHz, the MS parameters FWHM \hb\ and \rfe, along with a final classification code that includes radio-loudness class, radio morphology and optical spectral type. Borderline and intriguing cases of non-compact radio morphology are discussed in the footnotes of  Table \ref{tab:CDphys}. FRII required a careful identification following the method described in Sect. \ref{datanrad}, with an a-posteriori vetting to exclude spurious features.  Table \ref{tab:FRII} identifies the optical nucleus and the radio features (lobes) that are not coincident with the optical nucleus for the FRII radio morphology. It provides, in the following order, the SDSS identification, the right ascension and declination of the optical  core (Col. 2 -- 3), right ascension, declination, and separation of the first radio lobe (Col. 4 -- 6). The same information is repeated for the second radio feature in Col. 7 -- 9. Col. 10 provides the angle between the two lobes, with vertex on the nucleus, computed as described in Sect. \ref{datanrad}. An Atlas of the FRII sources identified in this work is provided in Appendix \ref{atlas}.

\subsection{Trends along the MS}

The numbers of sources in each MS spectral bin, for each radio-morphology class (CD, FRII), and for each radio-loudness emission range (RD, RI, RL) are reported in Table \ref{tab:E1-sample}. There are a total of 38 RD, 139 RI and 178 RL objects.

\subsubsection{Source distribution along the MS}

 Table \ref{tab:E1-sample} shows that the most populated bins are the Pop. B ones, especially B1 and B1$^{+}$ bins, at variance with the optically-selected sample in Fig. \ref{fig:e1r} where the most populated bins are B1 and A2.   The occupation of the spectral bins decreases from type A1/A2 to A3/A4, with a fraction of A3+A4,  $\sim 25$\% of the Pop. A ( $\sim 29$ \%\ if A5 sources are included), a prevalence than the one  measured on the optically-selected sample of Fig. \ref{fig:e1r}: the A3+A4 fraction of Pop. A sources is $\sim$0.085/0.457 $\approx$ 19 \%.  The distribution of sources in the various STs depends on both radio loudness and radio morphology. 


\begin{itemize}
\item  RD sources are present only with a CD radio morphology. Population A and B are similar in number (21 A and 17 B) but Population B sources with a R$_\mathrm{FeII}$ parameter higher than 0.5 are very rare (only three objects are in B2 and B3) in contrast with Population A objects. 
\item   RI sources also are present only with a CD radio morphology. They are more numerous in Population B than in A (96 in B and 43 in A) in contrast to the RD class. However, xA bins are populated in a significant way by CD objects, and 63\% of all the xA CD sources are RI.   
\item  RL objects are more numerous in population B for the CD radio morphology and they dominate in the FRII one (81 CD in B and 31 CD in A, all FRII in the sample are RL). RL (CD+FRII) in B1 and B1$^{+}$ accounts for the wide majority of RL sources (2/3 of the total, Tab. \ref{tab:E1-sample}). The FRII radio morphological class appears only in the RL domain in contrast with the CD one where all the three radio-loudness classes are significantly populated.
\end{itemize}

In summary, considering ST B1 and B1$^{+}$,  all FRII, and the wide majority of CDs ($\sim 90 \%$)  are in the RI and RL ranges only. Moreover, FRII are a sizable number  in ST B1 and B1$^{+}$, but almost completely absent in Population A. Only three RL FRII sources appear in bin A1. 

\begin{table}[htbp]
\centering
\caption[Sources binning in EV1 classes.]{Number of sources in each spectral type and radio-loudness  range. Bold numbers refer to classes for which a model fit was created. \label{tab:E1-sample}}
\begin{tabular}{lcccccc}
\hline\hline
\multicolumn{1}{c}{\textbf{}} & \multicolumn{3}{c}{\textbf{CD}} & \multicolumn{3}{c}{\textbf{FRII}}\\ 
\hline
\multicolumn{1}{c}{\textbf{}} & 
\multicolumn{1}{c}{\textbf{RD}} & \textbf{RI} & \textbf{RL} &
\multicolumn{1}{c}{\textbf{RD}} & \textbf{RI} & \textbf{RL}\\
\hline
A5     & -           & 2           & -           & - & - & -  \\
A4     & 1           & \textbf{5}  & 1           & - & - & -  \\
A3     & \textbf{4}  & \textbf{10} & \textbf{4}  & - & - & -  \\
A2     & \textbf{12} & \textbf{14} & \textbf{10} & - & - & -  \\
A1     & \textbf{4}  & \textbf{12} & \textbf{16} & - & - & 3  \\
B1     & \textbf{7}  & \textbf{35} & \textbf{47} & - & - & \textbf{39} \\
B1$^{+}$    & \textbf{7}  & \textbf{34} & \textbf{21} & - & - & \textbf{15}  \\
B1$^{++}$   & -           & \textbf{9}  & 2           & - & - & \textbf{8}  \\
B1$^{+++}$  & -           & 2           & 2           & - & - & 1  \\
B2     & 2           & \textbf{12} & \textbf{6}  & - & - & -  \\
B2$^{+}$    & -           & -           & 1           & - & - & -  \\
B2$^{++}$   & -           & 1           & 1           & - & - & -  \\
B3     & 1           & 3           & 1           & - & - & -  \\
Total  & 38          & 139         & 112         & - & - & 66 \\

\hline
\label{tab:1}
\end{tabular}
\end{table}%

\subsubsection{H$\beta$}

Figures \ref{fig:sp} and \ref{fig:spa} shows the composite spectra fits, organized along the MS, for each radio morphology and loudness class. The spectra of the E1 classes (from left to right) confirm the systematic differences between Pop. A and B summarized in Sect. \ref{intro}. Regarding the H$\beta$ line profile, we note that Pop. A sources have a broad H$\beta$ component best fit by a Lorentz function whereas the Pop. B ones are best fit with two Gaussians.  This was expected and confirms results of previous studies as far as the Lorentzian profiles of the sources with the narrowest profiles are concerned \citep[e.g., ][]{veroncettyetal01,sulenticetal02,craccoetal16}.

In addition, looking at Pop. B sources, going towards higher values of FWHM(H$\beta$) we note that the peaks of the broad and very broad component of H$\beta$ become increasingly more  separated. The trend is very clear passing from B1 to B1$^{+}$ for CD RD and RI, as well as for FRII RL.   Several physical processes could be at the origin of this effect: an infalling gas component, gravitational redshift, or a binary BLR associated with a binary black hole (BBH), with each black hole giving rise to a single and distinct broad component. These possibilities will  be briefly discussed in Section \ref{hbtrends}.

Complementary information independent from line profile decomposition can be inferred from the lines centroid shifts and equivalent width behavior shown in Table \ref{tab:4} and in Figures \ref{fig:cd-hb-shifts} for \hb. Figure \ref{fig:cd-hb-shifts} shows that in the CD class the value of $c(9/10)$, $c(1/2)$ and $c(1/4)$ generally increases going from the Pop. A bins to the B1$^{++}$. This translates in a redshifted line for Pop. B sources, more redshifted for larger FWHM(H$\beta$) values, and a slightly blue shifted or unshifted line for Pop. A sources. The only exception to this trend is CD RD B1$^{+}$, with \hb\ showing a small peak blueshift ($ c(0.9) \lesssim -100$ \kms).  Moreover, while in Pop. B bins  $W$(H$\beta$) does not vary significantly, at the boundary between Pop. B and Pop. A $W$(H$\beta$) begins to decrease towards bins with higher $R_\mathrm{FeII}$, reaching a minimum in correspondence of spectral type A3.

\begin{figure}
\begin{center}
\includegraphics[width=.4\textwidth]{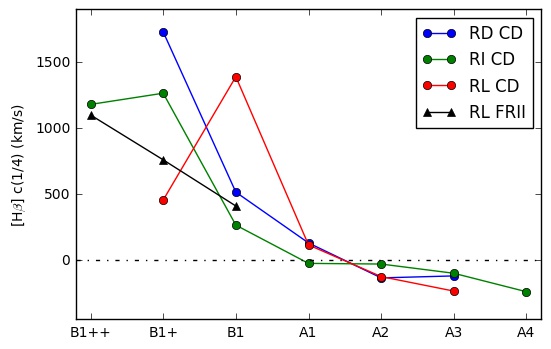}
\quad
\includegraphics[width=.4\textwidth]{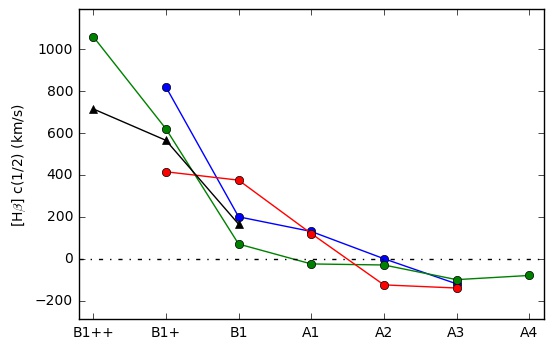}\\
\includegraphics[width=.4\textwidth]{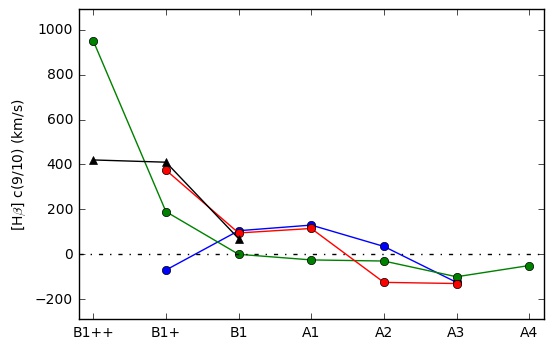}
\quad
\includegraphics[width=.39\textwidth]{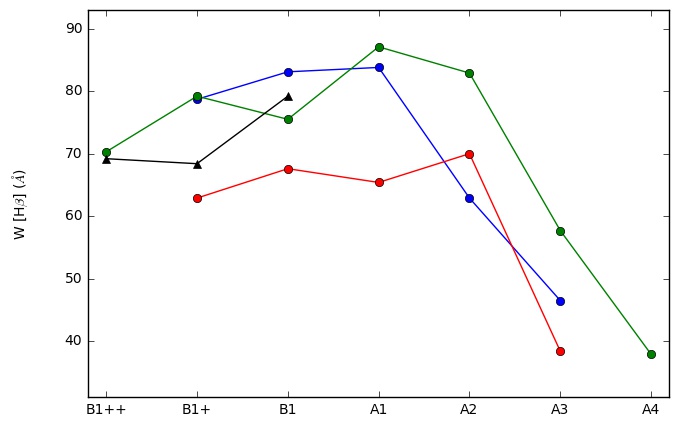}
\caption{CD H$\beta$ centroid shifts (km/s) for different E1 classes and for different line height. From top to bottom: $x=1/4$, $x=1/2$, $x=9/10$. Bottom: equivalent width (\textup{\AA}) distribution. Blue refers to the RD class, green to the RI one, red to the RL one and black to the RL FRII class.}
\label{fig:cd-hb-shifts}
\end{center}
\end{figure}

\subsubsection{[OIII]: equivalent width anti-correlation only for RD}

The [OIII] doublet becomes increasingly weaker in Pop. A going towards bins with  higher R$_\mathrm{FeII}$ values until the line almost disappears in A3 and A4 bins (Fig. \ref{fig:sp}, \ref{fig:spa} and Fig. \ref{fig:cd-oiii-shifts}}). This anti-correlation is one of the well-known and strongest trends of the quasar MS \citep{borosongreen92,bolleretal96,sulenticetal00a}.  However,  only the RD class   follows the anti-correlation, with the [OIII] lines becoming indistinguishable from the FeII blend in the xA bins. RD sources behave, in this respect, similar to RQ sources.  The $W$([OIII]) of Pop. B composites does not vary strongly, whereas, after an increase in the A1 bin, the $W$([OIII]) decreases going towards bins with higher $R_\mathrm{FeII}$.  Fig. \ref{fig:oiiibasecore} shows that the W(\oiii) behavior is somewhat different for the base semi-broad component and for the core component. The base component changes within a limited range of values (a factor $\sim 2$ in the Pop. A ST for RL and RD composites that show $W$([OIII]) in the range  3 --  7; a factor $\sim$ 5 in ST A2 and $\sim$ 2 in ST A1 for the CD RI);  the core component, after a sudden jump for ST A1 that is less pronounced in the RD and RI classes with respect to the RL one, systematically decreases in all radio loudness classes. The core component is practically null from ST A3 in the RD class whereas in the RI and RL ones is still present. This creates the impression of a fairly constant \oiii\ in RL (Fig.  \ref{fig:spa}).   The large $W$([OIII]$\gtrsim 10$ \AA) of the base component in spectral  types B1$^{+}$ and B1$^{++}$ is apparently surprising, but the base component has a small shift with respect to the core component, so that the decomposition core/base is, in this case, ill-defined.   The dependence on radio classes implies a significant strength of the NC among RL and RI CD compared to RD. This may be a genuine  effect associated with RLness, not detected earlier because of the small number of RL if ST types are isolated along the sequence. 

Figure \ref{fig:cd-oiii-shifts} shows that in the CD class the values of    $c(1/4)$\ become increasingly negative from Pop. B to A,  reflecting an increasing blueshift toward the line base. The blueshifting of the [OIII] wing component, represented by  $c(1/4)$, is very common in AGNs but not every AGNs shows a fully blueshifted line, that is, a negative $c(x)$ for all $x$. CD RL  A2 and A3 show a systematic blueshift at 1/2 and 0.9 fractional intensity, although the blueshift amplitude is modest $\sim 100$ \kms.  The most significant is the one of the RI A4 bin with the strongest \feii. 


\begin{figure}
\begin{center}
\includegraphics[width=.4\textwidth]{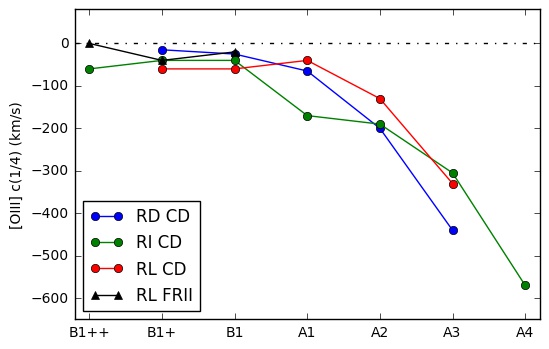}
\quad
\includegraphics[width=.4\textwidth]{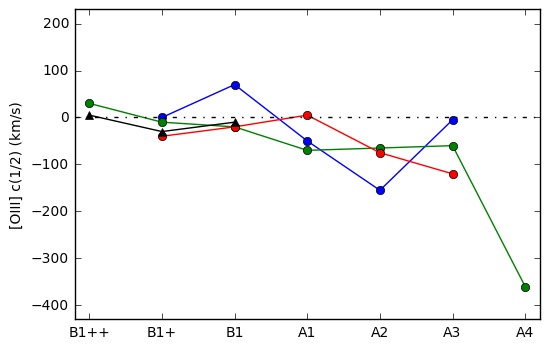}\\
\includegraphics[width=.4\textwidth]{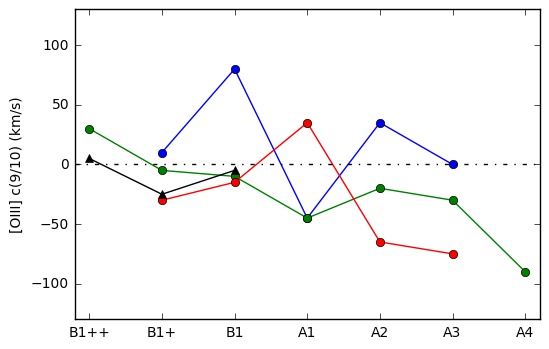}
\quad
\includegraphics[width=.39\textwidth]{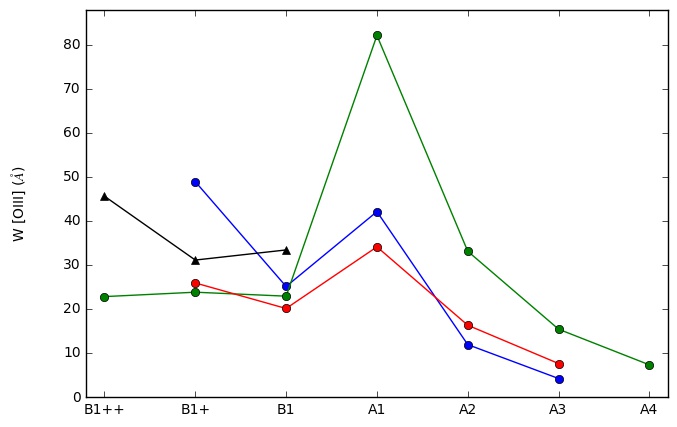}
\caption{CD [OIII] centroid shifts (km/s) for different E1 classes and for different line height. From top to bottom: $x=1/4$, $x=1/2$, $x=9/10$. Bottom: equivalent width (\textup{\AA}) distribution. Blue refers to the RD class, green to the RI one and red to the RL one.}
\label{fig:cd-oiii-shifts}
\end{center}
\end{figure}

\begin{figure}[htbp!]
\centering
\vspace{0cm}
\includegraphics[width=0.4\textwidth]{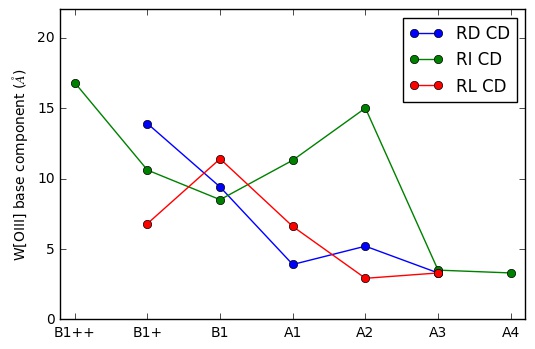}
\includegraphics[width=0.4\textwidth]{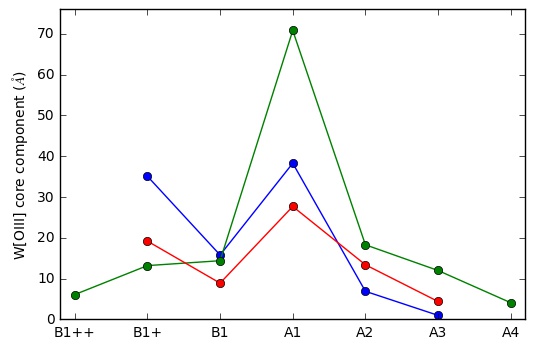}
\caption{Equivalent width of \oiii\ as a function of spectral types for CD RD, RI, RL sources. The upper panel refers to the base, semi-broad component, the bottom one to the narrower core of the \oiii\ line. } 
\label{fig:oiiibasecore}
\end{figure}


  \begin{table}[!htbp]
\begin{tiny}
\begin{center}
\caption{xA sources \label{tab:xA}}
\medskip
\begin{tabular}{ll}
\hline\hline
\multicolumn{1}{l}{\textbf{SDSS ID}} 
& \multicolumn{1}{l}{\textbf{Class}} \\
\hline

J080037.62+461257.9 	& RI CD A3 \\
J010123.42+005013.3 	& RL CD A3 \\
J024817.69+024128.4 	& RI CD A3 \\
J071933.35+403253.0 	& RI CD A3 \\
J081929.48+522345.2 	& RI CD A5 \\
J083558.43+261444.4 	& RL CD A4 \\
J094248.08+112934.3 	& RI CD A3 \\
J095150.49-025545.5 	& RI CD A3 \\
J095633.93+562216.0 	& RD CD A4 \\
J101952.59+073050.8 	& RI CD A4 \\
J102818.15+535113.6 	& RI CD A4 \\
J103346.39+233220.0 	& RL CD A3 \\
J104011.18+452125.9 	& RD CD A3 \\
J114201.83+603030.4 	& RI CD A4 \\
J114339.53+205921.1 	& RI CD A4 \\
J114915.30+393325.4      & RI CD A3 \\
J120910.61+561109.2 	& RI CD A4 \\
J121231.47+251429.1 	& RI CD A3 \\
J123640.35+563021.4 	& RI CD A5 \\
J124511.25+335610.1 	& RD CD A3 \\
130631.63+435100.4 	& RD CD A3 \\
J132146.53+265150.1 	& RD CD A3 \\
J132819.23+442432.9 	& RL CD A3 \\
J142549.19+394655.0 	& RL CD A3 \\
J163345.22+512748.4 	& RI CD A3 \\
J170300.48+410835.8 	& RI CD A3 \\
J171749.62+253908.7	& RI CD A3 \\
\hline
\end{tabular}
\end{center}
\end{tiny}
\end{table}

\subsection{Dependence on radio classes for a fixed spectral type/population}

\subsubsection{H$\beta$}

\begin{figure}
\begin{center}
\includegraphics[width=.40\textwidth]{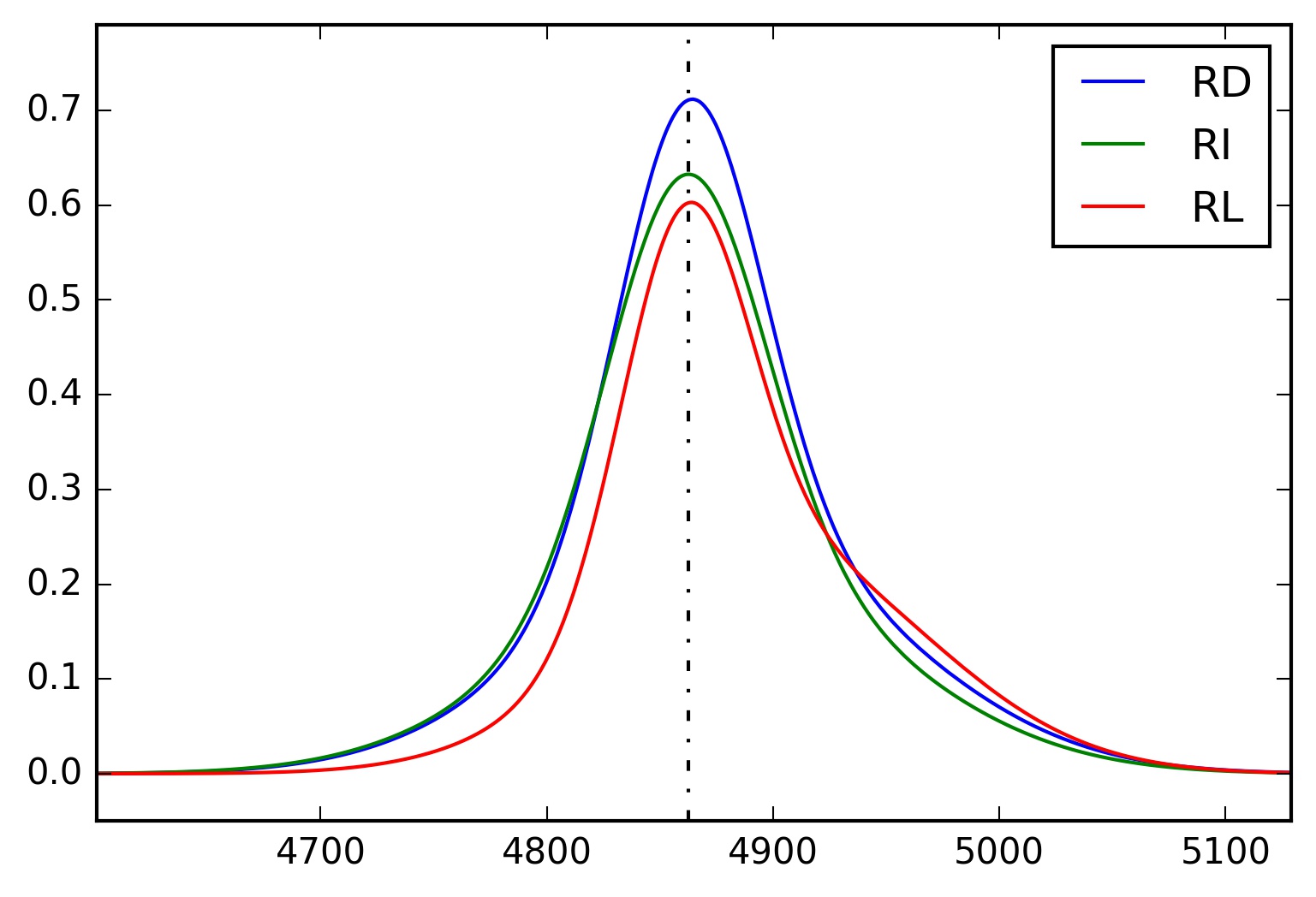}\qquad
\includegraphics[width=.41\textwidth]{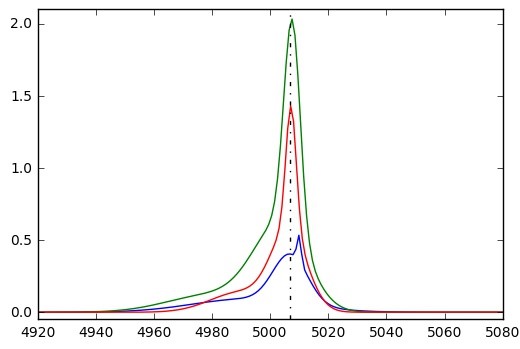}
\quad
\caption{CD B1 H$\beta$ (top) and CD A2 [OIII] profiles (bottom) for different radio power classes. Blue refers to the RD class, green to the RI one and red to the RL one.}
\label{fig:cd-shifts}
\end{center}
\end{figure}

Interesting findings can be seen by comparing the different radio-loudness emission classes within each ST or at least in Pop. A and B.  Generally speaking, for all spectral types, going from the RD to the RL class,  the line shape and intensity do not vary significantly, as can be seen from the spectra in Fig. \ref{fig:sp} and \ref{fig:spa}.  The Pop. B \hb\ profiles are always Gaussian-like. Pop. A \hb\ profiles can be well-fit by a Lorentzian function independently from  radio class. In all STs, the \feii\  template reproduces equally well the \feii\ emission across all radio classes, within the limits in S/N and dispersion. An important inference is that the properties of the low-ionization BLR emitting most of \hb\ are little affected by radio loudness.  


\paragraph{Pop. A} In  bin A1, the \hb\ line profiles are intrinsically symmetric independently on radio loudness (Table \ref{tab:4}), and the apparent redward excess can be explained by the combined effect of optical \feii\ blends and \oiiiopt\ semi-broad emission. Modest shifts to the blue appear in bin A2, and are apparently stronger for RL CD.  In the A3 bin, both the RD and RI \hb\ profiles show a blueshift. A blue shifted excess for \hb\ in the RL CD A3 composite is reflected by the increase in blueshift amplitude toward the line base. The RI CD A4 \hb\ profile is also consistent with a blueshifted excess toward the line base (Fig. \ref{fig:spa}), although the line peak for all radio classes and ST remains unshifted.  

\paragraph{Pop. B} From Figure \ref{fig:cd-hb-shifts} we note that line centroids do not vary qualitatively in the different CD radio-loudness classes with the exception of the B1$^{+}$ bin where the RD class shows a  H$\beta$ profile redshifted at the base and at half height and slightly  blueshifted at the peak.  RD, RI and RL show a prominent red wing, as can be seen  in Fig. \ref{fig:cd-shifts} where the profiles of \hb\ for the CD B1 radio classes are shown. The redward displacement of $c(1/4)$ is maximum for spectral type B1 for RL CD and for RD CD B1$^{+}$. In the FRII class, whose objects mostly belong to Pop. B, the H$\beta$\ line  is redshifted in all composites (Fig. \ref{fig:cd-hb-shifts}). The redshift increases with increasing FWHM(H$\beta$), as also found  for the CD radio morphology.

\subsubsection{\oiiiopt}


\paragraph{Pop. A}  The \oiiiopt\ lines  are usually asymmetric in correspondence of the line base, from the A1 to the xA bins.   The \oiii\ line is slightly blueshifted and asymmetric in spectral type A1 if RD and RI classes are considered.  Table  \ref{tab:4} reports a small blueshift $\sim -50$ \kms\ that apparently disappears for the CD RL class ST A1, the only case where the \oiii\ profile it is perfectly symmetric.  The blueward asymmetry of the ST A2 and A3 composite \oiii\ profiles is instead large due to centroid at 1/4 shifts of $\sim 400 $ \kms\ for the three radio classes  (Table \ref{tab:4}  and Fig. \ref{fig:cd-oiii-shifts}) . 

Sources that shows a blueshift higher than 250 \kms\ at peak have been termed  blue outliers and are common in the xA class \citep{zamanovetal02,komossaetal08,negreteetal18}. In the xA bins almost the whole line is blueshifted, although the blueshift is modest at peak and none of the composite spectra  would qualify as a blue outlier following \citet{zamanovetal02}: shifts are large close to the line base ($\lesssim -300$ \kms) but tend to 0 or to a small blueshift value in correspondence of the line peak. In A2 and A3 there might be a trend as a function of the radio class as far as the amplitude of the \oiii\ blueshift close to line peak is concerned (Fig. \ref{fig:spa}), with the \oiii\ peak blueshift increasing from RD to RL, but with a shift amplitude $< 100$ \kms.  Only RI CD A4 comes close to a blue outlier ($c(1/4) \approx -600$ \kms, $c(1/2) \approx -350$ \kms, $c(1/4) \approx -100$ \kms)  below the conventional limit of $-250$ \kms at the peak, which may imply that some of the sources of this ST are indeed blue outliers.  The RI CD A4 sources are listed in Table \ref{tab:xA} that identifies all the xA sources of the sample and provides their classification.   A study of individual   spectra is required, as the sources listed in Table \ref{tab:xA} with classification CD RI A4 should be considered  blue outlier candidates. 


 In the present sample, the \oiii\ blueshift at quarter maximum is only slightly larger in RD than in RI and RL for ST A3 and A4. The shift and to some extent the equivalent width of base component in [OIII] is not strongly dependent on the radio-class. Especially at 1/4 fractional intensity the shifts remain relatively high amplitude due to an appreciable semibroad component. (Tab. \ref{tab:4} and Fig. \ref{fig:spa}). This behavior is at variance with the one observed in the \oiiiopt\ lines in Pop. B, as far as the amplitude of the shifts is concerned. In Pop. B, we observe a  slight increase of the blueshifts toward the line bases, but the amplitude is $\lesssim 100$ \kms.  It is interesting to note that, for the  resonance high ionization  \civ\ line (whose blueshift is related to \oiii\ \citealt[e.g.,][and references therein]{coatmanetal19}),  RL samples show systematically lower blueshift amplitudes than RQ \citep{richardsetal11,marzianietal16a}.  A possible interpretation resides in the interplay between the accretion disk wind and the relativistic jet, with the latter hampering the development of a high-speed wind. 

\citet{bertonetal16} found that RL NLSy1s show more prominent \oiii\ blueward displacements than RQ ones. Considering that NLSy1s show the narrowest \hb\ profiles among Population A (and are completely absent by definition in bin B1 where the powerful jetted CDs are found), their results is not necessarily in contradiction with ours, as large amplitude shifts close to line base are also found in our sample of RI and  RL CDs. The issue is whether the sources in our sample could be assimilated to the RL Pop. A of  \citet{bertonetal16}.  Unfortunately only 1 source out of 61 is in common, \object{SDSSJ135845.38+265808.4}, from the \citet{foschinietal15} sample of flat-spectrum NLSy1s. Its classification is CD RI A2, shows a Lorentzian H$\beta$\ broad profile, and relatively strong \oiiiopt. The \oiii\ line has a prominent semi-broad component, whose   shift (tentatively measured on the DR12 Science Archive Server) is $\sim -1000$ \kms. The properties are consistent with the  properties of RQ sources in this spectral bin, and are akin to the xA RI. The WISE color index $W3-W4 >$2.5 indicates the presence of a starburst component \citep{caccianigaetal15}, suggesting a ``thermal'' origin for the radio power (Sect. \ref{radfir}). 
\paragraph{Pop. B} The \oiii\ line can be considered almost symmetric and unshifted all across the radio classes in Pop. B for both CD and FRII classes; the trend of an increase in blueshift toward the line base is preserved although Table \ref{tab:4} reveals that now the shifts at the line base are $\gtrsim -100$ \kms. In other words, CD RQ, RL and RI Pop. B show a similar behavior at 1/4 maximum, with a modest blueshift, and no significant shift at peak.

\begin{table*}[htbp]
\begin{tiny}
\begin{center}
\caption[H$\beta$ and \[OIII\] properties of each spectral type]{H$\beta$ and [OIII] properties of each spectral type, radio-power and radio morphology class. \label{tab:5}}
\medskip
\begin{tabular}{llcccccccc}
\hline\hline
\multicolumn{1}{c}{} & \multicolumn{1}{c}{H$\beta$} & \multicolumn{1}{c}{\textbf{B1++}} & \multicolumn{1}{c}{\textbf{B1+}} & \multicolumn{1}{c}{\textbf{B1}} & \multicolumn{1}{c}{\textbf{A1}} & \multicolumn{1}{c}{\textbf{A2}} & \multicolumn{1}{c}{\textbf{A3}} & \multicolumn{1}{c}{\textbf{A4}} & \multicolumn{1}{c}{\textbf{B2}} \\
\hline
\multirow{5}*{\textbf{CD RD}}   & EW         & - & 78.7 & 83.1 & 83.8 & 62.9 & 46.5 & -  & -                        \\
                                & FWHM       & - & 8585 & 5575 & 2780 & 2180 & 3500 & - & -          \\
                                & c($1/4$)   & - & 1730 & 515 & 130 & -135 & -120 & -   & -          \\
							    & c($1/2$)   & - & 820 & 200 & 130 & 0 & -120 & -   & -               \\
                                & c($9/10$)  & - & -70 & 105 & 130 & 35 & -125 & -    & -             \\
\hline					                                                                        
\multirow{5}*{\textbf{CD RI}}   & EW         & 70.3 & 79.2 & 75.5 & 87.1 & 82.9 & 57.7 & 37.9  & 58.4                       \\
                                & FWHM       & 11215 & 9650 & 6165 & 2905 & 2255 & 2275 & 1545 & 5280  \\
                                & c($1/4$)   & 1180 & 1265 & 265 & -25 & -30 & -100 & -240 & 40        \\
							    & c($1/2$)   & 1060 & 620 & 70 & -25 & -30 & -100 & -80 & -20           \\
							    & c($9/10$)  & 950 & 190 & 0 & -25 & -30 & -100 & -50 & -50             \\
\hline					                                                                        
\multirow{5}*{\textbf{CD RL}}   & EW         & - & 62.9 & 67.6 & 65.4 & 70.0 & 38.4 & -    & 71.3                   \\
                                & FWHM       & - & 10030 & 5485 & 2605 & 2390 & 1430 & - & 4170         \\
                                & c($1/4$)   & - & 455 & 1390 & 115 & -125 & -235 & - & 1420             \\
							    & c($1/2$)   & - & 415 & 375 & 120 & -125 & -140 & - & -100            \\
							    & c($9/10$)  & - & 375 & 95 & 115 & -125 & -130 & - & -265              \\
\hline                                                                                                                                                                              
\multirow{5}*{\textbf{FRII RL}} & EW         & 69.2 & 68.4 & 79.2 & - & - & - & -      & -                 \\
                                & FWHM       & 11660 & 9610 & 6650 & - & - & - & -  & -        \\
                                & c($1/4$)   & 1100 & 760 & 410 & - & - & - & -    & -           \\
							    & c($1/2$)   & 715 & 565 & 165 & - & - & - & -  & -              \\
							    & c($9/10$)  & 420 & 410 & 70 & - & - & - & -   & -              \\
\hline
\hline	
\multicolumn{1}{c}{} & \multicolumn{1}{c}{OIII} & \multicolumn{1}{c}{\textbf{B1++}} & \multicolumn{1}{c}{\textbf{B1+}} & \multicolumn{1}{c}{\textbf{B1}} & \multicolumn{1}{c}{\textbf{A1}} & \multicolumn{1}{c}{\textbf{A2}} & \multicolumn{1}{c}{\textbf{A3}} & \multicolumn{1}{c}{\textbf{A4}} & \multicolumn{1}{c}{\textbf{B2}} \\
\hline
\multirow{5}*{\textbf{CD RD}}      & EW         & - & 49.0 & 25.2 & 42.2 & 12.0 & 4.3 & - & - \\
                                   & FWHM       & - & 710 & 520 & 530 & 790 & 455 & - & - \\							
                                   & c($1/4$)   & - & -15 & -25 & -65 & -200 & -440 & - & - \\	
                                   & c($1/2$)   & - & 0 & 70 & -50 & -155 & -5 & - & - \\		
                                   & c($9/10$)  & - & 10 & 80 & -45 & 35 & 0 & - & - \\
\hline	                                                                                      
\multirow{5}*{\textbf{CD RI}}      & EW         & 22.9 & 23.9 & 23.0 & 82.2 & 33.2 & 15.5 & 7.4 & 13.3 \\
                                   & FWHM       & 290 & 470 & 480 & 550 & 515 & 520 & 840 & 535  \\
                                   & c($1/4$)   & -60 & -40 & -40 & -170 & -190 & -305 & -570 & -55  \\
            		               & c($1/2$)   & 30 & -10 & -20 & -70 & -65 & -60 & -360 & -20 \\
                                   & c($9/10$)  & 30 & -5 & -10 & -45 & -20 & -30 & -90 & -10  \\
\hline                                                          
\multirow{5}*{\textbf{CD RL}}      & EW         & - & 26.0 & 20.2 & 34.2 & 16.4 & 7.7 & - & 12.5 \\
                                   & FWHM       & - & 515 & 435 & 440 & 440 & 490 & - & 580 \\
                                   & c($1/4$)   & - & -60 & -60 & -40 & -130 & -330 & - & -115 \\
            			           & c($1/2$)   & - & -40 & -20 & 5 & -75 & -120 & - & -75  \\
                                   & c($9/10$)  & - & -30 & -15 & 35 & -65 & -75 & - & -65 \\
\hline                                                           				                           
\multirow{5}*{\textbf{FRII RL}}      & EW         & 45.7 & 31.2 & 33.5 & - & - & - & - & - \\
                                     & FWHM       & 430 & 555 & 440 & - & - & - & - & - \\
                                     & c($1/4$)   & 0 & -40 & -20 & - & - & - & -  & -\\
                                     & c($1/2$)   & 5 & -30 & -10 & - & - & - & - & - \\
                                     & c($9/10$)  & 5 & -25 & -5 & - & - & - & - & - \\
								
\hline
\label{tab:4}
\end{tabular}
\end{center}
\end{tiny}
\end{table*}

\begin{table*}[htbp]
\begin{tiny}
\begin{center}
\caption[Physical parameters of each class.]{Physical parameters of each E1, radio-power and radio morphology class.}
\begin{tabular}{llcccccccc}
\hline\hline
\multicolumn{1}{c}{} & \multicolumn{1}{c}{} & \multicolumn{1}{c}{\textbf{B1++}} & \multicolumn{1}{c}{\textbf{B1+}} & \multicolumn{1}{c}{\textbf{B1}} & \multicolumn{1}{c}{\textbf{A1}} & \multicolumn{1}{c}{\textbf{A2}} & \multicolumn{1}{c}{\textbf{A3}} & \multicolumn{1}{c}{\textbf{A4}} & \multicolumn{1}{c}{\textbf{B2}} \\
\hline
\multirow{6}*{\textbf{CD RD}}    & $L_{5100}$              & - & 3.0$\times 10^{45}$ & 1.0$\times 10^{45}$ & 3.3$\times 10^{44}$ & 6.9$\times 10^{44}$ & 3.85$\times 10^{45}$ & - & - \\
                                 & $\lambda_{E}$           & - & 0.035 & 0.052 & 0.147 & 0.345 & 0.609 & - & - \\
                                 & $M_{BH}$                & - & 3.5$\times 10^{9}$ & 1.0$\times 10^{9}$ & 1.6$\times 10^{8}$ & 1.1$\times 10^{8}$ & 3.2$\times 10^{8}$ & - & - \\
								 & $P_{\nu}$               & - & 3.2$\times 10^{24}$ & 8.9$\times 10^{23}$ & 1.2$\times 10^{24}$ & 8.8$\times 10^{23}$ & 4.0$\times 10^{24}$ & - & - \\
								 & log$R_{K}$              & - & 0.9 & 0.73 & 0.66 & 0.74 & 0.71 & - & - \\
								 & $SFR_\mathrm{r}$               & - & 1.8$\times 10^{3}$ & 5.2$\times 10^{2}$ & 7.2$\times 10^{2}$ & 5.1$\times 10^{2}$ & 2.3$\times 10^{3}$ & - & - \\
\hline						
\multirow{6}*{\textbf{CD RI}}    & $L_{5100}$              & 5.9$\times 10^{44}$ & 7.5$\times 10^{44}$ & 1.0$\times 10^{45}$ & 5.6$\times 10^{44}$ & 2.9$\times 10^{44}$ & 8.9$\times 10^{44}$ & 2.9$\times 10^{44}$ & 9.0$\times 10^{44}$ \\
                                 & $\lambda_{E}$           & 0.012 & 0.023 & 0.06 & 0.211 & 0.266 & 0.339 & 0.449 & 0.1 \\
                                 & $M_{BH}$                & 3.5$\times 10^{9}$ & 2.3$\times 10^{9}$ & 8.6$\times 10^{8}$ & 1.7$\times 10^{8}$ & 7.1$\times 10^{7}$ & 1.3$\times 10^{8}$ & 3.7$\times 10^{7}$ & 8.4$\times 10^{8}$ \\
								 & $P_{\nu}$               & 5.7$\times 10^{24}$ & 4.5$\times 10^{24}$ & 5.5$\times 10^{24}$ & 1.5$\times 10^{24}$ & 1.3$\times 10^{24}$ & 2.7$\times 10^{24}$ & 1.6$\times 10^{24}$ & 3.1$\times 10^{24}$ \\
								 & log$R_{K}$              & 1.39 & 1.46 & 1.45 & 1.49 & 1.27 & 1.43 & 1.42 & 1.36 \\
								 & SFR$_{r}$               & 3.3$\times 10^{3}$ & 2.7$\times 10^{3}$ & 3.2$\times 10^{3}$ & 8.7$\times 10^{2}$ & 8.0$\times 10^{2}$ & 1.6$\times 10^{3}$ & 9.2$\times 10^{2}$ & 1.8$\times 10^{3}$ \\
\hline						
\multirow{6}*{\textbf{CD RL}}    & $L_{5100}$              & - & 4.3$\times 10^{44}$ & 7.7$\times 10^{44}$ & 8.0$\times 10^{44}$ & 5.0$\times 10^{44}$ & 4.4$\times 10^{44}$ & - & 7.7$\times 10^{44}$ \\
                                 & $\lambda_{E}$           & - & 0.019 & 0.089 & 0.326 & 0.197 & 0.592 & - & 0.064 \\
                                 & $M_{BH}$                & - & 1.8$\times 10^{9}$ & 8.5$\times 10^{8}$ & 1.9$\times 10^{8}$ & 2.0$\times 10^{8}$ & 4.6$\times 10^{7}$ & - & 1.0$\times 10^{9}$ \\
								 & $P_{\nu}$               & - & 8.2$\times 10^{24}$ & 6.2$\times 10^{25}$ & 5.8$\times 10^{25}$ & 3.0$\times 10^{25}$ & 4.3$\times 10^{24}$ & - & 1.8$\times 10^{25}$ \\
								 & log$R_{K}$              & - & 2.12 & 2.44 & 2.91 & 2.47 & 1.98 & - & 2.27 \\
								 & SFR$_{r}$               & - & 4.8$\times 10^{3}$ & 3.6$\times 10^{4}$ & 3.4$\times 10^{4}$ & 1.7$\times 10^{4}$ & 2.5$\times 10^{3}$ & - & 1.0$\times 10^{4}$ \\
\hline       
\multirow{6}*{\textbf{FRII RL}}  & $L_{5100}$              & 7.5$\times 10^{44}$ & 6.6$\times 10^{44}$ & 7.5$\times 10^{44}$ & - & - & - & - & - \\
                                 & $\lambda_{E}$           & 0.011 & 0.027 & 0.070 & - & - & - & - & - \\
                                 & $M_{BH}$                & 4.6$\times 10^{9}$ & 2.0$\times 10^{9}$ & 8.9$\times 10^{8}$ & - & - & - & - & - \\
								 & $P_{\nu}$               & 2.2$\times 10^{26}$ & 1.6$\times 10^{26}$ & 1.8$\times 10^{26}$ & - & - & - & - & - \\
								 & log$R_{K}$              & 3.30 & 3.28 & 2.96 & - & - & - & - & - \\
								 & SFR$_{r}$               & 1.3$\times 10^{5}$ & 9.1$\times 10^{4}$ & 1.0$\times 10^{5}$ & - & - & - & - & - \\
\hline
\label{tab:3}
\end{tabular}
\end{center}
\textbf{Notes:} L$_{5100}$ [erg/s], $M_\mathrm{BH}$ [M$_{\odot}$/yr], $P_{\nu}$ [W/s], SFR$_{r}$ [M$_{\odot}$/yr].
\end{tiny}
\end{table*}

\subsection{Relation to physical parameters}	

Table \ref{tab:3} reports several physical parameters for each E1 ST related to radio-power and radio-morphology class: the luminosity at 5100 \textup{\AA}, $L_\mathrm{5100}$, the black hole mass, $M_\mathrm{BH}$, the Eddington ratio, $\lambda_\mathrm{E}$, the radio power at 1.4 GHz, $P_\mathrm{1.4GHz}$,  $\log R_\mathrm{K'}$, and $pSFR_\mathrm{r}$. The reported values are the median of the parameters of all the sources belonging in a given E1, radio-morphology and radio power class. 

Following the E1 trend from the B1$^{++}$ to the A4 ST, there is an increase of Eddington ratio and a decrease of the black hole mass, in agreement with past works \citep[e.g.,][and references therein]{fraix-burnetetal17}.  Populations A and B as a whole also show difference in terms of the \mbh\ and Eddington ratios, with Population B showing more massive BH having a lower Eddington ratio with respect to the Population A. The systematic difference in \lledd\ and \mbh\ are likely to account for the trends along the E1 MS \citep[e.g.,][]{sunshen15}. Around the critical value of $\lambda_\mathrm{E} \approx 0.2 \pm 0.1$ the active nucleus may undergo a fundamental change in accretion mode and BLR structure \citep{wangetal14}.

Focusing the attention on the different radio-loudness classes,  black hole masses and the Eddington ratios do not vary greatly along the radio classes: the B1 ST involves \mbh $\approx 10^{9}$ M$_{\odot}$, whereas the A2 ST \mbh $\approx 10^{8}$ M$_{\odot}$ along the RD, RI, RL  classes. The trends related to RL appear to be ``orthogonal'' with respect to the E1 MS trends concerning the low-ionization part of the BLR. The \mbh\ and \lledd\ values are, on average, comparable in each ST for the different radio classes. Differences in each ST for different radio classes described earlier  might have been lost in large samples of AGNs including RQ and RL quasars because RL quasars are a minority.

\subsection{Radio and FIR $SFR$}

\label{radfir}

\begin{figure*}[htbp!]
\centering
\vspace{0cm}
\includegraphics[width=0.45\textwidth]{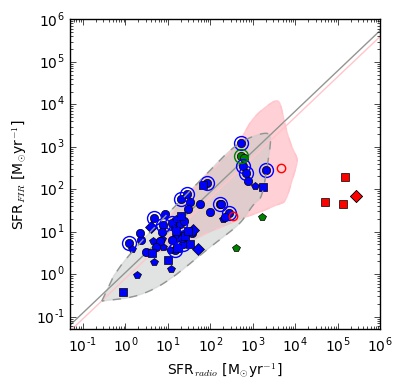}
\includegraphics[width=0.45\textwidth]{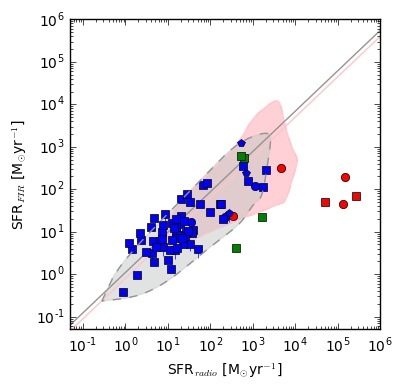}
\caption{FIR luminosity vs. radio power. The shaded areas trace the occupation zone of star-forming galaxies (gray) and RQ quasars (pink) following \citet{bonzinietal15}. Color codes are: red (RL), green (RI), blue (RD).  Left: sources identified on the basis of spectral types; xA: circled symbols; A2: circles; A1: pentagons; B1: squares; B2: rhomboids; not defined: open symbols. Right: sources identified on the basis of samples. Circles: this work; rhomboids: \citealt{marzianisulentic14}, squares: \citealt{sanietal10} + \citealt{marzianietal03a}. } 
\label{fig:bonz}
\end{figure*}

 Table \ref{tab:bonz} lists  SDSS identification and properties  of seven objects of the sample considered in the present paper for which FIR data are available. The properties listed are, in the following order:  the SDSS ID, the radio and FIR specific flux densities, radio power at 20 cm, the luminosity at 70$\mu m$ or 100$\mu m$, the $SFR$(FIR), the $pSFR$ from the radio power, and the classification code.  Since only seven sources are found to have available FIR data to be used for the  \citet{bonzinietal15}, we considered additional samples to ease their interpretation: the union of the low-$z$  \citet{sanietal10}, and  \citet{marzianietal03a} samples, along with the sample of  \citet{marzianisulentic14}. A search in the NASA/IPAC Infrared Science Archive (IRSA)\footnote{https://irsa.ipac.caltech.edu/frontpage/} for Herschel, IRAS and Spitzer data at 70, 100, and 160 $\mu m$ yields 73 sources in total: seven from the present work, three from  \citet{marzianisulentic14}, and 63 from the joint \citet{sanietal10}, and  \citet{marzianietal03a} papers. The samples were vetted to preferentially select Pop. A and especially xA ST, although some well-studied Pop. B quasars were included.

\begin{table*}[!htbp]
\begin{tiny}
\begin{center}
\caption[FIR data and Bonzini classification]{FIR and radio data \label{tab:bonz}}
\medskip
\begin{tabular}{llllllllll}
\hline\hline
{{SDSS ID}} & {{$f_\mathrm{1.4GHz}$}} &  {{$f_\mathrm{FIR}$}} & $\lambda_\mathrm{FIR}$ &   {{$P_\mathrm{1.4GHz}$}} & {{L$_\mathrm{FIR}$}} & {{SFR$_\mathrm{1.4GHz}$}} & {{SFR$_\mathrm{FIR}$}} & Class + ST \\
{} & {[mJy]} & {[mJy]} & [$\mu$m] &  [W Hz$^{-1}$]& [\ergss] & {[M$_{\odot}$/yr]} & {[M$_{\odot}$/yr]} & {}\\
\hline
J021640.73-044404.8	&	61.70	&	18.654	&	100  &	2.33$\times 10^{26}$	&	4.76$\times 10^{44}$	&	1.35$\times 10^{5}$	&	0.45$\times 10^{2}$   &	RL CD B1	\\
J100943.55+052953.8	&	81.45	&	19.09   &	 70  &	3.71$\times 10^{26}$	&	2.01$\times 10^{45}$	&	2.15$\times 10^{5}$	&	1.89$\times 10^{2}$   &	RL FRII B1	\\
J111908.67+211917.9	&	4.28	&	85.466	&    70  &  3.75$\times 10^{23}$	&	2.28$\times 10^{44}$	&	2.17$\times 10^{2}$	&	0.22$\times 10^{2}$   &	RD CD A2	\\
J122449.90+062917.1	&	2.33	&	140.571	&	100  &	8.10$\times 10^{24}$	&	3.37$\times 10^{45}$	&	4.69$\times 10^{3}$	&	3.17$\times 10^{2}$   &	RL CD	\\
J123547.98+090801.0	&	1.74	&	55.484	&	100  &	5.74$\times 10^{23}$	&	2.46$\times 10^{44}$	&	3.33$\times 10^{2}$	&	0.23$\times 10^{2}$   &	RL CD	\\
J143157.94+341650.2	&	0.84	&	69.287	&	100  &	1.93$\times 10^{24}$	&	1.23$\times 10^{45}$	&	1.12$\times 10^{3}$	&	1.16$\times 10^{2}$   &	RD CD A1	\\
J162052.59+540058.6	&	1.03	&	164.687	&	100  &	5.91$\times 10^{22}$	&	1.86$\times 10^{44}$	&	0.34$\times 10^{2}$	&	0.17$\times 10^{2}$   &	RD CD A2	\\
\hline
\label{tab:B}
\end{tabular}
\end{center}

\end{tiny}
\end{table*} 


Following the FIR-radio correlation for star forming galaxies and quasars as shown by \citet{bonzinietal15}, we assume that the maximum possible $SFR$\ is $SFR_\mathrm{max} = few \,10^{3}$ M$_{\odot}$\, yr$^{-1}$. The $SFR$\ is $\lesssim$ 6000 M$_\odot$/yr in the most luminous sub-mm galaxies, with a turn-down at $\sim 2000$ M$_{\odot}$/yr \citep{bargeretal14}. Higher $SFR$ are most likely unphysical, since the feedback effect due to stellar winds and supernova explosions may bring star formation to a halt on a  short timescale.   

The goal is to gain information about the dominating process in the radio AGNs activity, that is, to distinguish between star formation or relativistic jet even if we cannot exclude that, in any class we identified, there is a contribution from the non-thermal activity of the nucleus. The shaded area in Fig. \ref{fig:bonz} delimits the region of star forming galaxies and of RQ quasars.  Comparing the radio and FIR $SFR$\  shows that five out of seven sources of the sample presented in this work fall in the region of RQ quasars, including two RL and three RD. None of the sources that fall in the RL region (outside the shaded area in Fig. \ref{fig:bonz}) is an xA source. 

The left panel of Fig. \ref{fig:bonz} identifies the spectral type of the 73 sources. The most luminous xA sources can  reach $\sim 10^{39}$ \ergss\ Hz$^{-1}$\ (comparable to powerful RL “jetted” sources), although low-$z$ xA sources understandably populate the lower part of the RQ zone. However, of the ten RQ sources above $SFR \sim $ few $100 $ \msol\ yr$^{-1}$, five are xA, two borderline xA (A2 with \rfe\ slightly below 1), one B1, one A1, and one has no defined spectral type due to poor data.  Therefore, the majority of sources at the highest radio powers and $SFR$ shown in Fig. \ref{fig:bonz} are apparently xA, and their placement in the \citet{bonzinietal15}  diagram is consistent with them being RQ.

xA sources are mainly classified as RD. The xA sources are predominantly blueish quasars, as testified by their high prevalence in the PG survey. Several of them are also borderline RI with $\log R_\mathrm{K}$ slightly below 1. Had  they been  affected by some internal extinction, they would have been classified as RI. FIR data for CD RL A3 and A4 of our sample are unfortunately missing.   The only xA source classified as radio-intermediate is Mark 231, a high-$L$, low-$z$\ BAL QSOs suffering some internal extinction \citep{sulenticetal06a,veilleuxetal16}. Mark 231 has been described by Sulentic \& Veilleux and collaborators as a prototypical high-$L$, highly accreting quasar placed at low-$z$. xA properties are extreme, in terms of \feii\ emission, \civ\ blueshift, and blueward asymmetry of \hb. Mark 231 is known to possess an unresolved core, highly variable and of high brightness temperature\ \citep{condonetal91}. Superluminal radio  components of  Mark 231 have been detected, and the relation between BALs, radio ejections, and continuum change \citep{reynoldsetal17}  illustrate the complex interplay of thermal/nonthermal nuclear emission that can be perhaps typical of xA sources. Still, according to the location in the diagram of Fig. \ref{fig:bonz}, the dominant emission mechanism is thermal. This inference is consistent with the enormous CO luminosity of the AGN host  \citep{rigopoulouetal96}.


\section{Discussion}
\label{disc}


Our sample sources follows the MS, and in Sect. \ref{hbtrends} and  \ref{oiii} we discuss the \hb\ and [OIII] trends with reference to the radio class.   We  want to stress again that the distribution of sources in the various spectral types is different from the one of optically-selected samples mainly consisting  of RQ sources, and depends on both radio loudness and radio  morphology (Table \ref{tab:E1-sample}). As mentioned, the majority of objects in the present sample belong to the B1 and B1$^{+}$ bins, whereas the number of sources goes down in bins with higher FWHM(H$\beta$), such as B1$^{++}$, and higher $R_\mathrm{FeII}$, that is, toward Pop A2 ST and onward. The highest values of the radio power are seen in Pop. B for FRII and CD objects. All these objects are most likely “jetted” RL quasars. The \hb\ FWHM may be lower than the Pop. A upper limit if the CD are FRII sources seen at relatively small viewing angles (Sect. \ref{orient}). On the converse, broadest profiles (bin B1$^{++}$) may indicate a favorable orientation for the detection of a double structure, possibly associated with sub-pc supermassive binary black holes (Sect. \ref{bbh}). 

The CD class exhibits a significant number of xA sources, $\sim 30 \%$ of all Pop. A are CD objects. Interestingly,  $\sim 18 \%$ of xA CD AGNs have a radio power emission in the loud range, a prevalence even higher than the prevalence of RL sources in optically-selected samples.  The high prevalence in sources whose appearance is ``very thermal'' in the optical and UV ranges \citep{negreteetal18,martinez-aldamaetal18} calls into question the origin of their radio power (Sect. \ref{radiosfr}). 

\begin{figure}[htbp!]
\centering
\includegraphics[width=0.45\textwidth]{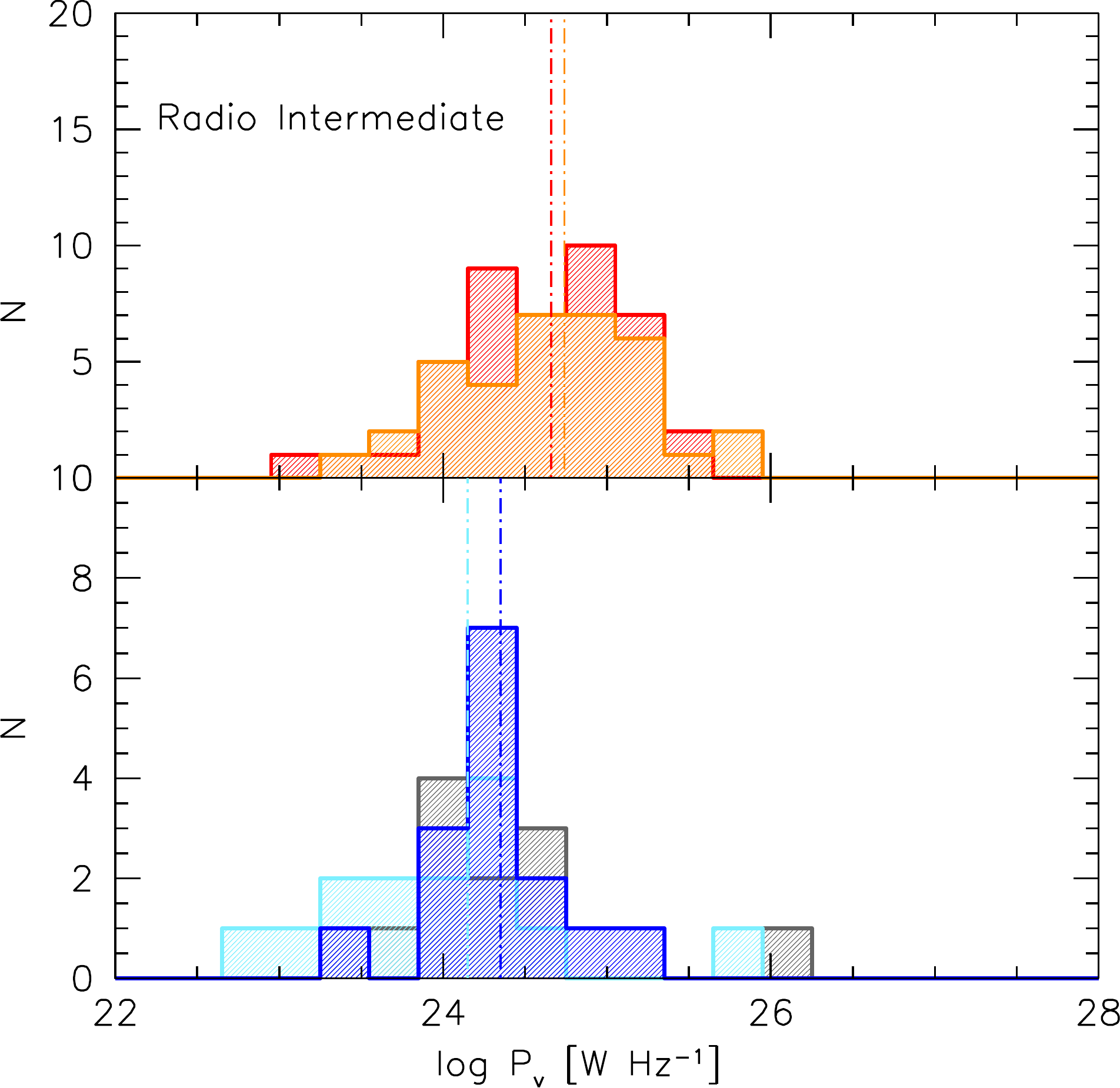}\\
\vspace{-0.cm}
\includegraphics[width=0.45\textwidth]{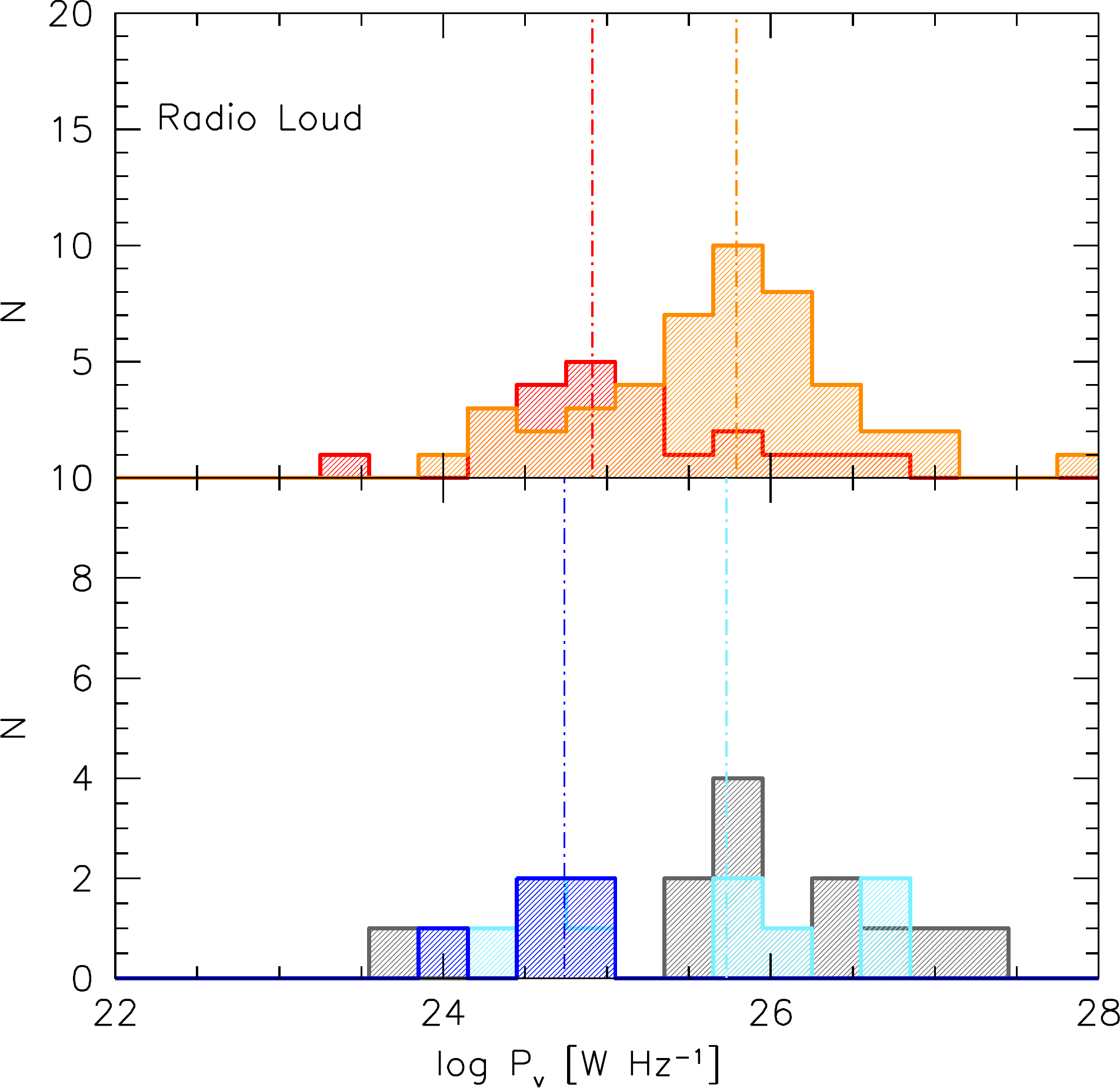}
\vspace{-0.cm}
\caption{Distribution of radio power $P_\nu$ in units of W for RI (top) and RL (bottom), for different spectral types along the main sequence. The top panels show the distributions for the union of  B1$^{++}$ and B1$^{+}$ (red), B1 (orange) spectral types, the bottom ones for A1 (gray), pale blue (A2), and xA (blue). The vertical dot-dashed lines mark the  medians of the Pop. B (red and orange), A2 (pale blue) and xA (blue) distributions.} 
\label{fig:pnudistr}
\end{figure}

\begin{figure}[htbp!]
\centering
\includegraphics[width=0.45\textwidth]{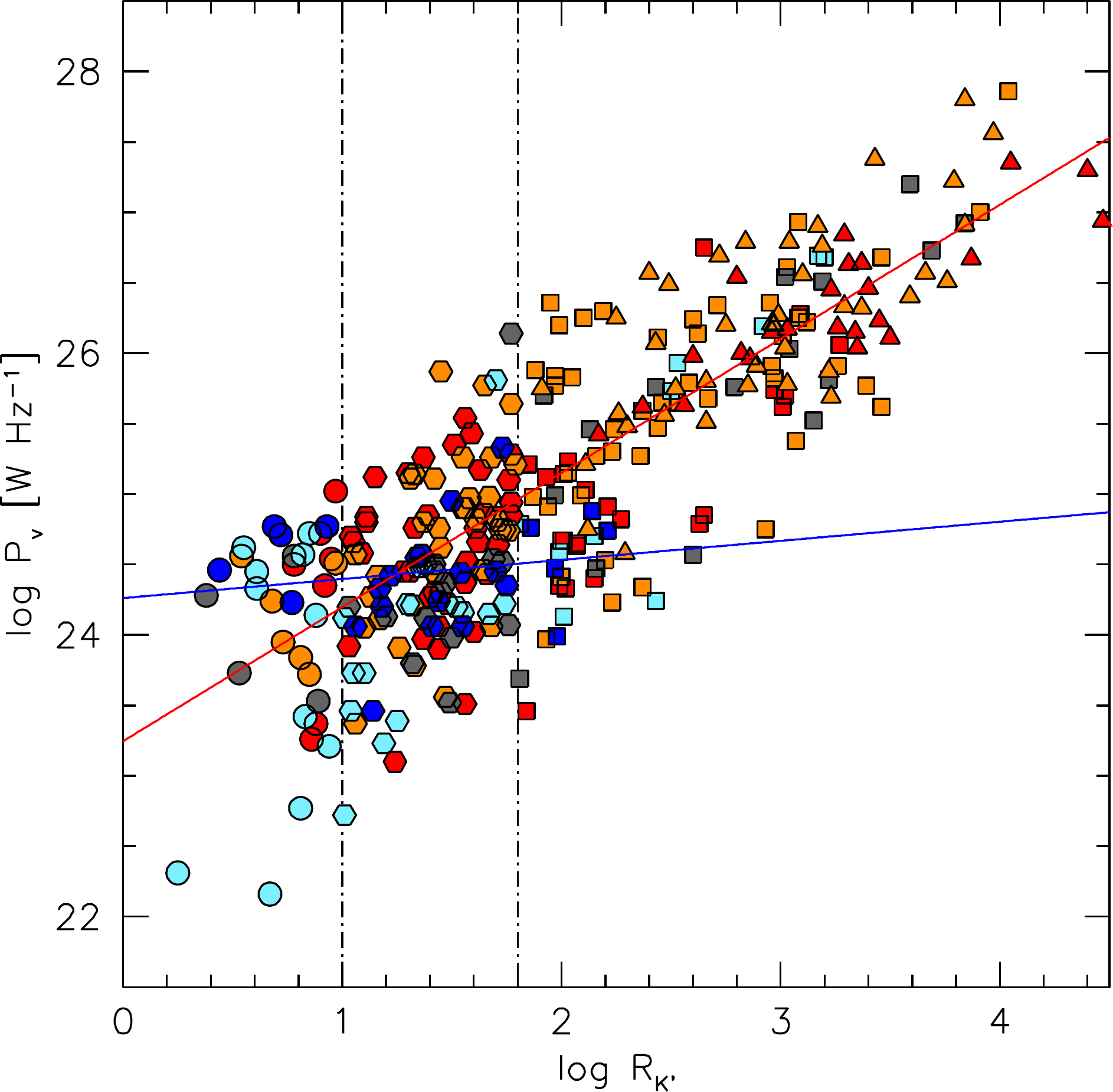}
\vspace{-0.1cm}
\caption{Behavior of radio power $P_\nu$ in units of W as a function of the Kellermann's parameter \rk. Symbol coding corresponds to different radio morphologies and to different STs. Triangles: RL FRII; squares: RL CD, exagons: CD RI, circles: CD RQ. Red: union of  B1$^{++}$ and B1$^{+}$, orange: B1,  gray: A1,  pale blue: A2, xA: blue. The lines trace unweighted least square fits for   Pop. B sources (red) and for  xA sources  considered over the three radio loudness classes.} 
\label{fig:rkpower}
\end{figure}

\subsection{Spectroscopic trends: H$\beta$}
\label{hbtrends}

The radial velocity dispersion (in the following $\delta v_\mathrm{obs} =$FWHM) can be written as 

\begin{equation}
\frac{\mathrm{FWHM}^{2}}{4} = \delta v^{2}_\mathrm{iso} + \delta v^{2}_\mathrm{K}\sin^{2}\theta, 
\label{eq:flat}
\end{equation} 

where $\kappa =  \delta v_\mathrm{iso} / \delta v_\mathrm{K}$, the ratio between an isotropic $\delta v_\mathrm{iso}$ and the true Keplerian velocity $\delta v_\mathrm{K}$. $\kappa \approx 0.1 \ll 1$\ implies a highly-flattened configuration, and a strong dependence on the line width of the viewing angle $\theta$. The effect of the viewing angle appears to be most important in the low-\rfe\ region of the MS (i.e., spectral types B1$^{+}$, B1 and A1), and will be discussed with reference to the  RL FRII/CD relation (Sect. \ref{orient}).

Orientation effects ideally leave a symmetric profile. The \hb\ line  however shows systematic trends of asymmetry and centroid shifts along the MS: in Pop. A, from A2 to A4, blueshifts become more prominent, such as the case of \oiiiopt. The physical  basis of the blueshift is most likely Doppler-shifted emission by outflowing gas approaching the observer, with the receding side of the outflow hidden from view  \citep[][and references therein]{marzianietal16}. 

The prominence of the outflow is related to the Eddington ratio, which is believed to increase with increasing \rfe\ along the sequence \citep[e.g., ][]{grupeetal99,sunshen15,duetal16,sulenticetal17}. The frequency of detection of a blue shifted excess over the symmetric Lorentzian profile assumed for Pop. A, is 0 in A1, very low in A2 (the blue excess is detected rarely), sizeable in A3, and high in A4.  

xA sources are the ones expected to provide the maximum feedback effect on the host galaxy. In other words, they are the AGNs most likely to affect the dynamical and structural properties of the host, and may be ultimately lead to the well-known correlation between BH mass and bulge mass \citep[e.g.,][and references therein]{magorrianetal98,kingpounds15}. Radio observation offer an unobscured view of star formation, and can be directly compared to FIR measurements (which are, at present, available only for a small subset of our sample).   This may have profound consequences in our understanding of galaxy evolution, as a proper estimate of the $SFR$\ at high redshift and for sources fainter than the ones considered in this analysis may allow to analyze to which extent  feedback effects may be produced by the AGN itself or by induced star formation. 

On the converse, in Pop. B, the \hb\ profile  appears predominantly redward asymmetric toward the line base. The asymmetry affects the FWHM by about $20$ \%\ \citep[][and references therein]{marzianietal13a}. There is ample evidence that the Pop. B \hb\ broadening is dominated by virial motions. However, the origin of the redward asymmetry is unclear. Two main possibilities are considered: infall (+obscuration) and gravitational redshift. It is at least conceivable that some line emitting gas is drifting toward the central black hole at a fraction $f$\ of the free fall velocity \citep[e.g., ][]{wangetal17}. In this case, the centroid displacement can be $c(x) = \frac{1}{2} \delta v_\mathrm{K} f \cos \theta$. The real difficulty in this scenario is to envisage a dynamical configuration that allows for a steady inflow with the receding part hidden from view in a large fraction of quasars. The alternative mechanism is gravitational + transverse redshift \citep[e.g.,][and references therein]{bonetal15}. In this case, the net effect is expected to be a line shift by $\frac{3}{2} c z_\mathrm{g}$, where $z_\mathrm{g} = GM_\mathrm{BH}/c^2 r_\mathrm{BLR}$. The centroid displacement should be  $c(x) = \frac{3}{4} c z_\mathrm{g}$, independent on orientation. The problems with gravitational redshift stem from the compactness of the emitting region that is required to produce a shift as large as $2000-3000$ \kms. The issue is open at the time of writing but the relevant result  of the present paper  is that all radio classes show a prominent redward asymmetry. Redward asymmetries were revealed as a systemic properties of the \hb\ profiles of Pop. B RQ quasars \citep{marzianietal13a} and in RL alike \citep{punsly10}.  The implication is that this feature is unlikely to be a consequence of the “jetted” nature of RL sources. 

\subsection{Spectroscopic trends: [OIII]} 
\label{oiii}


 In typical SDSS spectra, the [OIII] profile can be well-represented by two components: a narrower one (NC or core compoent) and a blueshifted semibroad ``base'' component \citep[e.g.,][]{komossaetal08,zhangetal11,marzianietal16}. The relative intensity of the two components is dependent on the ST along the MS. The presence of  blueshifts and blue asymmetries among RL and RI indicates emission of \oiiiopt\ by outflowing gas, in analogy to the cases of the RD class in this sample, and of the broader RQ class in previously-analyzed samples \citep[e.g.,][ the RD sample is at the high end of the RQ class radio emission]{zamanovetal02,bianetal05,komossaetal08,zakamskaetal16}. The fairly constant value of the W(\oiii) base component has been found also in previous investigations spanning a wide range in luminosity \citep{marzianietal16}. The proportionality between continuum and line luminosity then  supports a nuclear origin for the blueshifted line component.


\subsection{Radio-power and radio-loudness}

\subsubsection{Population A and B intrinsic differences}

FRII sources are the parent population of all powerful RL “jetted” AGNs \citep{urrypadovani95}. RL FRII have typical $P_\nu \approx 10^{26}$ W Hz$^{-1}$, and Pop. B CD RI and RL share similar  $P_\nu$ values with them in bin B1$^{+}$. Pop. B CD RI show a lower $P_\nu$ than RL, but, even if lower, the RI CD sources show radio powers that are systematically higher than the RI CD sources in Population A, at least by a factor of several. The radio-power values reported in Table \ref{tab:3} and the distribution of  radio powers for RL and RI for the STs along the MS (Fig. \ref{fig:pnudistr}) show  that, although there is a considerable overlap, the distributions of the xA $P_\nu$\ is significantly different from the one of the Pop. B sources, for both RI and RL: a Kolmogorov-Smirnov test yields $P_\mathrm{KS} \approx 0.01$\ for the RI, and  $P_\mathrm{KS} \approx 0.004$\ for RL. The behavior of the $P_\nu$\ vs. \rk\ for the various STs (Fig. \ref{fig:rkpower})  adds further evidence in favor of an intrinsic difference between the two populations:  Pop. B sources of the three radio loudness classes are significantly correlated, with a unweighted least square fit yielding $\log P_\nu \approx (0.952 \pm 0.041) R_\mathrm{K'} + (23.245 \pm 0.097)$. The trend is much shallower for the xA sources and the slope of the best-fit is not significantly different from  0  $\log P_\nu \approx (0.136 \pm 0.168) \log R_{K'} + (24.261 \pm 0.246)$. The Pop. B relation is consistent with the idea that all sources owe their radio power to the same mechanism: specifically, we expect that ``jetted'' sources are exclusively synchrotron radiation in the radio, and that the synchrotron emission is also giving an important contribution in the optical. On the converse, the xA sources show a large $P_\nu$\ scatter for a given \rk. 

In addition,  there is a significant range of FIR luminosity for a given radio power  ($\sim 1$ order of magnitude, Fig.  \ref{fig:bonz}). The origin of the range in optical and FIR properties for the xA remains to be clarified. While extinction could be at the origin of the dispersion in \rk, the FIR range for a given $P_\nu$\ instead suggests a spread in the star formation properties of the host galaxy.

\subsubsection{The Kellermann’s parameter as an identifier
of RL sources}

\begin{figure}[htbp!]
\centering
\includegraphics[width=0.45\textwidth]{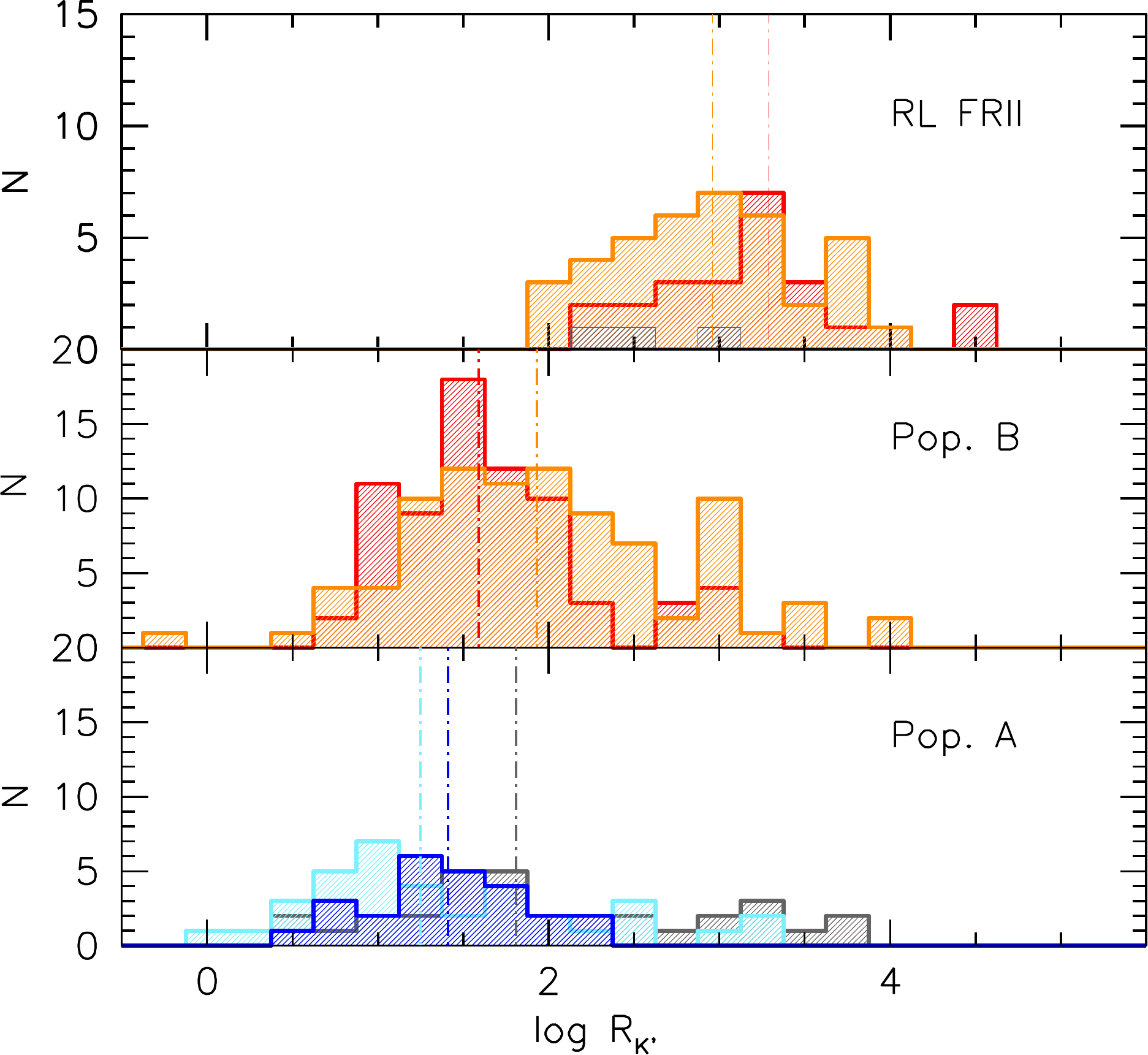}
\caption{Behavior of $\log$ \rk\ Top: Pop. A and B FRII (red: union of B1$^{+++}$, B1$^{++}$ and B1$^{+}$, orange: B1; gray: A1): Middle: Pop. B CD with the same color coding. Bottom: Pop. A   (gray: A1,  pale blue: A2, xA): blue. The dot-dashed lines indicate the median values for the samples shown.} 
\label{fig:distribu_rk}
\end{figure}

 Fig. \ref{fig:distribu_rk} shows the \rk\ index distribution for FRII, Pop. A and B CD sources. The FRII distribution validates the limit at $\log$ \rk $ \approx 1.8$ suggested by \citet{zamfiretal08} who proposed the minimum FRII \rk\ to identify ``jetted'' sources. Albeit simple, this limit may not be applicable to CDs. The basic reason is that lobes are due to extended emission representing emission integrated over the history of the activity cycle, whereas measurement of the core are ``instantaneous'' \citep[e.g.,][and references therein]{deyoung02}.  The distribution of CD Pop. B (middle panel of Fig. \ref{fig:distribu_rk}) shows that CD are mainly distributed with $\log$ \rk $ \gtrsim 1.$, which instead suggests a validation of the classical condition. The continuous distribution of radio powers (Fig. \ref{fig:pnudistr}) indicates that it is perhaps not meaningful to look for a bimodal distribution. Even if this sort of distribution is sometime seen \citep[e.g.,][]{guptaetal18}, the appearance may be dependent on selection effects involving frequency and selection criteria \citep{lafrancaetal10,cirasuoloetal03}. For instance, Fig. \ref{fig:distribu_rk} suggests a bimodal distribution if RL FRII and Pop. B CD are merged. The implication is that ``jetted'' sources might be associated with $\log$ \rk $ \gg 1.$ However, in Pop. B truly jetted sources are found in the range  $1 \lesssim \log$ \rk $ \lesssim 1.8$.  

 The limit $\log R_\mathrm{K} \gtrsim 2.4$\ proposed by \citet{falckeetal96} would give rise to a bimodal distribution of $\log R_\mathrm{K}$ for CD Pop. B sources implying that among CD only a relatively small fraction could be truly jetted.  In relatively old samples RL sources were confined to very massive black holes. These samples even suggested a \mbh\ ``threshold'' for radio-loudness \citep[see the recent analysis of][and references therein]{fraix-burnetetal17}. This view might have become outdated because of the discovery of RL NLSy1s \citep{komossaetal06}, and because of  systematic comparisons between \mbh\ in RL and RQ samples \citep{woourry02}. Results support the presence of ``jetted" sources with small \mbh, with RL sources  spanning a broad range of masses. While the CD Pop. B sources with $\log R_\mathrm{K} \gtrsim 2.4$\ could be likely considered the ``face-on'' counterpart of the FRII in our sample, low \mbh\ RL Pop. B sources could be ``jetted'' as well, since the power of a radio jet depends on three factors: \mbh, the magnetic field intensity, and the spin angular momentum of the black hole \citep{blandfordznajek77}. Classical FRII along with the FRII identified in this work are extended sources, whose ages are estimated around several hundreds of million years \citep[e.g.,][]{hirotakaetal06}. The lobe extension has an evolutionary meaning akin to \mbh\ that can only grow over cosmic time \citep{fraix-burnetetal17}.  So, whether the \rk\ distribution may appear unimodal or bimodal\ may depend on the \mbh\ distribution of the sample. 

FRII sources do not include RD and RI objects in contrast to the CD class. They show a small prevalence in Population A, limited exclusively to ST A1.  The existence of Pop. A  FRII objects may be explained in term of the angle between the line of sight and the radio axis, as discussed below (Sect. \ref{orient}). The same effect is expected to operate also for RI and RL CD in Pop. B. The largest \rk\ in ST A1 and A2 might be similarly ascribed to sources seen preferentially face-on. The high radio power that implies  extremely large $pSFR$, and the consistency with the expectation of unification models for RL sources, make it unquestionable to consider both RI CD along with RL CD as “jetted” sources, and  the condition $\log R_\mathrm{K'} \gtrsim 1$\  sufficient to identify truly “jetted” sources in Population B and in bin A1. Considering the radio-power of the CD RD B1$^+$\ that is comparable to the one of CD RI B1$^+$\ (Tab. \ref{tab:3}), it is even possible that some sources at RD loudness level belonging to Pop. B might be intrinsically jetted.   For the remaining ST of Population A,  the situation is expected to be fundamentally different (Sect. \ref{radiosfr}), although in the samples used for Fig.  \ref{fig:bonz} we have no indication of thermal xA  sources entering into the RL radio-loudness range, and thus no direct evidence that the condition $\log R_\mathrm{K} \gtrsim 1$\ does not select truly ``jetted'' sources.  

 \subsection{Orientation effects among RL sources}
\label{orient}

\begin{figure}[htbp!]
\centering
\includegraphics[width=0.4\textwidth]{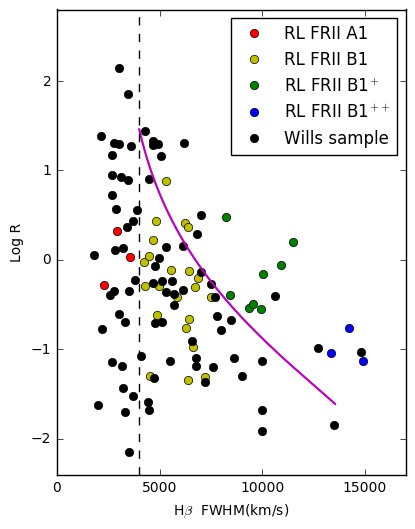}
\includegraphics[width=0.4\textwidth]{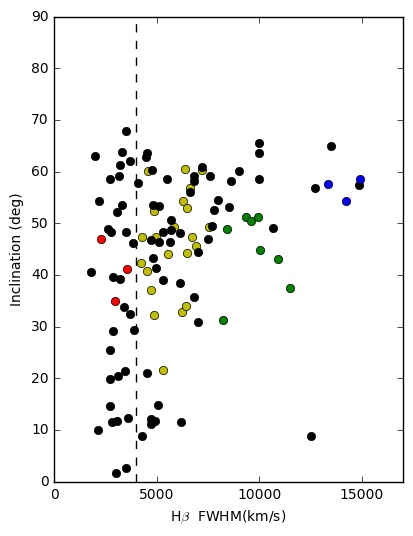}
\caption{Top:  Core-to-lobe radio flux density ratio  vs. FWHM(H$\beta$), adapted  from \citet{willsbrowne86}. Colored dots represent sources of our sample: red dots represent A1 RL FRII sources, yellow dots B1 RL FRII, green B1$^{+}$ RL FRII and blue ones B1$^{++}$ RL FRII. Black dots are the original data from \citet{willsbrowne86}. The magenta line represents the change of $R$ with FWHM($\beta$), predicted by the Orr \& Browne (1982) beaming model with $R_{T}=0.024$ and $\gamma = 5$. Bottom: FWHM($\beta$) vs. inclination, derived from $R_{c}$ through Eq. \ref{eq:ma}.}
\label{fig:wills-marin}
\end{figure}

The existence of the A1 FRII sources may be explained in term of the angle between the radio axis and the line of sight. In this view,  they should be simply end-on extended double  sources with a Doppler-boosted core \citep[e.g.,][]{blandfordrees78,deyoung84}.  To better understand  this effect, \citet{willsbrowne86} investigated how the FWHM(H$\beta$) changes as a function of a parameter $R_\mathrm{c}$, the ratio between the radio flux density of the core and that of the extended lobes. The parameter $R_\mathrm{c}$\ reflects the angle between the radio axis and the line of sight in the relativistic beaming model for radio sources. The top plot of Fig. \ref{fig:wills-marin}  shows a continuous distribution in both $R_\mathrm{c}$ and FWHM(H$\beta$) that goes against the idea of  two distinct classes characterized by a flat and a steep spectrum. Instead, the trend may translate into an intrinsic difference between sources with high and low $R_\mathrm{c}$ values or, if  relativistic beaming models are correct, it may translate in a different viewing angle of the emission line regions. In this case, the motions of the emission-line gas would be predominantly in a plane perpendicular to the radio axis. Then, the gas velocity field can be expressed as in Eq. \ref{eq:flat}.  In the top plot of Figure \ref{fig:wills-marin} the relation between observed \hb\ FWHM and $R_\mathrm{c}$ is showed by the magenta line, for $v_\mathrm{iso} = 4000$ \kms, $v_{p} = 13000$ \kms. The beaming model discussed by \citet{orrbrowne82} is used to relate $\theta$ and $R_\mathrm{c}$ as:  
\begin{equation} \label{eq:core}
\cos \theta =\dfrac{1}{\beta} \left\langle \dfrac{1}{2R_\mathrm{c}} \left\lbrace 2R_\mathrm{c}+R_{T}-[R_{T}(8R_\mathrm{c}+R_{T})]^{1/2} \right\rbrace \right\rangle ^{1/2}, 
\end{equation}
where $R_{T}=R_\mathrm{c}(90^{\circ})$ and $\beta=v/c$.  
Sources of the present samples are shown by red circles (A1 FRII), yellow (B1 FRII), green (B1$^{+}$) and blue (B1$^{++}$) whereas black circles refer to the \citet{willsbrowne86} sample.  Our sources follow the trend as the ones of  \citet{willsbrowne86}. On average, the narrower FRII A1 sources have a higher $R_\mathrm{c}$, whereas the broader B1, B1$^{+}$ and B1$^{++}$ sources have lower values suggesting that the orientation is an important factor affecting the line width. The bottom plot of Fig. \ref{fig:wills-marin} shows the trend between the FWHM(H$\beta$) and the inclination estimated following the work of \citet{marinantonucci16} through the equation: \begin{equation} \label{eq:ma}
i = \mathrm{g} + \mathrm{h} \, (LR_\mathrm{c}) + \mathrm{j} \, (LR_\mathrm{c})^{2} + \mathrm{k} \, (LR_\mathrm{c})^{3} + \mathrm{l} \, (LR_\mathrm{c})^{4} + \mathrm{m} \, (LR_\mathrm{c})^{5}, 
\end{equation}
where $LR_\mathrm{c} =$ log(R$_{5 \mathrm{GHz}}$), g $ = 41.799 \pm 1.190$, h $ = -20.002 \pm 1.429$, j $ = -4.603 \pm 1.347$, k $ = 0.706 \pm 0.608$, l $ = 0.663 \pm 0.226$, m $ = 0.062 \pm 0.075$. The range of inclinations a source can have gets smaller going towards higher FWHM($\beta$) (sources with FWHM(H$\beta)\gtrsim 7000$ \kms are constrained by $\theta \gtrsim 50^{\circ}$), but no source shows an inclination higher than $\sim 70^{\circ}$ and only sources with FWHM(H$\beta) \lesssim 8000$ km/s show an inclination smaller than $\sim 40^{\circ}$. There is no restriction on $\theta$ at low FWHM(H$\beta$); highly inclined source can be Pop. A, implying that the viewing angle, although important, is not the sole factor affecting the H$\beta$\ width. The inclination values for the three FRII RL A1 in our sample are intermediate ($35 \lesssim \theta \lesssim 45$\ degrees), that is, located in the lower part of the $\theta$ vs. FWHM(\hb) diagram (Fig. \ref{fig:wills-marin}) where orientation effects are expected to be relevant. We conclude that the FRII RL sources, the parent population of all jetted sources, is confined in Pop. B but these sources seen at smaller viewing angle may enter spectral type A1. Considering that the probability of observing  a source at angle $\theta$\ in an idealized sample of randomly-oriented sources is $\propto \sin \theta$, ST A1 is expected to be infrequently occupied by truly jetted sources of radio classes  RL FRII and RL CD. Actually, a non-negligible occupation in A1  from RL CD occurs because real radio flux-limited sample are biased  in favor of smaller viewing angle due to relativistic beaming.  As a matter of fact, flux-limited, optical samples of quasars  contain a large fraction of  their RL population that are flat-spectrum radio sources  \citep{falckeetal96}.     


\subsection{Candidate binary black hole systems in extreme Population B} 
\label{bbh}

Super-massive binary black holes (BBHs) are thought to be produced during the merging of two galaxies.  Their detection and number estimates can give us constraints on formation and evolution of galaxies  \citep{hopkinsetal06}. The presence of a binary black hole (BBH) in the nucleus of a RL AGN was first suggested by \citet{begelmanetal80}. Super-massive BBHs (SMBBHs) may lead  to an increase of the AGN activity and trigger starburst processes \citep{gaskell85}. More important to our topic, ideas have been proposed that see the formation of radio jets associated with the merger of a BBHs \citep{wilsoncolbert95}.   BBHs are thought to be a possible cause of the different classes of RL AGNs and of the misalignments between the direction of radio jet and the accretion disk \citep{merrittekers02}.  In addition to these effects the presence of two BHs may affect the dynamics of the broad-line region gas marking the profiles of the emission lines \citep{begelmanetal80}, creating in some cases highly-shifted single-peaked emission lines (BBH ``Doppler'' detection). Then from the broad optical lines stems a way to detect BBHs system with sub-parsec separation \citep[e.g.,][]{bonetal12,lietal16}. 

Looking at at the composite spectra shown in Fig.  \ref{fig:sp}, we note that several  \hb\ profiles in Pop. B could reflect this situation. Especially striking appear the cases of RD CD B1$^{+}$\ (where a significantly blueward-displaced \hbbc\ provides the best fit), RI CD  B1$^{+}$, RL CD B1$^{+}$, RL FRII B1$^{+}$ and FRII RL  B1$^{++}$.   The H$\beta$ profile includes a component close to the rest frame, but slightly displaced toward the blue, and a component with a larger displacement to the red. This configuration is consistent with the expectation of two BLRs, one associated with a more massive BH (the less-shifted component) and one associated with a less massive BH orbiting around a common center of mass. However, the blue shifted peak may be also due to an unresolved outflow.  

\begin{figure*}[htbp!]
\begin{center}
\includegraphics[width=.23\textwidth]{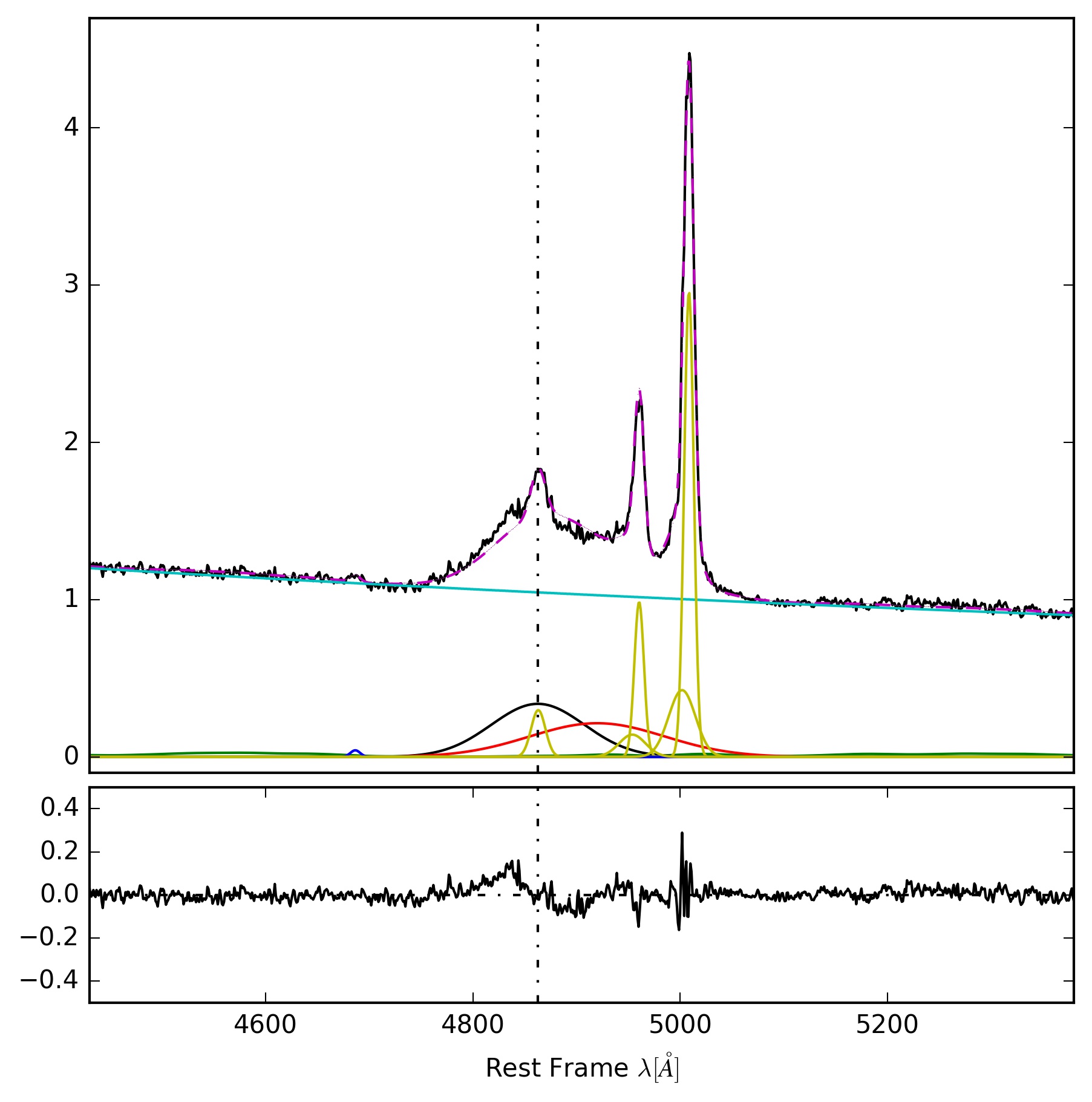}\;
\includegraphics[width=.23\textwidth]{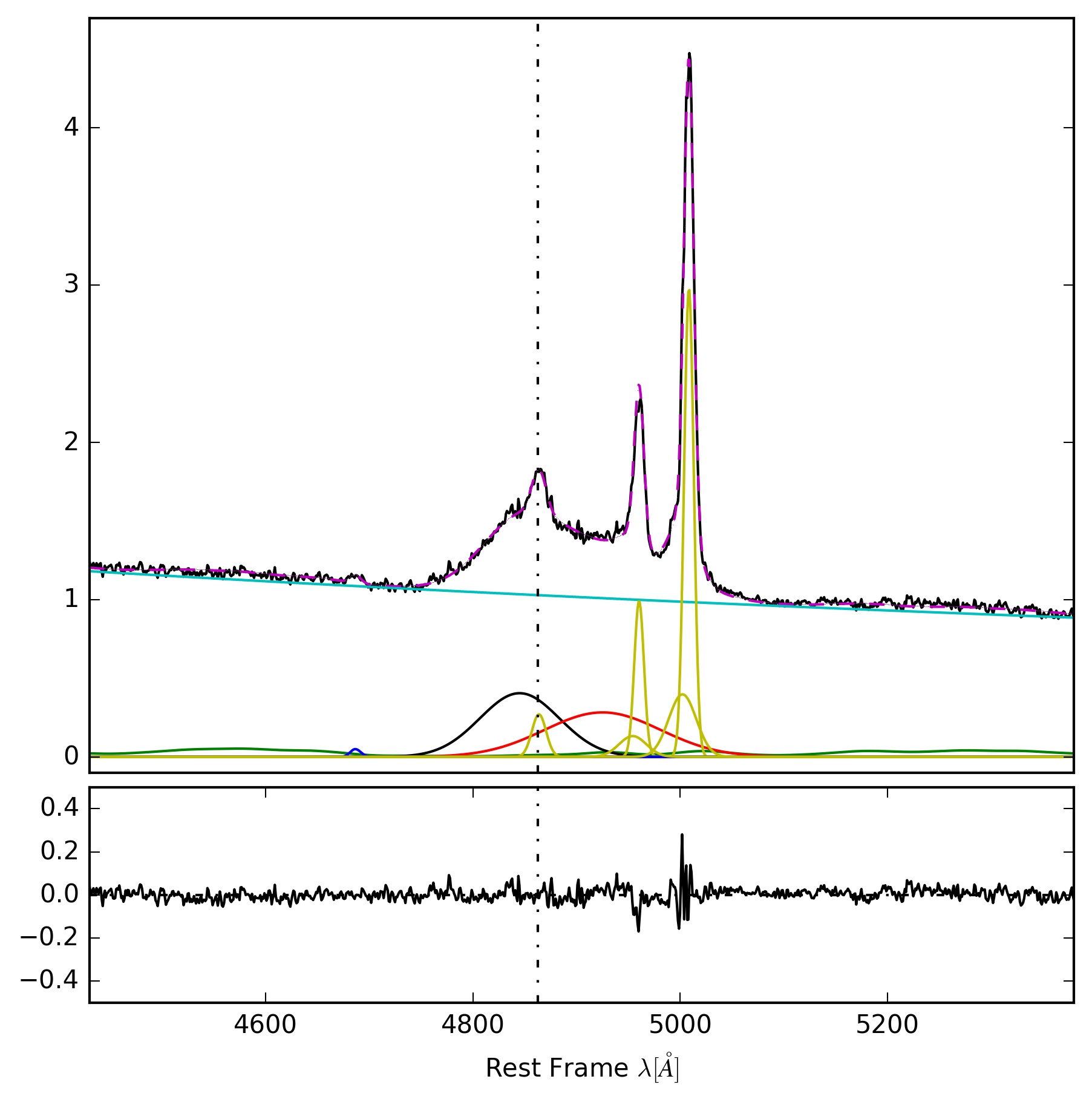} \quad
\includegraphics[width=.23\textwidth]{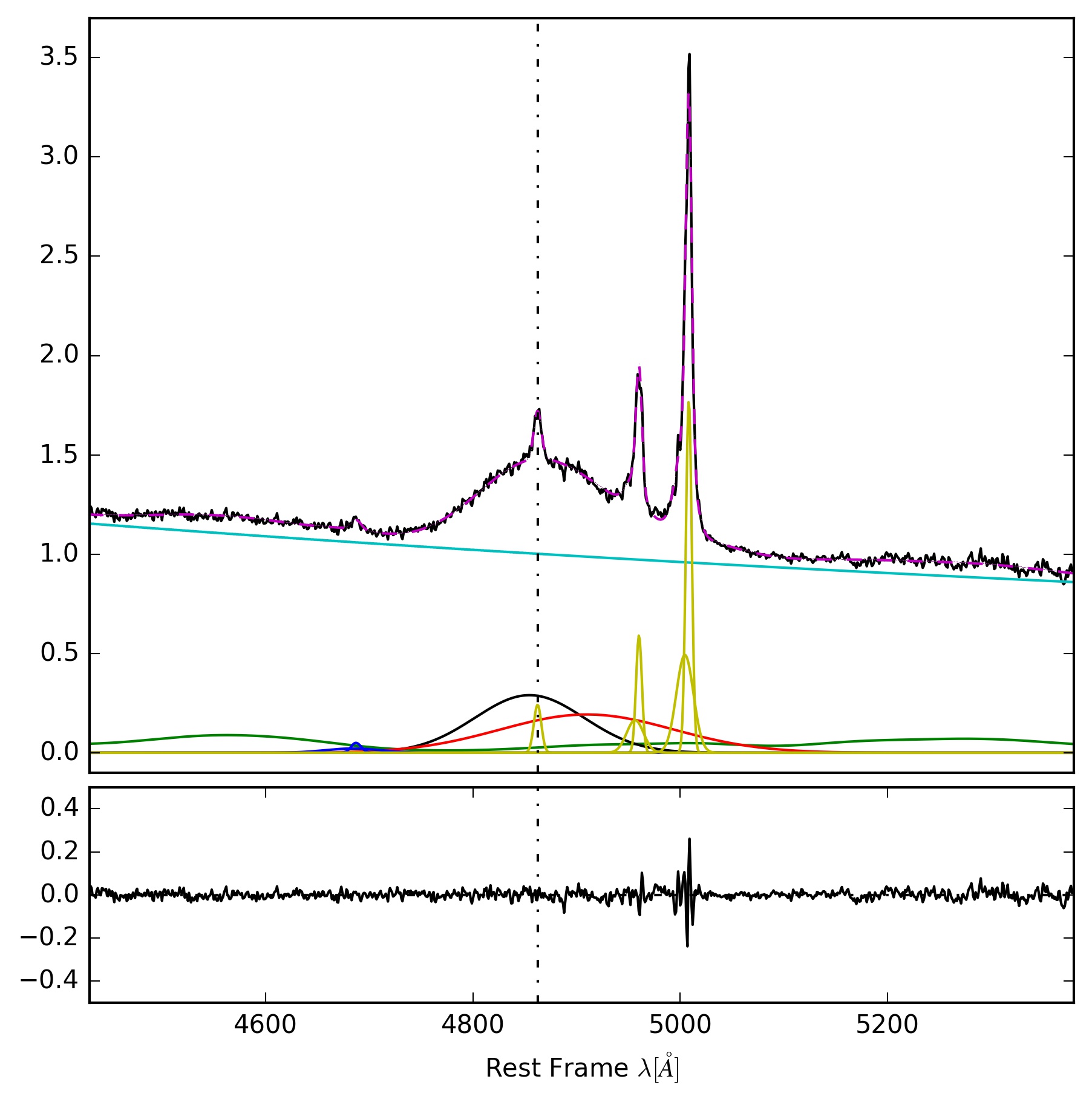}\;
\includegraphics[width=.23\textwidth]{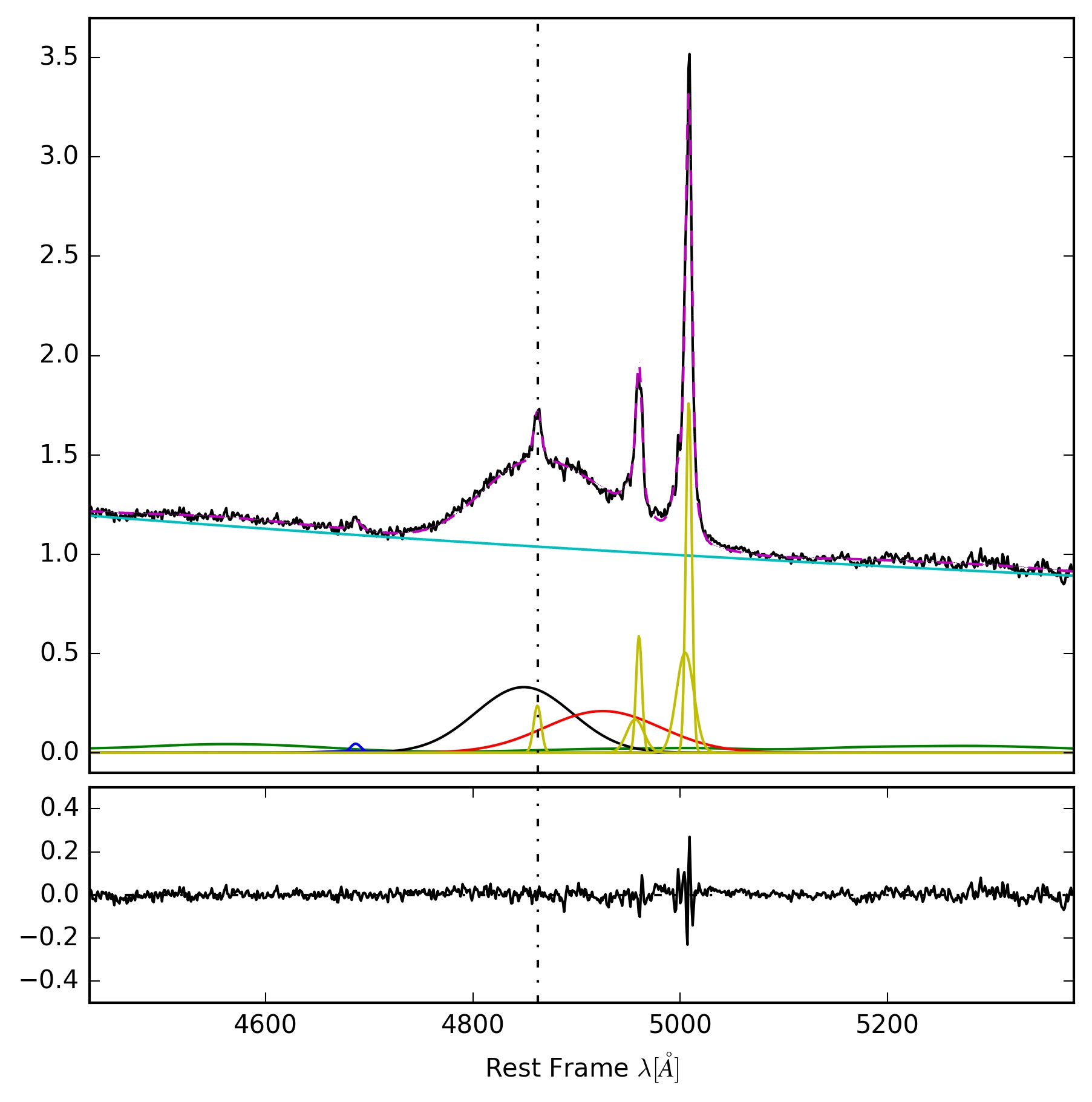} \\[1.em]
\includegraphics[width=.23\textwidth]{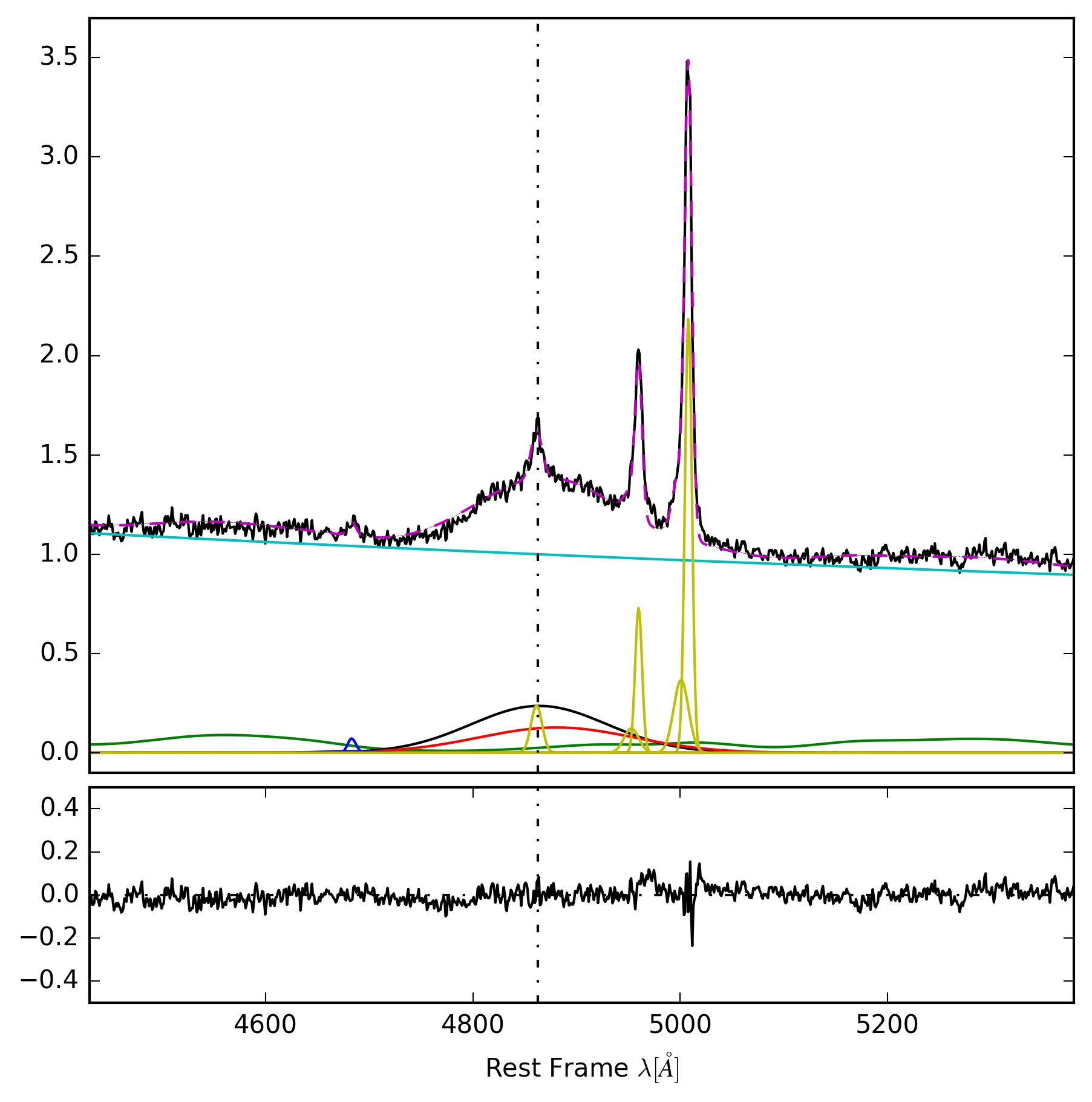}\;
\includegraphics[width=.23\textwidth]{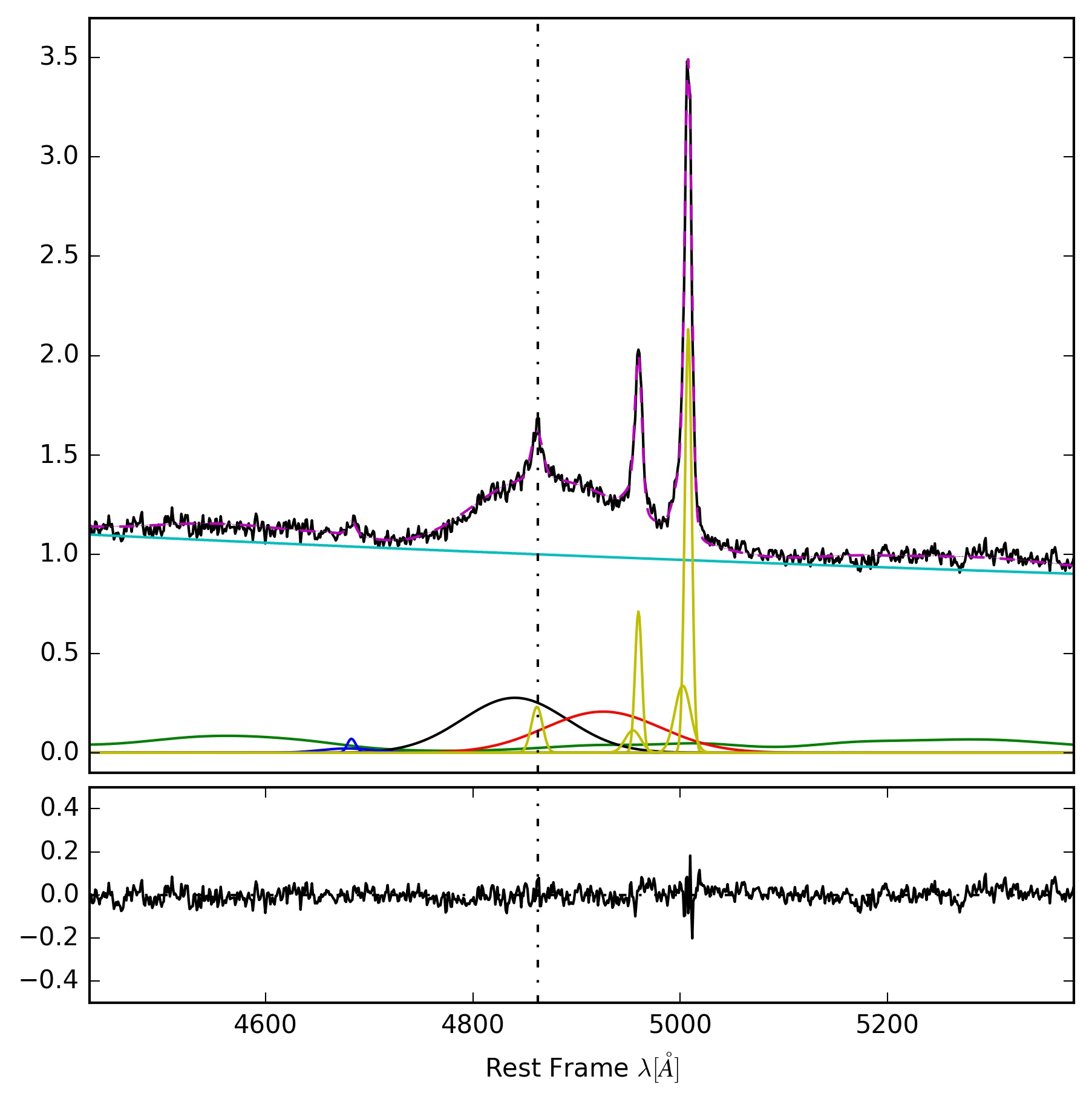} \quad
\includegraphics[width=.23\textwidth]{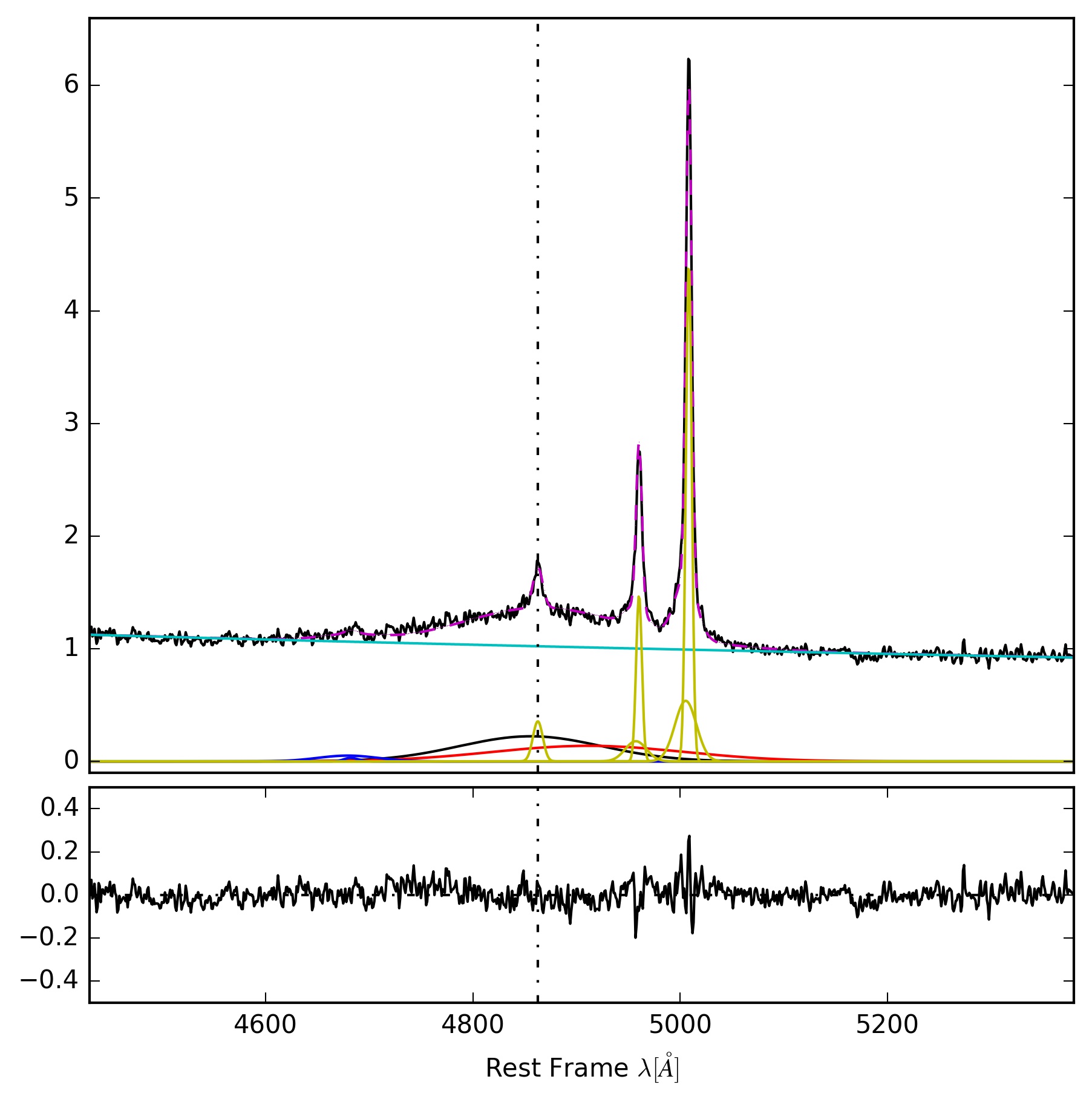}\;
\includegraphics[width=.23\textwidth]{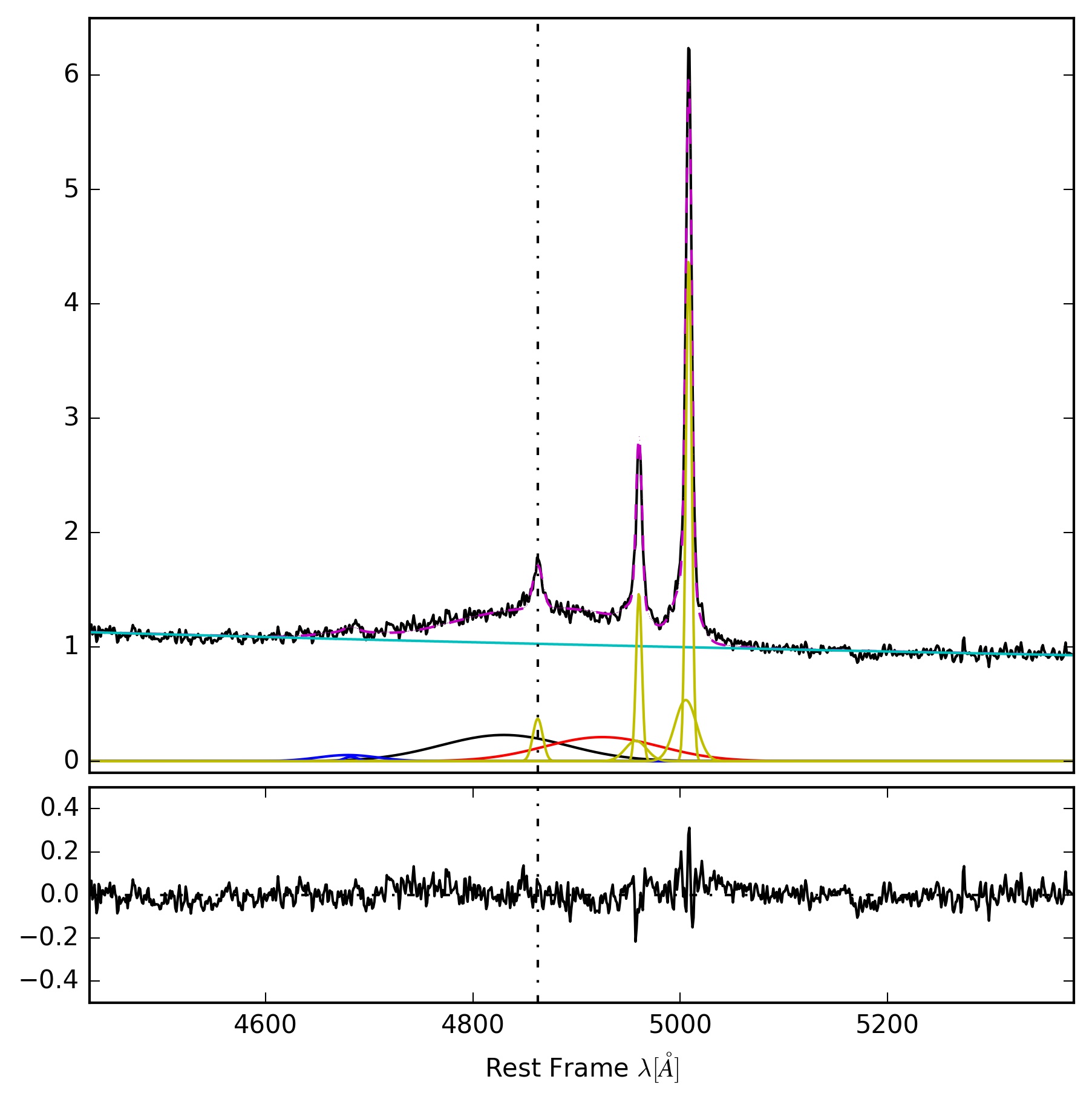}\\
 \quad
\caption{Comparison between broad + very broad and double broad H$\beta$ fitting models. From \textup{left to right}, top to bottom: CD-RD-B1$^{+}$, CD-RI-B1$^{+}$, CD-RL-B1$^{+}$ and FRII-RL-B1$^{++}$. For each pair the left plot represents the broad + very broad fitting BC+VBC model and the right plot represents the double broad component model BC+BD. Color scheme same as in Fig. \ref{fig:sp}}.
\label{fig:doublefit}
\end{center}
\end{figure*}

We tested  the conventional model BC+VBC (symmetric or slightly redshifted BC and redshifted broader VBC) of the \hb\ profile with a BC+BC model, in which one BC is significantly blueshifted and one is significantly redshifted. We considered cases in which the BC+VBC model fit yielded a blueshift for the BC. Four cases are shown in Fig. \ref{fig:doublefit}. In the case of  CD RD B1$^{+}$, the BC+BC model was tested against a BC+VBC model fit in which the peak of the BC was set to 0. Table \ref{tab:ftest} reports the results of an $F$-test between the two models, listing in the following order the ST identification, the fitting model, the standard deviation, the $\chi^2$, the ratio $F$\ and the probability of a chance occurence for values of $F$\ lower or equal than the observed one. The BC+BC fit is significantly better for RD CD B1$^{+}$ and RL CD B1$^{+}$, whereas there is almost no difference for CD RI B1$^{+}$. The BC+BC model for FRII RL  B1$^{++}$\ is apparently worse than the VBC+BC one although the difference is not significant. {In this latter case,  indications in support of a binary BLR  are provided by the analysis of the individual spectra.  Inspection of the eight  RL FRII B1$^{++}$ \hb\ profiles shows that, along with redward asymmetries or bumps, objects show symmetric and blue bumps or strong blueward asymmetries. This is the configuration naively expected in a sample in which a second feature radial velocity displacement is due to rotational motion around a more massive black hole that roughly sets the center of mass of the binary system.} B1$^{++}$ are believed to be massive or highly inclined sources that maximize the orbital velocity projection along the line-of-sight \citep[e.g., ][]{pandaetal19}. This is part of the evidence suggesting that RL sources might be preferentially associated with supermassive binary black holes, although the observation of single-epoch profiles and even extended temporal monitoring usually do not exclude other phenomena connected with accretion disk instabilities \citep[e.g.,][and references therein]{storchi-bergmannetal17}. Sources in the B1$^{++}$ ST are high-\mbh, low \lledd\ accretors. Relatively low accretion rate is known to be associated with extreme variability (even at a ``changing look'' level, \citealt{hirofumidone18}), giving rise to complex emission line profiles.  

While it is obviously not possible to demonstrate the presence of a sub-pc binary black hole even in the cases where a double  BC provides a better fit than the fit BC+VBC,  our analysis  identified spectral types with a higher prevalence of profiles that suggest   possible  SMBBH candidates. This means that Pop. A sources (which are always almost unshifted and symmetric) might be excluded in a systematic search for SMBBH candidates.

\begin{table}[!htbp]
\begin{tiny}
\setlength{\tabcolsep}{2pt}
\renewcommand{\arraystretch}{0.8}
\begin{center}
\caption[F test.]{Fitting model parameters and F test.$^a$}
\medskip
\begin{tabular}{llcccc}
\hline
\hline
\multicolumn{1}{c}{\textbf{}} & \multicolumn{1}{c}{\textbf{}} & \multicolumn{1}{c}{\textbf{$\sigma^{2}$}} & \multicolumn{1}{c}{\textbf{$\chi^{2}$}} &    \multicolumn{1}{c}{\textbf{F}} & \multicolumn{1}{c}{\textbf{P}}     \\
\hline
                                                           
\multirow{2}*{\textbf{CD RD B1$^{+}$}}    & bc + vbc &  1.357$\times 10^{-3}$ & 0.974  &   \multirow{2}*{1.460} & \multirow{2}*{8.267$\times 10^{-10}$}   \\
								     & bc + bc  & 9.261$\times 10^{-4}$ & 0.668       \\

\multirow{2}*{\textbf{CD RI B1$^{+}$}}    & bc + vbc & 6.217$\times 10^{-4}$ & 0.443  &   \multirow{2}*{1.123}  & \multirow{2}*{0.0319}     \\
								           & bc + bc  & 6.983$\times 10^{-4}$  &  0.517  \\
                                                            
\multirow{2}*{\textbf{CD RL B1$^{+}$}}    & bc + vbc & 1.289$\times 10^{-3}$ & 1.095  &   \multirow{2}*{1.220}  & \multirow{2}*{0.0008}     \\
								     & bc + bc  & 1.057$\times 10^{-3}$ & 0.896  \\
                                                           
\multirow{2}*{\textbf{FRII RL B1$^{++}$}} & bc + vbc & 2.024$\times 10^{-3}$ & 1.631  &   \multirow{2}*{1.066} & \multirow{2}*{0.154}     \\
										  & bc + bc  & 2.157$\times 10^{-3}$   & 1.726   &      \\
\hline 
\label{tab:ftest}                                                                              
\end{tabular} 
\end{center}
$^a$  For 1020 dof, the $F$\ value at a 1$\sigma$\ confidence limit is 1.109.
\end{tiny}                                                                                                                                                            
\end{table}


\subsection{Interpretation of the radio pseudo-SFR}
\label{radiosfr}

The origin of radio emission in RQ quasars is still debated \citep[e.g.,][and references therein]{panessaetal19}.  Star formation is considered an important agent especially at lower power (the fraction of star-forming galaxies is increasing at fluxes below 10 mJy \citealt{condon89}), although other mechanisms more closely related to the AGN have been proposed.  

An important result of the present investigation is the identification of the population of RI and RL sources in spectral type A2, A3 and A4. A3 and A4, or in other words xA sources, that are known to have high accretion rate, possibly at a super-Eddington level \citep{duetal16a}.  High \lledd\ (and the defining property of xA sources, $R_\mathrm{FeII}\gtrsim 1$) can coexist with powerful “jetted” radio emission ($\log P_{\nu} \gtrsim 10^{31}$ erg s$^{-1}$ Hz$^{-1}$, in the power ranges of the Population A sources). There is no  physical impossibility in this respect \citep[][and references therein]{czernyetal18}, and recent works consider the relativisitic jet and non-relativistic wide angle outflows as two aspects of an hydromagnetically-driven wind \citep{reynoldsetal17}. Concomitant high accretion and high radio power are observed in  CSS \citep{wu09b,sulenticetal15}. Our view is likely biased by the relative rarity of this sort of sources at low-$z$ \citep{odea98}, and  especially by the scarcity of high-quality optical and UV spectra.  The estimates of  $pSFR$s derived from the radio power should be interpreted carefully.  The basic issue is whether the RL and RI in the xA spectral bins are truly jetted  Pop. A (as RL NLSy1s are believed to be) or dominated by thermal emission in their radio power. 

Since RL jetted sources are expected to have FIR luminosity much below the one expected for their radio power if they were following the RQ FIR-radio relation, they are preferentially excluded from the Bonzini et al. diagram. This is expected to occur also for the RL xA Pop. A of the present sample.  On the one hand, the trend shown in Fig. \ref{fig:bonz} (all xA in agreement with the RQ correlation) suggests that RL and RI xA, {\em that are sources with  power comparable with the ones of RI and RD} ($\gtrsim 10^{31}$ \ergss/Hz), may be intrinsically RQ and with power due to thermal process. Even if they are intrinsically rare sources, thermal sources emitting $P_\nu \gtrsim 10^{31}$ \ergss/Hz are found in the local Universe  \citep{condon89}.  The  increase in prevalence of RL Pop. A objects after a deep minimum in correspondence of spectral type A2 calls into question the notion that the origin of the radio power in ST A3 and A4 might be necessarily the same of the jetted sources of Pop. B.   On the other hand, RL NLSy1s do exist, and may show significant FeII emission. They may be showing a large blueshift in the \oiiiopt\ profiles \citep{bertonetal16,komossaetal18a}. therefore RL NLSy1s may appear undistinguishable in the \hb\ spectral range from their RQ counterparts. The radio power may be associated with a relatively small black hole mass. They could be young systems for which a spin-up has been made possible by a recent accretion event \citep[e.g.,][]{mathur00}. However, a large fraction (75\%) of RL NLSy1s shows mid-IR emission properties consistent with star formation being the dominant component also in radio emission \citep{caccianigaetal15}. It is therefore legitimate to suggest that  most of the sources of the present samples that are xA and RI and RL might be truly ``thermal sources.'' In other words, the analysis of Fig. \ref{fig:bonz} supports the suggestion that RI and RL A2, A3, A4 sources could be thermal in origin but lack of data leaves open the possibility that they could be jetted RL NLSy1s of relatively low radio-power.  

It is out of the question that the RL and RI  sources in Pop. B and at least  part   of the sources in A1 owe their enormous power to the contribution of  relativistic jets, with the FRII being their parent population. The compact sources that are RL in the xA spectral types, however, do not  have a parent population of extended sources, at least in our sample: if the xA CDs were a sample of sources preferentially oriented almost face-on, we should observe a sample of FRII in bin B3, and B4 or at least B2. These sources either do not exist or are lost.  This finding provides circumstantial evidence that, even for RL sources, in spectral type A3 and A4 the compactness might not be associated with a preferential orientation, and could instead be related to age (as expected for CSS and GPS e.g., \citealt{odea98}).  If this is the case, FRII in bin B2 and B3 could be in their earliest stage of development and not well-resolvable in radio maps. The CSS could be the “parent population” of NLSy1s  \citep{bertonetal15} if understood as the more evolved counterpart of the low-power flat-spectrum NLSy1s.  Increase in FWHM leading to a population of B3 and B4 might be expected for larger masses radiating close to the Eddington limit if they were seen at relatively high viewing angles. However, their absence  may not be  related to orientation but instead to the global downsizing of nuclear activity at low-$z$ \citep{hopkinsetal06}.


In summary, the data of Table \ref{tab:4} suggest that A2+xAs RD and RI CD and even RL {\em could be possibly consistent} with $SFR$\  associated with powerful star formation phenomena, on average. This hypothesis is supported by the behavior of xA sources in Fig. \ref{fig:bonz}: especially striking is the result for xA sources that are known by the FIR data to be highly star-forming system. Our analysis has identified a small sample of sources that are the best candidates (RL CD A3) to test the $SFR$\ contribution to the total radio power. The nature of these systems (listed in Table \ref{tab:xA}) should be checked by high-resolution radio mapping and by FIR and optical observations on a source-by-source basis.  The sources listed in Table \ref{tab:xA} should be considered preferential targets to address the issue of whether their radio power (and the one of RL NLSy1s as a class) is “thermal,” due to a relativistic jet, or a mixture of both.

\section{Conclusion}

The analysis of $\sim 350$ type-1 AGNs selected on the basis of radio  detection allowed us to overcome a difficulty of optically selected samples, where RL and RI are a minority. In the sample of this paper, there is a significant number of RD, RI, RL sources in each spectral types along the MS. The effect of radio-loudness can be studied within each spectral type, that is, for sources that are believed to be otherwise in similar physical/orientation conditions \citep{sulenticetal02}.  The analysis allowed the identification of several trends involving the \hb\  and the \oiii\ line profiles associated with radio-loudness, or in other words, orthogonal with respect the E1 MS.

 More in detail, the results of the present investigation can be summarized as follows: 
\begin{itemize}
\item The prevalences of RL, RI, RD sources are different along the main sequence. RLs are more frequent in Pop. B and among the A1 spectral type. This result confirms \citet{zamfiretal08} who found that powerful jetted sources occupy the  Population B region of the AGNs MS. 
\item In the extreme Population A we found a relatively high prevalence of RI and RL; RI account for the majority of sources in CD A3 and A4 bins.  This result is somewhat unexpected.  The RI and even RL power is compatible with ``thermal'' origin of the radio emission. The origin of the radio power that leads to a classification of RI and even RL remains to be clarified. There are independent lines of evidence associating high \lledd\ as found in xA with high $SFR$; within the context of the present work,  RI  xA sources share similar properties in $R_\mathrm{FeII}$ and in \lledd\ as in the RD sample (which can be considered RQ). On the other hand, high \lledd\ and the occurrence of relativistic jets are not mutually exclusive, so that it is possible that a fraction of the sources might be “jetted” CSS or flat spectrum RL sources.  We present a sample of xA sources that deserves further observations to ascertain whether the dominant radio emission mechanism is thermal or due to a relativistic jet or a mixture of the two.  
\item To confirm the suspect of thermal emission, the FIR luminosity was computed for seven sources for which data are available, along with other sources for which FIR data from Herschel are available. The location in the plot $SFR$(FIR) vs $SFR_\mathrm{radio}$ confirms that xA sources are predominantly RQ sources even at the highest radio power.
\item By comparing RD, RI, RL in each spectral bin, we see that radio loudness affects the H$\beta$ and the [OIII] properties as well. [OIII] remains strong also in extreme Pop. A RL, whereas it becomes fainter in RD and RI. The appearance of the [OIII] profile is also more symmetric in RL sources of Pop. B. In spectral type A3 also RL show a prominent displacement to the blue at 1/4 and 1/2 maximum. RI in the A4 spectral bin show a high amplitude blueshift also at peak, and the composite spectrum almost qualifies as a blue outlier following \citet{zamanovetal02}. Since extreme blueshifts are possible also among RL sources \citep{bertonetal16,komossaetal18}, this results needs to be interpreted by detailed studies of individual sources. 
\item    Searches for BBHs are undergoing  and comparison between models and observations is becoming more refined  \citep{nguyenetal19} but is not yet clear how frequent sub-parsec BBHs may be, and even how frequent they might be detectable \citep{kelleyetal18}.  Alternative interpretations of emission line blueshifts such as  outflows associated with the accretion disk remain possible.  We suggest that an efficient search for Doppler detection of BBHs should be focused on the extreme Population B (B1$^{++}$) where the profile may become consistent with a double BLR. 
\end{itemize}







\begin{acknowledgements}
VG acknowledges the COST Action CA16104 “GWverse”, supported by COST (European Cooperation in Science and Technology), and is grateful for the hospitality offered at the Belgrade Astronomical Observatory where this work was advanced. PM and MDO acknowledge funding from  the INAF PRIN-SKA 2017 program 1.05.01.88.04. PM also acknowledges the Programa de Estancias de Investigaci\'on (PREI) No. DGAP/DFA/2192/2018 of UNAM. AdO acknowledges financial support from the Spanish Ministry of Economy and Competitiveness through grant AYA2016-76682-C3-1-P and from the State Agency for Research of the Spanish MCIU through the “Center of Excellence Severo Ochoa" award for the Instituto de Astrofísica de Andalucía (SEV-2017-0709). The relevant research is part of the project 176001 ''Astrophysical spectroscopy of extragalactic objects`` supported by the Ministry of Education, Science and Technological Development of the Republic of Serbia.

This research has made use of the NASA/ IPAC Infrared Science Archive, which is operated by the Jet Propulsion Laboratory, California Institute of Technology, under contract with the National Aeronautics and Space Administration.  

Funding for the Sloan Digital Sky Survey (SDSS) has been provided by the Alfred P. Sloan Foundation, the Participating Institutions, the National Aeronautics and Space Administration, the National Science Foundation, the U.S. Department of Energy, the Japanese Monbukagakusho, and the Max Planck Society. The SDSS Web site is http://www.sdss.org/.

The SDSS is managed by the Astrophysical Research Consortium (ARC) for the Participating Institutions. The Participating Institutions are The University of Chicago, Fermilab, the Institute for Advanced Study, the Japan Participation Group, The Johns Hopkins University, Los Alamos National Laboratory, the Max-Planck-Institute for Astronomy (MPIA), the Max-Planck-Institute for Astrophysics (MPA), New Mexico State University, University of Pittsburgh, Princeton University, the United States Naval Observatory, and the University of Washington.

\end{acknowledgements}
\vfill\eject\newpage\pagebreak
\vfill\eject\newpage\pagebreak

\bibliographystyle{aa} 
\bibliography{main}

\begin{thebibliography}{145}
\expandafter\ifx\csname natexlab\endcsname\relax\def\natexlab#1{#1}\fi

\bibitem[{{Ackermann} {et~al.}(2012){Ackermann}, {Ajello}, {Allafort},
  {Baldini}, {Ballet}, {Barbiellini}, {Bastieri}, {Bechtol}, {Bellazzini},
  {Berenji}, {Bloom}, {Bonamente}, {Borgland}, {Bregeon}, {Brigida}, {Bruel},
  {Buehler}, {Buson}, {Caliandro}, {Cameron}, {Caraveo}, {Casandjian},
  {Cavazzuti}, {Cecchi}, {Charles}, {Chekhtman}, {Cheung}, {Chiang}, {Ciprini},
  {Claus}, {Cohen-Tanugi}, {Conrad}, {Cutini}, {D'Ammando}, {de Angelis}, {de
  Palma}, {Dermer}, {Silva}, {Drell}, {Drlica-Wagner}, {Enoto}, {Favuzzi},
  {Fegan}, {Ferrara}, {Fortin}, {Fukazawa}, {Fusco}, {Gargano}, {Gasparrini},
  {Gehrels}, {Germani}, {Giglietto}, {Giommi}, {Giordano}, {Giroletti},
  {Godfrey}, {Grove}, {Guiriec}, {Hadasch}, {Hayashida}, {Hays}, {Hughes},
  {J{\'o}hannesson}, {Johnson}, {Kamae}, {Katagiri}, {Kataoka},
  {Kn{\"o}dlseder}, {Kuss}, {Lande}, {Llena Garde}, {Longo}, {Loparco}, {Lott},
  {Lovellette}, {Lubrano}, {Madejski}, {Mazziotta}, {Michelson}, {Mizuno},
  {Monte}, {Monzani}, {Morselli}, {Moskalenko}, {Murgia}, {Nishino}, {Norris},
  {Nuss}, {Ohno}, {Ohsugi}, {Okumura}, {Orlando}, {Ozaki}, {Paneque},
  {Pesce-Rollins}, {Pierbattista}, {Piron}, {Pivato}, {Porter}, {Rain{\`o}},
  {Rando}, {Razzano}, {Reimer}, {Reimer}, {Ritz}, {Roth}, {Sanchez}, {Sbarra},
  {Sgr{\`o}}, {Siskind}, {Spandre}, {Spinelli}, {Stawarz}, {Strong},
  {Takahashi}, {Takahashi}, {Tanaka}, {Thayer}, {Thompson}, {Tibaldo},
  {Tinivella}, {Torres}, {Tosti}, {Troja}, {Uchiyama}, {Usher},
  {Vandenbroucke}, {Vasileiou}, {Vianello}, {Vitale}, {Waite}, {Winer}, {Wood},
  {Wood}, {Yang}, \& {Zimmer}}]{ackermannetal12}
{Ackermann}, M., {Ajello}, M., {Allafort}, A., {et~al.} 2012, \apj, 747, 104

\bibitem[{{Adhikari} {et~al.}(2016){Adhikari}, {R{\'o}{\.z}a{\'n}ska},
  {Czerny}, {Hryniewicz}, \& {Ferland}}]{adhikarietal16}
{Adhikari}, T.~P., {R{\'o}{\.z}a{\'n}ska}, A., {Czerny}, B., {Hryniewicz}, K.,
  \& {Ferland}, G.~J. 2016, \apj, 831, 68

\bibitem[{{Ai} {et~al.}(2010){Ai}, {Yuan}, {Zhou}, {Wang}, {Dong}, {Wang}, \&
  {Lu}}]{aietal10}
{Ai}, Y.~L., {Yuan}, W., {Zhou}, H.~Y., {et~al.} 2010, \apjl, 716, L31

\bibitem[{{Assef} {et~al.}(2011){Assef}, {Denney}, {Kochanek}, {Peterson},
  {Koz{\l}owski}, {Ageorges}, {Barrows}, {Buschkamp}, {Dietrich}, {Falco},
  {Feiz}, {Gemperlein}, {Germeroth}, {Grier}, {Hofmann}, {Juette}, {Khan},
  {Kilic}, {Knierim}, {Laun}, {Lederer}, {Lehmitz}, {Lenzen}, {Mall}, {Madsen},
  {Mandel}, {Martini}, {Mathur}, {Mogren}, {Mueller}, {Naranjo}, {Pasquali},
  {Polsterer}, {Pogge}, {Quirrenbach}, {Seifert}, {Stern}, {Shappee}, {Storz},
  {Van Saders}, {Weiser}, \& {Zhang}}]{assefetal11}
{Assef}, R.~J., {Denney}, K.~D., {Kochanek}, C.~S., {et~al.} 2011, \apj, 742,
  93

\bibitem[{{Barger} {et~al.}(2014){Barger}, {Cowie}, {Chen}, {Owen}, {Wang},
  {Casey}, {Lee}, {Sanders}, \& {Williams}}]{bargeretal14}
{Barger}, A.~J., {Cowie}, L.~L., {Chen}, C.~C., {et~al.} 2014, \apj, 784, 9

\bibitem[{{Becker} {et~al.}(1995){Becker}, {White}, \&
  {Helfand}}]{beckeretal95}
{Becker}, R.~H., {White}, R.~L., \& {Helfand}, D.~J. 1995, \apj, 450, 559

\bibitem[{{Begelman} {et~al.}(1980){Begelman}, {Hatchett}, {McKee}, {Sarazin},
  \& {Arons}}]{begelmanetal80}
{Begelman}, M.~C., {Hatchett}, S.~P., {McKee}, C.~F., {Sarazin}, C.~L., \&
  {Arons}, J. 1980, \apj, 238, 722

\bibitem[{{Berton} {et~al.}(2016){Berton}, {Foschini}, {Ciroi}, {Cracco}, {La
  Mura}, {Di Mille}, \& {Rafanelli}}]{bertonetal16}
{Berton}, M., {Foschini}, L., {Ciroi}, S., {et~al.} 2016, \aap, 591, A88

\bibitem[{{Berton} {et~al.}(2015){Berton}, {Foschini}, {Ciroi}, {Cracco}, {La
  Mura}, {Lister}, {Mathur}, {Peterson}, {Richards}, \&
  {Rafanelli}}]{bertonetal15}
{Berton}, M., {Foschini}, L., {Ciroi}, S., {et~al.} 2015, \aap, 578, A28

\bibitem[{{Bian} {et~al.}(2005){Bian}, {Yuan}, \& {Zhao}}]{bianetal05}
{Bian}, W., {Yuan}, Q., \& {Zhao}, Y. 2005, MNRAS, 364, 187

\bibitem[{{Blandford} \& {Rees}(1978)}]{blandfordrees78}
{Blandford}, R.~D. \& {Rees}, M.~J. 1978, \physscr, 17, 265

\bibitem[{{Blandford} \& {Znajek}(1977)}]{blandfordznajek77}
{Blandford}, R.~D. \& {Znajek}, R.~L. 1977, \mnras, 179, 433

\bibitem[{{Boller} {et~al.}(1996){Boller}, {Brandt}, \& {Fink}}]{bolleretal96}
{Boller}, T., {Brandt}, W.~N., \& {Fink}, H. 1996, \aap, 305, 53

\bibitem[{{Bon} {et~al.}(2012){Bon}, {Jovanovi{\'c}}, {Marziani},
  {Shapovalova}, {Bon}, {Borka Jovanovi{\'c}}, {Borka}, {Sulentic}, \&
  {Popovi{\'c}}}]{bonetal12}
{Bon}, E., {Jovanovi{\'c}}, P., {Marziani}, P., {et~al.} 2012, \apj, 759, 118

\bibitem[{{Bon} {et~al.}(2006){Bon}, {Popovi{\'c}}, {Ili{\'c}}, \&
  {Mediavilla}}]{bonetal06}
{Bon}, E., {Popovi{\'c}}, L.~{\v{C}}., {Ili{\'c}}, D., \& {Mediavilla}, E.
  2006, \nar, 50, 716

\bibitem[{{Bon} {et~al.}(2015){Bon}, {Bon}, {Marziani}, \&
  {Jovanovi{\'c}}}]{bonetal15}
{Bon}, N., {Bon}, E., {Marziani}, P., \& {Jovanovi{\'c}}, P. 2015, \apss, 360,
  7

\bibitem[{{Bonzini} {et~al.}(2015){Bonzini}, {Mainieri}, {Padovani},
  {Andreani}, {Berta}, {Bethermin}, {Lutz}, {Rodighiero}, {Rosario}, {Tozzi},
  \& {Vattakunnel}}]{bonzinietal15}
{Bonzini}, M., {Mainieri}, V., {Padovani}, P., {et~al.} 2015, \mnras, 453, 1079

\bibitem[{{Boroson} \& {Green}(1992)}]{borosongreen92}
{Boroson}, T.~A. \& {Green}, R.~F. 1992, ApJS, 80, 109

\bibitem[{{Brotherton} {et~al.}(1994){Brotherton}, {Wills}, {Francis}, \&
  {Steidel}}]{brothertonetal94a}
{Brotherton}, M.~S., {Wills}, B.~J., {Francis}, P.~J., \& {Steidel}, C.~C.
  1994, \apj, 430, 495

\bibitem[{{Caccianiga} {et~al.}(2015){Caccianiga}, {Ant{\'o}n}, {Ballo},
  {Foschini}, {Maccacaro}, {Della Ceca}, {Severgnini}, {March{\~a}}, {Mateos},
  \& {Sani}}]{caccianigaetal15}
{Caccianiga}, A., {Ant{\'o}n}, S., {Ballo}, L., {et~al.} 2015, \mnras, 451,
  1795

\bibitem[{{Calzetti}(2013)}]{calzetti13a}
{Calzetti}, D. 2013, {Star Formation Rate Indicators} (Cambridge, UK: Cambridge
  University Press), 419

\bibitem[{{Cirasuolo} {et~al.}(2003){Cirasuolo}, {Magliocchetti}, {Celotti}, \&
  {Danese}}]{cirasuoloetal03}
{Cirasuolo}, M., {Magliocchetti}, M., {Celotti}, A., \& {Danese}, L. 2003,
  \mnras, 341, 993

\bibitem[{{Coatman} {et~al.}(2019){Coatman}, {Hewett}, {Banerji}, {Richards},
  {Hennawi}, \& {Xavier Prochaska}}]{coatmanetal19}
{Coatman}, L., {Hewett}, P.~C., {Banerji}, M., {et~al.} 2019, \mnras, 1122

\bibitem[{{Comparat} {et~al.}(2013){Comparat}, {Kneib}, {Bacon}, {Mostek},
  {Newman}, {Schlegel}, \& {Y{\`e}che}}]{comparatetal13}
{Comparat}, J., {Kneib}, J.-P., {Bacon}, R., {et~al.} 2013, \aap, 559, A18

\bibitem[{{Condon}(1989)}]{condon89}
{Condon}, J.~J. 1989, \apj, 338, 13

\bibitem[{{Condon}(1992)}]{condon92}
{Condon}, J.~J. 1992, \araa, 30, 575

\bibitem[{{Condon} {et~al.}(1991){Condon}, {Huang}, {Yin}, \&
  {Thuan}}]{condonetal91}
{Condon}, J.~J., {Huang}, Z.~P., {Yin}, Q.~F., \& {Thuan}, T.~X. 1991, \apj,
  378, 65

\bibitem[{{Corbin} \& {Boroson}(1996)}]{corbinboroson96}
{Corbin}, M.~R. \& {Boroson}, T.~A. 1996, \apjs, 107, 69

\bibitem[{{Cracco} {et~al.}(2016){Cracco}, {Ciroi}, {Berton}, {Di Mille},
  {Foschini}, {La Mura}, \& {Rafanelli}}]{craccoetal16}
{Cracco}, V., {Ciroi}, S., {Berton}, M., {et~al.} 2016, \mnras, 462, 1256

\bibitem[{{Cutini} {et~al.}(2014){Cutini}, {Ciprini}, {Orienti}, {Tramacere},
  {D'Ammando}, {Verrecchia}, {Polenta}, {Carrasco}, {D'Elia}, {Giommi},
  {Gonz{\'a}lez-Nuevo}, {Grandi}, {Harrison}, {Hays}, {Larsson},
  {L{\"a}hteenm{\"a}ki}, {Le{\'o}n-Tavares}, {L{\'o}pez-Caniego}, {Natoli},
  {Ojha}, {Partridge}, {Porras}, {Reyes}, {Recillas}, \&
  {Torresi}}]{cutinietal14}
{Cutini}, S., {Ciprini}, S., {Orienti}, M., {et~al.} 2014, \mnras, 445, 4316

\bibitem[{{Czerny} {et~al.}(2018){Czerny}, {Beaton}, {Bejger}, {Cackett},
  {Dall'Ora}, {Holanda}, {Jensen}, {Jha}, {Lusso}, {Minezaki}, {Risaliti},
  {Salaris}, {Toonen}, \& {Yoshii}}]{czernyetal18}
{Czerny}, B., {Beaton}, R., {Bejger}, M., {et~al.} 2018, Space Science Reviews,
  214, \#32

\bibitem[{{Dawson} {et~al.}(2013){Dawson}, {Schlegel}, {Ahn}, {Anderson},
  {Aubourg}, {Bailey}, {Barkhouser}, {Bautista}, {Beifiori}, {Berlind},
  {Bhardwaj}, {Bizyaev}, {Blake}, {Blanton}, {Blomqvist}, {Bolton}, {Borde},
  {Bovy}, {Brandt}, {Brewington}, {Brinkmann}, {Brown}, {Brownstein}, {Bundy},
  {Busca}, {Carithers}, {Carnero}, {Carr}, {Chen}, {Comparat}, {Connolly},
  {Cope}, {Croft}, {Cuesta}, {da Costa}, {Davenport}, {Delubac}, {de Putter},
  {Dhital}, {Ealet}, {Ebelke}, {Eisenstein}, {Escoffier}, {Fan}, {Filiz Ak},
  {Finley}, {Font-Ribera}, {G{\'e}nova-Santos}, {Gunn}, {Guo}, {Haggard},
  {Hall}, {Hamilton}, {Harris}, {Harris}, {Ho}, {Hogg}, {Holder}, {Honscheid},
  {Huehnerhoff}, {Jordan}, {Jordan}, {Kauffmann}, {Kazin}, {Kirkby}, {Klaene},
  {Kneib}, {Le Goff}, {Lee}, {Long}, {Loomis}, {Lundgren}, {Lupton}, {Maia},
  {Makler}, {Malanushenko}, {Malanushenko}, {Mandelbaum}, {Manera}, {Maraston},
  {Margala}, {Masters}, {McBride}, {McDonald}, {McGreer}, {McMahon}, {Mena},
  {Miralda-Escud{\'e}}, {Montero-Dorta}, {Montesano}, {Muna}, {Myers},
  {Naugle}, {Nichol}, {Noterdaeme}, {Nuza}, {Olmstead}, {Oravetz}, {Oravetz},
  {Owen}, {Padmanabhan}, {Palanque-Delabrouille}, {Pan}, {Parejko},
  {P{\^a}ris}, {Percival}, {P{\'e}rez-Fournon}, {P{\'e}rez-R{\`a}fols},
  {Petitjean}, {Pfaffenberger}, {Pforr}, {Pieri}, {Prada}, {Price-Whelan},
  {Raddick}, {Rebolo}, {Rich}, {Richards}, {Rockosi}, {Roe}, {Ross}, {Ross},
  {Rossi}, {Rubi{\~n}o-Martin}, {Samushia}, {S{\'a}nchez}, {Sayres}, {Schmidt},
  {Schneider}, {Sc{\'o}ccola}, {Seo}, {Shelden}, {Sheldon}, {Shen}, {Shu},
  {Slosar}, {Smee}, {Snedden}, {Stauffer}, {Steele}, {Strauss}, {Streblyanska},
  {Suzuki}, {Swanson}, {Tal}, {Tanaka}, {Thomas}, {Tinker}, {Tojeiro},
  {Tremonti}, {Vargas Maga{\~n}a}, {Verde}, {Viel}, {Wake}, {Watson}, {Weaver},
  {Weinberg}, {Weiner}, {West}, {White}, {Wood-Vasey}, {Yeche}, {Zehavi},
  {Zhao}, \& {Zheng}}]{dawsonetal13}
{Dawson}, K.~S., {Schlegel}, D.~J., {Ahn}, C.~P., {et~al.} 2013, \aj, 145, 10

\bibitem[{{de Vries} {et~al.}(2006){de Vries}, {Becker}, \&
  {White}}]{devriesetal06}
{de Vries}, W.~H., {Becker}, R.~H., \& {White}, R.~L. 2006, \aj, 131, 666

\bibitem[{{De Young}(1984)}]{deyoung84}
{De Young}, D.~S. 1984, \physrep, 111, 373

\bibitem[{{de Young}(2002)}]{deyoung02}
{de Young}, D.~S. 2002, {The physics of extragalactic radio sources} (The
  University of Chicago Press)

\bibitem[{{Dong} {et~al.}(2011){Dong}, {Wang}, {Ho}, {Wang}, {Fan}, {Wang},
  {Zhou}, \& {Yuan}}]{dongetal11}
{Dong}, X.-B., {Wang}, J.-G., {Ho}, L.~C., {et~al.} 2011, \apj, 736, 86

\bibitem[{{Du} {et~al.}(2016{\natexlab{a}}){Du}, {Lu}, {Hu}, {Qiu}, {Li},
  {Huang}, {Wang}, {Bai}, {Bian}, {Yuan}, {Ho}, {Wang}, \& {SEAMBH
  Collaboration}}]{duetal16}
{Du}, P., {Lu}, K.-X., {Hu}, C., {et~al.} 2016{\natexlab{a}}, \apj, 820, 27

\bibitem[{{Du} {et~al.}(2016{\natexlab{b}}){Du}, {Wang}, {Hu}, {Ho}, {Li}, \&
  {Bai}}]{duetal16a}
{Du}, P., {Wang}, J.-M., {Hu}, C., {et~al.} 2016{\natexlab{b}}, \apjl, 818, L14

\bibitem[{{Dubner} \& {Giacani}(2015)}]{dubnergiacani15}
{Dubner}, G. \& {Giacani}, E. 2015, Astronomy and Astrophysics Review, 23, 3

\bibitem[{{Duras} {et~al.}(2017){Duras}, {Bongiorno}, {Piconcelli}, {Bianchi},
  {Pappalardo}, {Valiante}, {Bischetti}, {Feruglio}, {Martocchia}, {Schneider},
  {Vietri}, {Vignali}, {Zappacosta}, {La Franca}, \& {Fiore}}]{durasetal17}
{Duras}, F., {Bongiorno}, A., {Piconcelli}, E., {et~al.} 2017, \aap, 604, A67

\bibitem[{{Falcke} {et~al.}(1996){Falcke}, {Sherwood}, \&
  {Patnaik}}]{falckeetal96}
{Falcke}, H., {Sherwood}, W., \& {Patnaik}, A.~R. 1996, \apj, 471, 106

\bibitem[{{Fernini}(2014)}]{fernini14}
{Fernini}, I. 2014, \apjs, 212, 19

\bibitem[{{Foschini } {et~al.}(2015){Foschini }, {Berton}, {Caccianiga},
  {Ciroi}, {Cracco}, {Peterson}, {Angelakis}, {Braito}, {Fuhrmann}, {Gallo},
  {Grupe}, {J\"arvel\"a}, {Kaufmann}, {Komossa}, {Kovalev}, {L\"ahteenm\"aki},
  {Lisakov}, {Lister}, {Mathur, S.}, {Richards, J. L.}, {Romano, P.}, {Sievers,
  A.}, {Tagliaferri, G.}, {Tammi, J.}, {Tibolla, O.}, {Tornikoski, M.},
  {Vercellone, S.}, {La Mura, G.}, {Maraschi, L.}, \& {Rafanelli,
  P.}}]{foschinietal15}
{Foschini }, L., {Berton}, M., {Caccianiga}, A., {et~al.} 2015, A\&A, 575, A13

\bibitem[{Fraix-Burnet {et~al.}(2017)Fraix-Burnet, Marziani, D'Onofrio, \&
  Dultzin}]{fraix-burnetetal17}
Fraix-Burnet, D., Marziani, P., D'Onofrio, M., \& Dultzin, D. 2017, Frontiers
  in Astronomy and Space Sciences, 4, 1

\bibitem[{{Gallimore} {et~al.}(2006){Gallimore}, {Axon}, {O'Dea}, {Baum}, \&
  {Pedlar}}]{gallimoreetal06}
{Gallimore}, J.~F., {Axon}, D.~J., {O'Dea}, C.~P., {Baum}, S.~A., \& {Pedlar},
  A. 2006, \aj, 132, 546

\bibitem[{{Gaskell}(1985)}]{gaskell85}
{Gaskell}, C.~M. 1985, \nat, 315, 386

\bibitem[{{Gawro{\'n}ski} {et~al.}(2006){Gawro{\'n}ski}, {Marecki},
  {Kunert-Bajraszewska}, \& {Kus}}]{gawronskietal06}
{Gawro{\'n}ski}, M.~P., {Marecki}, A., {Kunert-Bajraszewska}, M., \& {Kus},
  A.~J. 2006, \aap, 447, 63

\bibitem[{{Graham} {et~al.}(1996){Graham}, {Clowes}, \&
  {Campusano}}]{grahametal96}
{Graham}, M.~J., {Clowes}, R.~G., \& {Campusano}, L.~E. 1996, \mnras, 279, 1349

\bibitem[{{Grupe} {et~al.}(1999){Grupe}, {Beuermann}, {Mannheim}, \&
  {Thomas}}]{grupeetal99}
{Grupe}, D., {Beuermann}, K., {Mannheim}, K., \& {Thomas}, H.-C. 1999, \aap,
  350, 805

\bibitem[{{Gupta} {et~al.}(2018){Gupta}, {Sikora}, {Rusinek}, \&
  {Madejski}}]{guptaetal18}
{Gupta}, M., {Sikora}, M., {Rusinek}, K., \& {Madejski}, G.~M. 2018, \mnras,
  480, 2861

\bibitem[{{G{\"u}rkan} {et~al.}(2015){G{\"u}rkan}, {Hardcastle}, {Jarvis},
  {Smith}, {Bourne}, {Dunne}, {Maddox}, {Ivison}, \& {Fritz}}]{gurkanetal15}
{G{\"u}rkan}, G., {Hardcastle}, M.~J., {Jarvis}, M.~J., {et~al.} 2015, \mnras,
  452, 3776

\bibitem[{{Haas} {et~al.}(2003){Haas}, {Klaas}, {M{\"u}ller}, {Bertoldi},
  {Camenzind}, {Chini}, {Krause}, {Lemke}, {Meisenheimer}, {Richards}, \&
  {Wilkes}}]{haasetal03}
{Haas}, M., {Klaas}, U., {M{\"u}ller}, S.~A.~H., {et~al.} 2003, \aap, 402, 87

\bibitem[{{Hopkins} {et~al.}(2006){Hopkins}, {Hernquist}, {Cox}, {Di Matteo},
  {Robertson}, \& {Springel}}]{hopkinsetal06}
{Hopkins}, P.~F., {Hernquist}, L., {Cox}, T.~J., {et~al.} 2006, \apjs, 163, 1

\bibitem[{{Hu} {et~al.}(2008){Hu}, {Wang}, {Ho}, {Chen}, {Bian}, \&
  {Xue}}]{huetal08}
{Hu}, C., {Wang}, J.-M., {Ho}, L.~C., {et~al.} 2008, ApJL, 683, L115

\bibitem[{Ito {et~al.}(2006)Ito, Kino, Kawakatu, Isobe, \&
  Yamada}]{hirotakaetal06}
Ito, H., Kino, M., Kawakatu, N., Isobe, N., \& Yamada, S. 2006, Journal of
  Physics: Conference Series, 31, 201

\bibitem[{{Ivezi{\'c}} {et~al.}(2002){Ivezi{\'c}}, {Menou}, {Knapp}, {Strauss},
  {Lupton}, {Vand en Berk}, {Richards}, {Tremonti}, {Weinstein}, {Anderson},
  {Bahcall}, {Becker}, {Bernardi}, {Blanton}, {Eisenstein}, {Fan},
  {Finkbeiner}, {Finlator}, {Frieman}, {Gunn}, {Hall}, {Kim}, {Kinkhabwala},
  {Narayanan}, {Rockosi}, {Schlegel}, {Schneider}, {Strateva}, {SubbaRao},
  {Thakar}, {Voges}, {White}, {Yanny}, {Brinkmann}, {Doi}, {Fukugita},
  {Hennessy}, {Munn}, {Nichol}, \& {York}}]{ivezicetal02}
{Ivezi{\'c}}, {\v{Z}}., {Menou}, K., {Knapp}, G.~R., {et~al.} 2002, \aj, 124,
  2364

\bibitem[{{Kellermann} {et~al.}(1989){Kellermann}, {Sramek}, {Schmidt},
  {Shaffer}, \& {Green}}]{kellermannetal89}
{Kellermann}, K.~I., {Sramek}, R., {Schmidt}, M., {Shaffer}, D.~B., \& {Green},
  R. 1989, \aj, 98, 1195

\bibitem[{{Kelley} {et~al.}(2018){Kelley}, {Haiman}, {Sesana}, \&
  {Hernquist}}]{kelleyetal18}
{Kelley}, L.~Z., {Haiman}, Z., {Sesana}, A., \& {Hernquist}, L. 2018, arXiv
  e-prints [\eprint[arXiv]{1809.02138}]

\bibitem[{{Kimball} {et~al.}(2011){Kimball}, {Ivezi{\'c}}, {Wiita}, \&
  {Schneider}}]{kimballetal11}
{Kimball}, A.~E., {Ivezi{\'c}}, {\v Z}., {Wiita}, P.~J., \& {Schneider}, D.~P.
  2011, \aj, 141, 182

\bibitem[{{King} \& {Pounds}(2015)}]{kingpounds15}
{King}, A. \& {Pounds}, K. 2015, \araa, 53, 115

\bibitem[{{Komossa} {et~al.}(2006){Komossa}, {Voges}, {Xu}, {Mathur}, {Adorf},
  {Lemson}, {Duschl}, \& {Grupe}}]{komossaetal06}
{Komossa}, S., {Voges}, W., {Xu}, D., {et~al.} 2006, \aj, 132, 531

\bibitem[{{Komossa} {et~al.}(2008){Komossa}, {Xu}, {Zhou}, {Storchi-Bergmann},
  \& {Binette}}]{komossaetal08}
{Komossa}, S., {Xu}, D., {Zhou}, H., {Storchi-Bergmann}, T., \& {Binette}, L.
  2008, \apj, 680, 926

\bibitem[{{Komossa} {et~al.}(2018{\natexlab{a}}){Komossa}, {Xu}, \&
  {Wagner}}]{komossaetal18a}
{Komossa}, S., {Xu}, D.~W., \& {Wagner}, A.~Y. 2018{\natexlab{a}}, \mnras, 477,
  5115

\bibitem[{{Komossa} {et~al.}(2018{\natexlab{b}}){Komossa}, {Xu}, \&
  {Wagner}}]{komossaetal18}
{Komossa}, S., {Xu}, D.~W., \& {Wagner}, A.~Y. 2018{\natexlab{b}}, \mnras, 477,
  5115

\bibitem[{{Kriss}(1994)}]{kriss94}
{Kriss}, G. 1994, Astronomical Data Analysis Software and Systems III, A.S.P.
  Conference Series, 61, 437

\bibitem[{{Kuraszkiewicz} {et~al.}(2009){Kuraszkiewicz}, {Wilkes}, {Schmidt},
  {Smith}, {Cutri}, \& {Czerny}}]{kuraszkiewiczetal09}
{Kuraszkiewicz}, J., {Wilkes}, B.~J., {Schmidt}, G., {et~al.} 2009, \apj, 692,
  1180

\bibitem[{{La Franca} {et~al.}(2010){La Franca}, {Melini}, \&
  {Fiore}}]{lafrancaetal10}
{La Franca}, F., {Melini}, G., \& {Fiore}, F. 2010, \apj, 718, 368

\bibitem[{{Li} {et~al.}(2010){Li}, {Calzetti}, {Kennicutt}, {Hong},
  {Engelbracht}, {Dale}, \& {Moustakas}}]{lietal10}
{Li}, Y., {Calzetti}, D., {Kennicutt}, R.~C., {et~al.} 2010, \apj, 725, 677

\bibitem[{{Li} {et~al.}(2016){Li}, {Wang}, {Ho}, {Lu}, {Qiu}, {Du}, {Hu},
  {Huang}, {Zhang}, {Wang}, \& {Bai}}]{lietal16}
{Li}, Y.-R., {Wang}, J.-M., {Ho}, L.~C., {et~al.} 2016, \apj, 822, 4

\bibitem[{{Li} {et~al.}(2015){Li}, {Zhou}, {Hao}, {Wang}, {Ji}, {Shi}, {Liu},
  {Zhang}, {Liu}, {Pan}, \& {Jiang}}]{zietal15}
{Li}, Z., {Zhou}, H., {Hao}, L., {et~al.} 2015, \apj, 812, 99

\bibitem[{{Lipari} {et~al.}(1993){Lipari}, {Terlevich}, \&
  {Macchetto}}]{liparietal93}
{Lipari}, S., {Terlevich}, R., \& {Macchetto}, F. 1993, \apj, 406, 451

\bibitem[{{Magorrian} {et~al.}(1998){Magorrian}, {Tremaine}, {Richstone},
  {Bender}, {Bower}, {Dressler}, {Faber}, {Gebhardt}, {Green}, {Grillmair},
  {Kormendy}, \& {Lauer}}]{magorrianetal98}
{Magorrian}, J., {Tremaine}, S., {Richstone}, D., {et~al.} 1998, \aj, 115, 2285

\bibitem[{{Malizia} {et~al.}(2014){Malizia}, {Molina}, {Bassani}, {Stephen},
  {Bazzano}, {Ubertini}, \& {Bird}}]{maliziaetal14}
{Malizia}, A., {Molina}, M., {Bassani}, L., {et~al.} 2014, \apjl, 782, L25

\bibitem[{{Marconi} {et~al.}(2009){Marconi}, {Axon}, {Maiolino}, {Nagao},
  {Pietrini}, {Risaliti}, {Robinson}, \& {Torricelli}}]{marconietal09}
{Marconi}, A., {Axon}, D.~J., {Maiolino}, R., {et~al.} 2009, \apjl, 698, L103

\bibitem[{{Marin} \& {Antonucci}(2016)}]{marinantonucci16}
{Marin}, F. \& {Antonucci}, R. 2016, \apj, 830, 82

\bibitem[{Mart{\'\i}nez-Aldama {et~al.}(2018)Mart{\'\i}nez-Aldama, Del~Olmo,
  Marziani, Sulentic, Negrete, Dultzin, Perea, \&
  D'Onofrio}]{martinez-aldamaetal18}
Mart{\'\i}nez-Aldama, M.~L., Del~Olmo, A., Marziani, P., {et~al.} 2018,
  Frontiers in Astronomy and Space Sciences, 4, 65

\bibitem[{{Marziani} {et~al.}(2016{\natexlab{a}}){Marziani}, {Mart{\'{\i}}nez
  Carballo}, {Sulentic}, {Del Olmo}, {Stirpe}, \& {Dultzin}}]{marzianietal16a}
{Marziani}, P., {Mart{\'{\i}}nez Carballo}, M.~A., {Sulentic}, J.~W., {et~al.}
  2016{\natexlab{a}}, \apss, 361, 29

\bibitem[{{Marziani} \& {Sulentic}(2014)}]{marzianisulentic14}
{Marziani}, P. \& {Sulentic}, J.~W. 2014, \mnras, 442, 1211

\bibitem[{{Marziani} {et~al.}(2013){Marziani}, {Sulentic}, {Plauchu-Frayn}, \&
  {del Olmo}}]{marzianietal13a}
{Marziani}, P., {Sulentic}, J.~W., {Plauchu-Frayn}, I., \& {del Olmo}, A. 2013,
  AAp, 555, 89, 16pp

\bibitem[{{Marziani} {et~al.}(2016{\natexlab{b}}){Marziani}, {Sulentic},
  {Stirpe}, {Dultzin}, {Del Olmo}, \&
  {Mart{\'{\i}}nez-Carballo}}]{marzianietal16}
{Marziani}, P., {Sulentic}, J.~W., {Stirpe}, G.~M., {et~al.}
  2016{\natexlab{b}}, \apss, 361, 3

\bibitem[{{Marziani} {et~al.}(2003){Marziani}, {Sulentic}, {Zamanov},
  {Calvani}, {Dultzin-Hacyan}, {Bachev}, \& {Zwitter}}]{marzianietal03a}
{Marziani}, P., {Sulentic}, J.~W., {Zamanov}, R., {et~al.} 2003, ApJS, 145, 199

\bibitem[{{Mathur}(2000)}]{mathur00}
{Mathur}, S. 2000, \mnras, 314, L17

\bibitem[{{Mej{\'{\i}}a-Restrepo} {et~al.}(2018){Mej{\'{\i}}a-Restrepo},
  {Lira}, {Netzer}, {Trakhtenbrot}, \& {Capellupo}}]{mejia-restrepoetal18a}
{Mej{\'{\i}}a-Restrepo}, J.~E., {Lira}, P., {Netzer}, H., {Trakhtenbrot}, B.,
  \& {Capellupo}, D.~M. 2018, Nature Astronomy, 2, 63

\bibitem[{{Merritt} \& {Ekers}(2002)}]{merrittekers02}
{Merritt}, D. \& {Ekers}, R.~D. 2002, Science, 297, 1310

\bibitem[{{Middelberg} {et~al.}(2004){Middelberg}, {Roy}, {Nagar}, {Krichbaum},
  {Norris}, {Wilson}, {Falcke}, {Colbert}, {Witzel}, \&
  {Fricke}}]{middelbergetal04}
{Middelberg}, E., {Roy}, A.~L., {Nagar}, N.~M., {et~al.} 2004, \aap, 417, 925

\bibitem[{{Mu{\~n}oz} {et~al.}(2003){Mu{\~n}oz}, {Falco}, {Kochanek},
  {Leh{\'a}r}, \& {Mediavilla}}]{munozetal03}
{Mu{\~n}oz}, J.~A., {Falco}, E.~E., {Kochanek}, C.~S., {Leh{\'a}r}, J., \&
  {Mediavilla}, E. 2003, \apj, 594, 684

\bibitem[{{Negrete} {et~al.}(2018){Negrete}, {Dultzin}, {Marziani}, \& {et
  al.}}]{negreteetal18}
{Negrete}, C.~A., {Dultzin}, D., {Marziani}, P., \& {et al.} 2018, in
  preparation

\bibitem[{{Netzer}(2013)}]{netzer13}
{Netzer}, H. 2013, {The Physics and Evolution of Active Galactic Nuclei}
  (Cambridge University Press)

\bibitem[{{Nguyen} {et~al.}(2019){Nguyen}, {Bogdanovic}, {Runnoe}, {Eracleous},
  {Sigurdsson}, \& {Boroson}}]{nguyenetal19}
{Nguyen}, K., {Bogdanovic}, T., {Runnoe}, J.~C., {et~al.} 2019, arXiv e-prints,
  arXiv:1908.01799

\bibitem[{{Noda} \& {Done}(2018)}]{hirofumidone18}
{Noda}, H. \& {Done}, C. 2018, \mnras, 480, 3898

\bibitem[{{O'Dea}(1998)}]{odea98}
{O'Dea}, C.~P. 1998, \pasp, 110, 493

\bibitem[{{Orr} \& {Browne}(1982)}]{orrbrowne82}
{Orr}, M.~J.~L. \& {Browne}, I.~W.~A. 1982, \mnras, 200, 1067

\bibitem[{{Padovani}(2016)}]{padovani16}
{Padovani}, P. 2016, AApR, 24, 13

\bibitem[{{Padovani}(2017)}]{padovani17}
{Padovani}, P. 2017, Frontiers in Astronomy and Space Sciences, 4, 35

\bibitem[{Panda {et~al.}(2018)Panda, Czerny, Adhikari, Hryniewicz, Wildy,
  Kuraszkiewicz, \& {\'{S}}niegowska}]{pandaetal18}
Panda, S., Czerny, B., Adhikari, T.~P., {et~al.} 2018, The Astrophysical
  Journal, 866, 115

\bibitem[{{Panda} {et~al.}(2019){Panda}, {Marziani}, \& {Czerny}}]{pandaetal19}
{Panda}, S., {Marziani}, P., \& {Czerny}, B. 2019, arXiv e-prints,
  arXiv:1905.01729

\bibitem[{{Panessa} {et~al.}(2019){Panessa}, {Baldi}, {Laor}, {Padovani},
  {Behar}, \& {McHardy}}]{panessaetal19}
{Panessa}, F., {Baldi}, R.~D., {Laor}, A., {et~al.} 2019, Nature Astronomy, 3,
  387

\bibitem[{{P{\^a}ris} {et~al.}(2017){P{\^a}ris}, {Petitjean}, {Ross}, {Myers},
  {Aubourg}, {Streblyanska}, {Bailey}, {Armengaud}, {Palanque-Delabrouille},
  {Y{\`e}che}, {Hamann}, {Strauss}, {Albareti}, {Bovy}, {Bizyaev}, {Niel
  Brandt}, {Brusa}, {Buchner}, {Comparat}, {Croft}, {Dwelly}, {Fan},
  {Font-Ribera}, {Ge}, {Georgakakis}, {Hall}, {Jiang}, {Kinemuchi},
  {Malanushenko}, {Malanushenko}, {McMahon}, {Menzel}, {Merloni}, {Nandra},
  {Noterdaeme}, {Oravetz}, {Pan}, {Pieri}, {Prada}, {Salvato}, {Schlegel},
  {Schneider}, {Simmons}, {Viel}, {Weinberg}, \& {Zhu}}]{parisetal17}
{P{\^a}ris}, I., {Petitjean}, P., {Ross}, N.~P., {et~al.} 2017, \aap, 597, A79

\bibitem[{{Punsly}(2010)}]{punsly10}
{Punsly}, B. 2010, \apj, 713, 232

\bibitem[{{Reynolds} {et~al.}(2017){Reynolds}, {Punsly}, {Miniutti}, {O'Dea},
  \& {Hurley-Walker}}]{reynoldsetal17}
{Reynolds}, C., {Punsly}, B., {Miniutti}, G., {O'Dea}, C.~P., \&
  {Hurley-Walker}, N. 2017, \apj, 836, 155

\bibitem[{{Richards} {et~al.}(2011){Richards}, {Kruczek}, {Gallagher}, {Hall},
  {Hewett}, {Leighly}, {Deo}, {Kratzer}, \& {Shen}}]{richardsetal11}
{Richards}, G.~T., {Kruczek}, N.~E., {Gallagher}, S.~C., {et~al.} 2011, \aj,
  141, 167

\bibitem[{{Rigopoulou} {et~al.}(1996){Rigopoulou}, {Lawrence}, {White},
  {Rowan-Robinson}, \& {Church}}]{rigopoulouetal96}
{Rigopoulou}, D., {Lawrence}, A., {White}, G.~J., {Rowan-Robinson}, M., \&
  {Church}, S.~E. 1996, \aap, 305, 747

\bibitem[{{Sandage}(1965)}]{sandage65}
{Sandage}, A. 1965, \apj, 141, 1560

\bibitem[{{Sanders} {et~al.}(2009){Sanders}, {Kartaltepe}, {Kewley}, {U},
  {Yuan}, {Evans}, {Armus}, \& {Mazzarella}}]{sandersetal09}
{Sanders}, D.~B., {Kartaltepe}, J.~S., {Kewley}, L.~J., {et~al.} 2009, in
  Astronomical Society of the Pacific Conference Series, Vol. 408, The
  Starburst-AGN Connection, ed. W.~{Wang}, Z.~{Yang}, Z.~{Luo}, \& Z.~{Chen}, 3

\bibitem[{{Sanders} \& {Mirabel}(1996)}]{mirabelsanders96}
{Sanders}, D.~B. \& {Mirabel}, I.~F. 1996, \araa, 34, 749

\bibitem[{{Sanders} {et~al.}(1988){Sanders}, {Soifer}, {Elias}, {Madore},
  {Matthews}, {Neugebauer}, \& {Scoville}}]{sandersetal88}
{Sanders}, D.~B., {Soifer}, B.~T., {Elias}, J.~H., {et~al.} 1988, \apj, 325, 74

\bibitem[{{Sani} {et~al.}(2010){Sani}, {Lutz}, {Risaliti}, {Netzer}, {Gallo},
  {Trakhtenbrot}, {Sturm}, \& {Boller}}]{sanietal10}
{Sani}, E., {Lutz}, D., {Risaliti}, G., {et~al.} 2010, MNRAS, 403, 1246

\bibitem[{{Shen}(2013)}]{shen13}
{Shen}, Y. 2013, Bulletin of the Astronomical Society of India, 41, 61

\bibitem[{{Shen} \& {Ho}(2014)}]{shenho14}
{Shen}, Y. \& {Ho}, L.~C. 2014, \nat, 513, 210

\bibitem[{{Shen} \& {Liu}(2012)}]{shenliu12}
{Shen}, Y. \& {Liu}, X. 2012, \apj, 753, 125

\bibitem[{{Sikora} {et~al.}(2007){Sikora}, {Stawarz}, \&
  {Lasota}}]{sikoraetal07}
{Sikora}, M., {Stawarz}, {\L}., \& {Lasota}, J.-P. 2007, \apj, 658, 815

\bibitem[{{Smith} {et~al.}(2014){Smith}, {Jarvis}, {Hardcastle}, {Vaccari},
  {Bourne}, {Dunne}, {Ibar}, {Maddox}, {Prescott}, {Vlahakis}, {Eales},
  {Maddox}, {Smith}, {Valiante}, \& {de Zotti}}]{smithetal14}
{Smith}, D.~J.~B., {Jarvis}, M.~J., {Hardcastle}, M.~J., {et~al.} 2014, \mnras,
  445, 2232

\bibitem[{{Snedden} \& {Gaskell}(2007)}]{sneddengaskell07}
{Snedden}, S.~A. \& {Gaskell}, C.~M. 2007, ApJ, 669, 126

\bibitem[{{Storchi-Bergmann} {et~al.}(2017){Storchi-Bergmann}, {Schimoia},
  {Peterson}, {Elvis}, {Denney}, {Eracleous}, \&
  {Nemmen}}]{storchi-bergmannetal17}
{Storchi-Bergmann}, T., {Schimoia}, J.~S., {Peterson}, B.~M., {et~al.} 2017,
  \apj, 835, 236

\bibitem[{{Sulentic} \& {Marziani}(2015)}]{sulenticmarziani15}
{Sulentic}, J. \& {Marziani}, P. 2015, Frontiers in Astronomy and Space
  Sciences, 2, 6

\bibitem[{{Sulentic} {et~al.}(2011){Sulentic}, {Marziani}, \&
  {Zamfir}}]{sulenticetal11}
{Sulentic}, J., {Marziani}, P., \& {Zamfir}, S. 2011, Baltic Astronomy, 20, 427

\bibitem[{{Sulentic} {et~al.}(2001{\natexlab{a}}){Sulentic}, {Calvani}, \&
  {Marziani}}]{sulenticetal01a}
{Sulentic}, J.~W., {Calvani}, M., \& {Marziani}, P. 2001{\natexlab{a}}, The
  Messenger, 104, 25

\bibitem[{{Sulentic} {et~al.}(2017){Sulentic}, {del Olmo}, {Marziani},
  {Mart{\'{\i}}nez-Carballo}, {D'Onofrio}, {Dultzin}, {Perea},
  {Mart{\'{\i}}nez-Aldama}, {Negrete}, {Stirpe}, \& {Zamfir}}]{sulenticetal17}
{Sulentic}, J.~W., {del Olmo}, A., {Marziani}, P., {et~al.} 2017, \aap, 608,
  A122

\bibitem[{{Sulentic} {et~al.}(2006){Sulentic}, {Dultzin-Hacyan}, {Marziani},
  {Bongardo}, {Braito}, {Calvani}, \& {Zamanov}}]{sulenticetal06a}
{Sulentic}, J.~W., {Dultzin-Hacyan}, D., {Marziani}, P., {et~al.} 2006, Revista
  Mexicana de Astronomia y Astrofisica, 42, 23

\bibitem[{{Sulentic} {et~al.}(2015){Sulentic}, {Mart{\'{\i}}nez-Carballo},
  {Marziani}, {del Olmo}, {Stirpe}, {Zamfir}, \&
  {Plauchu-Frayn}}]{sulenticetal15}
{Sulentic}, J.~W., {Mart{\'{\i}}nez-Carballo}, M.~A., {Marziani}, P., {et~al.}
  2015, \mnras, 450, 1916

\bibitem[{{Sulentic} {et~al.}(2001{\natexlab{b}}){Sulentic}, {Marziani}, \&
  {Calvani}}]{sulenticetal01}
{Sulentic}, J.~W., {Marziani}, P., \& {Calvani}, M. 2001{\natexlab{b}}, in AIP
  CP, Vol. 599, X-ray Astronomy: Stellar Endpoints, AGN, and the Diffuse X-ray
  Background, 963--966

\bibitem[{{Sulentic} {et~al.}(2000{\natexlab{a}}){Sulentic}, {Marziani}, \&
  {Dultzin-Hacyan}}]{sulenticetal00a}
{Sulentic}, J.~W., {Marziani}, P., \& {Dultzin-Hacyan}, D. 2000{\natexlab{a}},
  ARA\&A, 38, 521

\bibitem[{{Sulentic} {et~al.}(2002){Sulentic}, {Marziani}, {Zamanov}, {Bachev},
  {Calvani}, \& {Dultzin-Hacyan}}]{sulenticetal02}
{Sulentic}, J.~W., {Marziani}, P., {Zamanov}, R., {et~al.} 2002, ApJL, 566, L71

\bibitem[{{Sulentic} {et~al.}(2008){Sulentic}, {Zamfir}, {Marziani}, \&
  {Dultzin}}]{sulenticetal08}
{Sulentic}, J.~W., {Zamfir}, S., {Marziani}, P., \& {Dultzin}, D. 2008, in
  Revista Mexicana de Astronomia y Astrofisica Conference Series, Vol.~32,
  51--58

\bibitem[{{Sulentic} {et~al.}(2000{\natexlab{b}}){Sulentic}, {Zwitter},
  {Marziani}, \& {Dultzin-Hacyan}}]{sulenticetal00c}
{Sulentic}, J.~W., {Zwitter}, T., {Marziani}, P., \& {Dultzin-Hacyan}, D.
  2000{\natexlab{b}}, ApJL, 536, L5

\bibitem[{{Sun} \& {Shen}(2015)}]{sunshen15}
{Sun}, J. \& {Shen}, Y. 2015, \apjl, 804, L15

\bibitem[{{Ulvestad} {et~al.}(2005){Ulvestad}, {Antonucci}, \&
  {Barvainis}}]{ulvestadetal05}
{Ulvestad}, J.~S., {Antonucci}, R.~R.~J., \& {Barvainis}, R. 2005, \apj, 621,
  123

\bibitem[{{Urry} \& {Padovani}(1995)}]{urrypadovani95}
{Urry}, C.~M. \& {Padovani}, P. 1995, PASP, 107, 803

\bibitem[{{Veilleux} {et~al.}(2016){Veilleux}, {Mel{\'e}ndez}, {Tripp},
  {Hamann}, \& {Rupke}}]{veilleuxetal16}
{Veilleux}, S., {Mel{\'e}ndez}, M., {Tripp}, T.~M., {Hamann}, F., \& {Rupke},
  D.~S.~N. 2016, \apj, 825, 42

\bibitem[{{V{\'e}ron-Cetty} {et~al.}(2001){V{\'e}ron-Cetty}, {V{\'e}ron}, \&
  {Gon{\c c}alves}}]{veroncettyetal01}
{V{\'e}ron-Cetty}, M.-P., {V{\'e}ron}, P., \& {Gon{\c c}alves}, A.~C. 2001,
  AAp, 372, 730

\bibitem[{{Vollmer}(2009)}]{vollmer09}
{Vollmer}, B. 2009, VizieR Online Data Catalog, VIII/85A

\bibitem[{{Wang} \& {Li}(2011)}]{wangli11}
{Wang}, J. \& {Li}, Y. 2011, \apjl, 742, L12

\bibitem[{{Wang} {et~al.}(2017){Wang}, {Du}, {Brotherton}, {Hu}, {Songsheng},
  {Li}, {Shi}, \& {Zhang}}]{wangetal17}
{Wang}, J.-M., {Du}, P., {Brotherton}, M.~S., {et~al.} 2017, Nature Astronomy,
  1, 775

\bibitem[{{Wang} {et~al.}(2014){Wang}, {Du}, {Li}, {Ho}, {Hu}, \&
  {Bai}}]{wangetal14}
{Wang}, J.-M., {Du}, P., {Li}, Y.-R., {et~al.} 2014, \apjl, 792, L13

\bibitem[{{Wills} \& {Browne}(1986)}]{willsbrowne86}
{Wills}, B.~J. \& {Browne}, I.~W.~A. 1986, \apj, 302, 56

\bibitem[{{Wilson} \& {Colbert}(1995)}]{wilsoncolbert95}
{Wilson}, A.~S. \& {Colbert}, E.~J.~M. 1995, \apj, 438, 62

\bibitem[{{Woo} \& {Urry}(2002)}]{woourry02}
{Woo}, J.-H. \& {Urry}, C.~M. 2002, \apjl, 581, L5

\bibitem[{{Wu}(2009)}]{wu09b}
{Wu}, Q. 2009, \mnras, 398, 1905

\bibitem[{{Yun} {et~al.}(2001){Yun}, {Reddy}, \& {Condon}}]{yunetal01}
{Yun}, M.~S., {Reddy}, N.~A., \& {Condon}, J.~J. 2001, \apj, 554, 803

\bibitem[{{Zakamska} {et~al.}(2016){Zakamska}, {Hamann}, {P{\^a}ris}, {Brandt},
  {Greene}, {Strauss}, {Villforth}, {Wylezalek}, {Alexandroff}, \&
  {Ross}}]{zakamskaetal16}
{Zakamska}, N.~L., {Hamann}, F., {P{\^a}ris}, I., {et~al.} 2016, \mnras, 459,
  3144

\bibitem[{{Zamanov} {et~al.}(2002){Zamanov}, {Marziani}, {Sulentic}, {Calvani},
  {Dultzin-Hacyan}, \& {Bachev}}]{zamanovetal02}
{Zamanov}, R., {Marziani}, P., {Sulentic}, J.~W., {et~al.} 2002, ApJL, 576, L9

\bibitem[{{Zamfir} {et~al.}(2008){Zamfir}, {Sulentic}, \&
  {Marziani}}]{zamfiretal08}
{Zamfir}, S., {Sulentic}, J.~W., \& {Marziani}, P. 2008, MNRAS, 387, 856

\bibitem[{{Zamfir} {et~al.}(2010){Zamfir}, {Sulentic}, {Marziani}, \&
  {Dultzin}}]{zamfiretal10}
{Zamfir}, S., {Sulentic}, J.~W., {Marziani}, P., \& {Dultzin}, D. 2010, \mnras,
  403, 1759

\bibitem[{{Zhang} {et~al.}(2011){Zhang}, {Dong}, {Wang}, \&
  {Gaskell}}]{zhangetal11}
{Zhang}, K., {Dong}, X.-B., {Wang}, T.-G., \& {Gaskell}, C.~M. 2011, \apj, 737,
  71

\bibitem[{{Zickgraf} {et~al.}(2003){Zickgraf}, {Engels}, {Hagen}, {Reimers}, \&
  {Voges}}]{zickgrafetal03}
{Zickgraf}, F.~J., {Engels}, D., {Hagen}, H.~J., {Reimers}, D., \& {Voges}, W.
  2003, \aap, 406, 535

\end{thebibliography}
\vfill\eject\newpage\pagebreak


\onecolumn
\scriptsize
\begin{appendix}
\section{Tables}
\label{CDphys}
\setcounter{table}{0}
\begin{longtable}{lllllllllll}
\caption{CD AGNs sample physical and spectral parameters. \label{tab:CDphys}}
\\
\hline\hline
{{SDSS ID}} & {{z}} & {{$f_\mathrm{g}$}} & {{$f_\mathrm{1.4GHz, P}$}} & {{$f_\mathrm{1.4GHz, I}$}} & {{$\log R_\mathrm{K}$}} & {{$\log L_\mathrm{5100}$}} & {{$\log P_\mathrm{1.4GHz, P}$}} & {{FWHM[H$\beta$]}} & {{R$_\mathrm{FeII}$}} & {}\\
{} & {} & {[mJy]} & {[mJy]} & {[mJy]} & {} & {[erg/s]} & {[W/Hz]} & {[km/s]} & {} & {}\\
\hline
\endfirsthead
\caption{continued.}\\
\hline\hline
{{SDSS ID}} & {{z}} & {{$f_\mathrm{g}$}} & {{$f_\mathrm{1.4GHz, P}$}} & {{$f_\mathrm{1.4GHz, I}$}} & {{$\log R_\mathrm{K}$}} & {{$\log L_\mathrm{5100}$}} & {{$\log P_\mathrm{1.4GHz, P}$}} & {{FWHM[H$\beta$]}} & {{R$_\mathrm{FeII}$}} & {}\\
{} & {} & {[mJy]} & {[mJy]} & {[mJy]} & {} & {[erg/s]} & {[W/s]} & {[km/s]} & {} & {}\\
\hline
\endhead
\hline
\endfoot
J000652.00+061509.2 & 0.873061 & 0.1125 &  1.9076    &	-	  & 1.22 &   45.42  &   24.85  &  4484.2 & $\sim$ 0.7    & RD CD B2 \\
J000727.89+055701.3 & 0.791175 & 0.0765 &  5.5325    &	5.8822	  & 1.85 &   44.96  &   25.21  &  8677.3 & $\leq 0.5$    & RL CD B1$^{+}$ \\
J000804.16-012917.0 & 0.312733 & 0.0238 &  1.5704    &	-	  & 1.81 &   44.50  &   23.69  &  2104.2 & $\leq 0.5$    & RL CD A1 \\
J001741.45+082755.7 & 0.678577 & 0.0775 &  22.3868   &	23.2705	  & 2.46 &   44.61  &   25.65  &  6981.3 & $\leq 0.5$    & RL CD B1 \\
J001932.63-070302.6 & 0.773397 & 0.0500 &  0.6830    &	1.4193	  & 1.13 &   44.83  &   24.27  &  6103.7 & $\leq 0.5$    & RI CD B1 \\
J002312.69+072455.4 & 0.660799 & 0.1705 &  1.9841    &	-	  & 1.06 &   45.16  &   24.57  &  4753.7 & $\leq 0.5$    & RI CD B1 \\
J002554.26+083711.3 & 0.814390 & 0.1473 &  90.5472   &	94.7792	  & 2.78 &   45.31  &   26.45  &  5425.3 & $\sim$ 0.6     & RL CD B2 \\
J003706.29-050111.3 & 0.718323 & 0.0514 &  3.0737    &	3.3855	  & 1.77 &   44.77  &   24.85  &  13993.6 & $\leq 0.5$   & RI CD B1$^{++}$ \\
\end{longtable}


\setcounter{table}{1}
\begin{longtable}{llllllllll}
\caption{FRII AGNs sample  \label{tab:FRII}}\\
\hline\hline
\multicolumn{1}{l}{\textbf{SDSS ID}} & \multicolumn{1}{l}{\textbf{RA}} & \multicolumn{1}{l}{\textbf{DEC}} & \multicolumn{1}{l}{\textbf{RA$_{1}$}} & \multicolumn{1}{l}{\textbf{DEC$_{1}$}} & \multicolumn{1}{l}{\textbf{Sep.$_{1}$}} & \multicolumn{1}{l}{\textbf{RA$_{2}$}} & \multicolumn{1}{l}{\textbf{DEC$_{2}$}} & \multicolumn{1}{l}{\textbf{Sep.$_{2}$}} & \multicolumn{1}{l}{\textbf{$\Psi_{1,2}$}}\\
\multicolumn{1}{l}{} & \multicolumn{1}{l}{(deg)} & \multicolumn{1}{l}{(deg)} & \multicolumn{1}{l}{(deg)} & \multicolumn{1}{l}{(deg)} & \multicolumn{1}{l}{(arcsec)} & \multicolumn{1}{l}{(deg)} & \multicolumn{1}{l}{(deg)} & \multicolumn{1}{l}{(arcsec)} & \multicolumn{1}{l}{(deg)}\\
\hline
\endfirsthead
\caption{continued.}\\
\hline\hline
\multicolumn{1}{l}{\textbf{SDSS ID}} & \multicolumn{1}{l}{\textbf{RA}} & \multicolumn{1}{l}{\textbf{DEC}} & \multicolumn{1}{l}{\textbf{RA$_{1}$}} & \multicolumn{1}{l}{\textbf{DEC$_{1}$}} & \multicolumn{1}{l}{\textbf{Sep.$_{1}$}} & \multicolumn{1}{l}{\textbf{RA$_{2}$}} & \multicolumn{1}{l}{\textbf{DEC$_{2}$}} & \multicolumn{1}{l}{\textbf{Sep.$_{2}$}} & \multicolumn{1}{l}{\textbf{$\Psi_{1,2}$}}\\
\multicolumn{1}{l}{} & \multicolumn{1}{l}{(deg)} & \multicolumn{1}{l}{(deg)} & \multicolumn{1}{l}{(deg)} & \multicolumn{1}{l}{(deg)} & \multicolumn{1}{l}{(arcsec)} & \multicolumn{1}{l}{(deg)} & \multicolumn{1}{l}{(deg)} & \multicolumn{1}{l}{(arcsec)} & \multicolumn{1}{l}{(deg)}\\
\hline
\endhead
\hline
\endfoot
J010255.68+103702.9	&   15.7320 	&   10.6174	&   15.7306 	&   10.6108	&   24.38	&   15.7355 	&   10.6255 &   31.46	& 168.4  \\
J014022.68+075357.0	&   25.0945 	&   7.8991	&   25.0957 	&   7.8979	&   6.26	&   25.0930 	&   7.9004  &   6.89	& 171.9  \\
J014239.99+073239.4	&   25.6666 	&   7.5442	&   25.6700 	&   7.5449	&   12.52	&   25.6595 	&   7.5448  &   25.25	& 165.3  \\
J015635.70+014816.1	&   29.1487 	&   1.8044	&   29.1434 	&   1.8048	&   19.11	&   29.1550 	&   1.8029  &   23.43	& 170.3  \\
J023322.16-045506.8	&   38.3423 	&   -4.9185	&   38.3332 	&   -4.9169 &   33.00	&   38.3519 	&   -4.9179 &   34.64	& 166.1  \\
J024500.70-074736.4	&   41.2529 	&   -7.7934	&   41.2532 	&   -7.7912 &   7.93	&   41.2533 	&   -7.7968 &   12.19	& 163.6  \\
J081703.10+162259.7	&   124.2629   	&   16.3832	&   124.2643 	&   16.3855 &   9.32	&   124.2620 	&   16.3779 &   19.34	& 158.9  \\
J081735.07+223717.7	&   124.3961   	&   22.6215	&   124.3953 	&   22.6247 &   11.48	&   124.3965 	&   22.6183 &   11.89	& 173.9  \\
\end{longtable}

\begin{longtable}{llllllllll}
\caption{FRII AGNs sample physical and spectral parameters.\label{tab:FRIIphys}}\\
\hline\hline
{{SDSS ID}} & {{z}} & {{$f_\mathrm{g}$}} & {{$f_\mathrm{1.4GHz, I}$}} & {{$\log R_\mathrm{K}$}} & {{$\log L_\mathrm{5100}$}} & {{$\log P_\mathrm{1.4GHz, I}$}} & {{FWHM[H$\beta$]}} & {{R$_\mathrm{FeII}$}} & {{}}\\
{} & {} & {[mJy]} & {[mJy]} & {} & {[erg/s]} & {[W/s]} & {[km/s]} & {} & {}\\
\hline
\endfirsthead
\caption{continued.}\\
\hline\hline
{{SDSS ID}} & {{z}} & {{$f_\mathrm{g}$}} & {{$f_\mathrm{1.4GHz, I}$}} & {{$\log R_\mathrm{K}$}} & {{$\log L_\mathrm{5100}$}} & {{$\log P_\mathrm{1.4GHz, I}$}} & {{FWHM[H$\beta$]}} & {{R$_\mathrm{FeII}$}} & {{}}\\
{} & {} & {[mJy]} & {[mJy]} & {} & {[erg/s]} & {[W/s]} & {[km/s]} & {} & {}\\
\hline
\endhead
\hline
\endfoot
J010255.68+103702.9 &	 0.910985 &	   0.0679 &	38.24	& 2.75	  & 45.02   &	26.20 &	 4858.5  &	$\leq 0.5$ & RL FRII B1 \\
J014022.68+075357.0 &	 0.899535 &	   0.0717 &	90.13	& 3.10	  & 44.81   &	26.56 &	 7283.5  &	$\leq 0.5$ & RL FRII B1 \\
J014239.99+073239.4 &	 0.906408 &	   0.0372 &	34.03	& 2.96	  & 44.70   &	26.15 &	 18945.9 &	$\leq 0.5$ & RL FRII B1$^{+++}$ \\
J015635.70+014816.1 &	 0.864186 &	   0.0410 &	27.24	& 2.82	  & 44.89   &	26.00 &	 9571.6  &	$\leq 0.5$ & RL FRII B1$^{+}$ \\
J023322.16-045506.8 &	 0.780403 &	   0.3583 &	109.69	& 2.49	  & 45.49   &	26.49 &	 4843.1  &	$\leq 0.5$ & RL FRII B1 \\
J024500.70-074736.4 &	 0.887057 &	   0.0321 &	23.44	& 2.86	  & 45.08   &	25.96 &	 9923.1  &	$\leq 0.5$ & RL FRII B1$^{+}$ \\
J081703.10+162259.7 &	 0.699275 &	   0.0527 &	132.08	& 3.40	  & 44.90   &	26.46 &	 12982.1 &	$\leq 0.5$ & RL FRII B1$^{++}$ \\
J081735.07+223717.7 &	 0.981346 &	   0.1805 &	1252.64	& 3.84	  & 45.82   &	27.80 &	 6457.1  &	$\leq 0.5$ & RL FRII B1 \\
\end{longtable}

\vfill\newpage\eject

\newpage\pagebreak\pagebreak

\onecolumn

\section{FRII Atlas}
\label{atlas}

The Figures of the Appendix show an atlas of the 66 FRII identified from the matching of the BOSS and the FIRST (Sect. \ref{sample}).

\begin{figure*}[htbp!]
\begin{center}
\includegraphics[width=.75\textwidth]{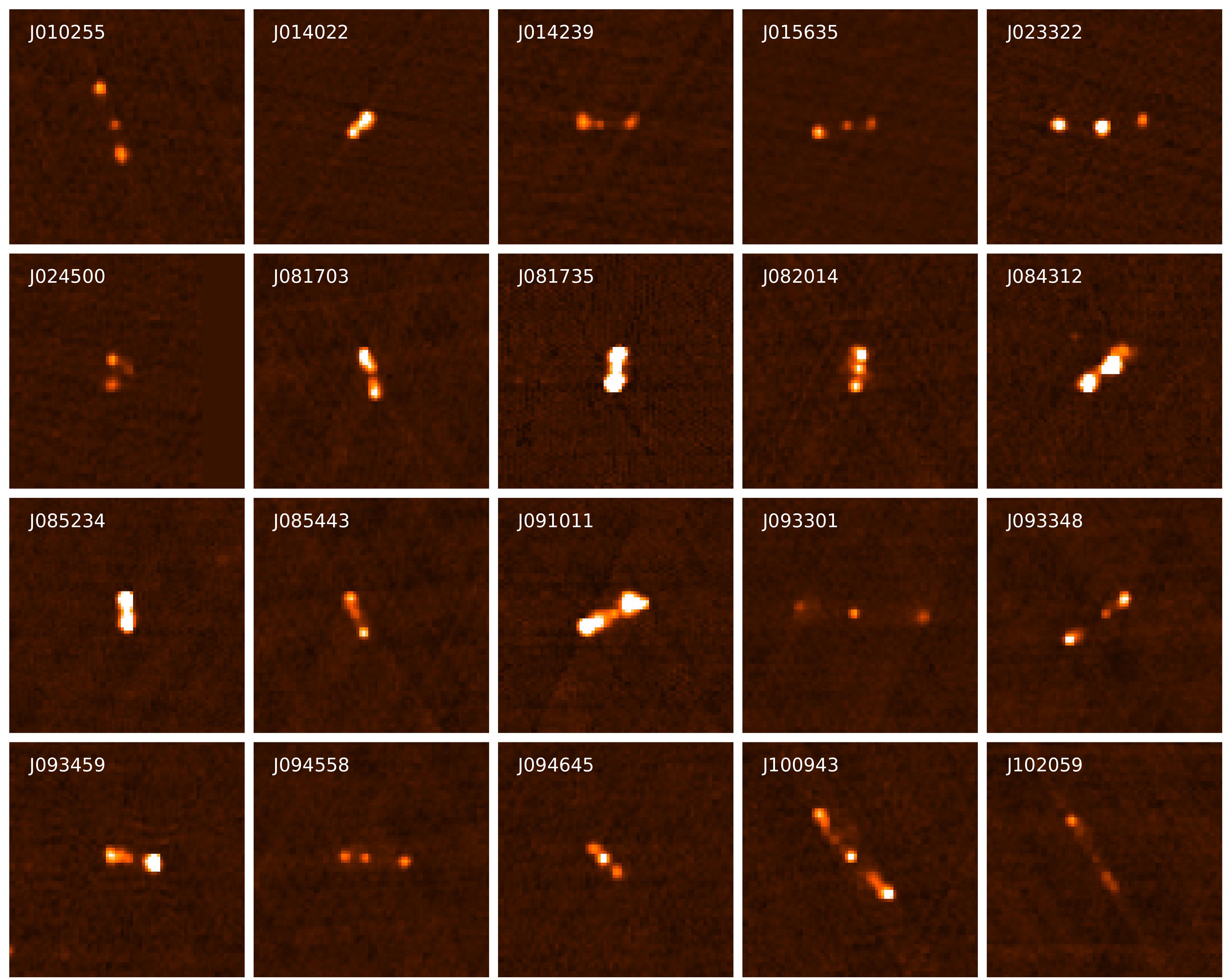}\\
\includegraphics[width=.75\textwidth]{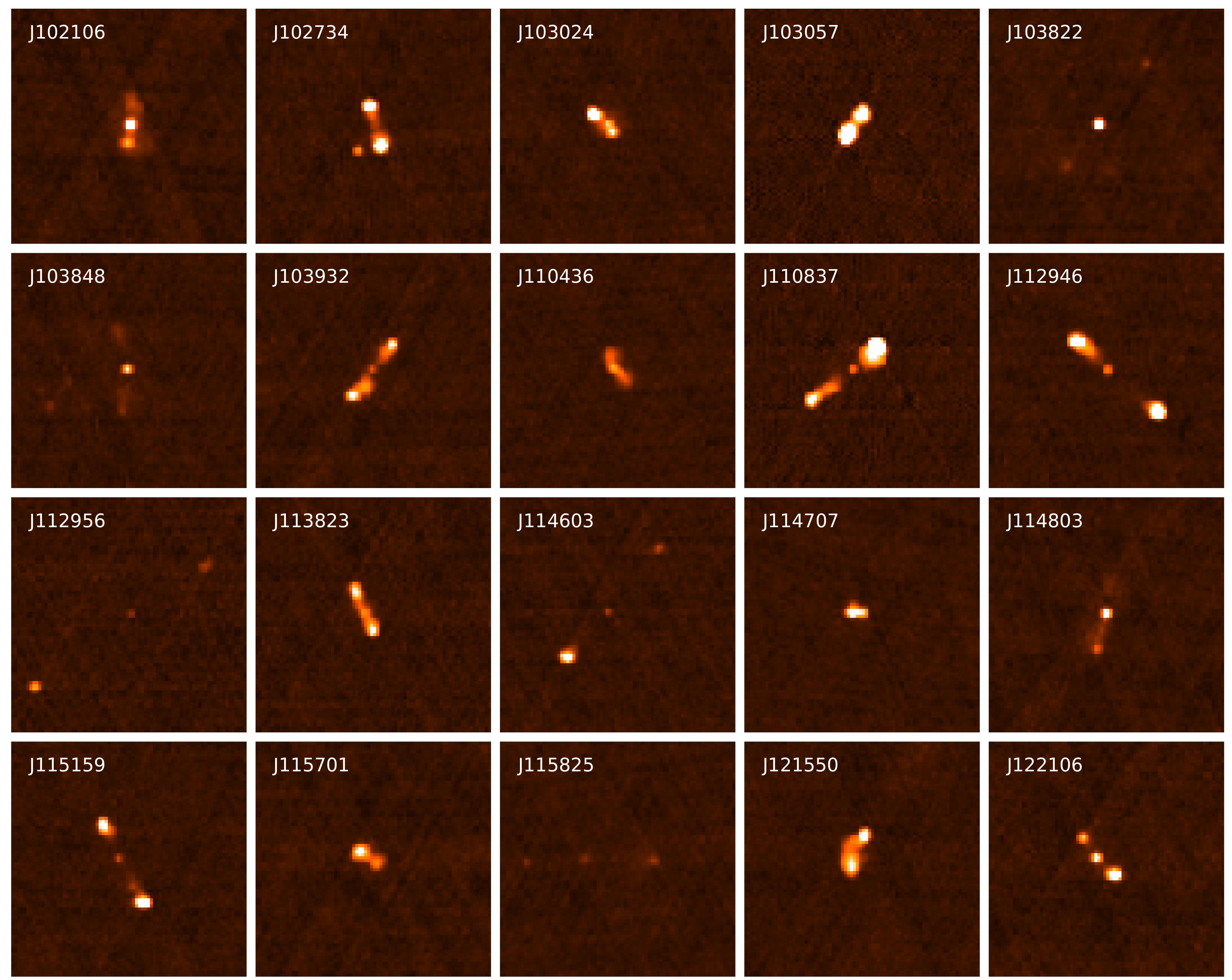}
\label{fig:atla}
\end{center}
\end{figure*}

\begin{figure*}[htbp!]
\begin{center}
\includegraphics[width=.75\textwidth]{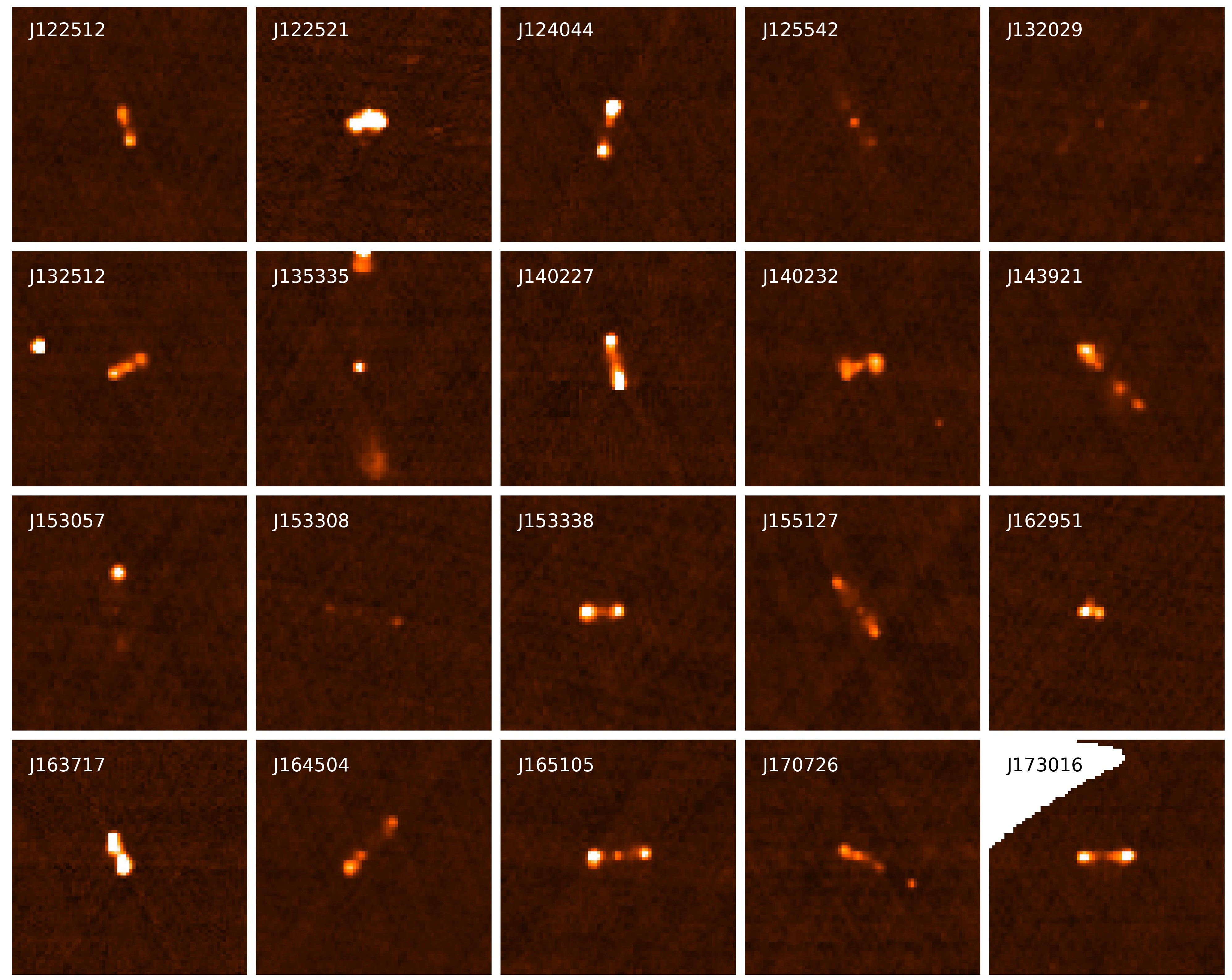}\\
\includegraphics[width=.75\textwidth]{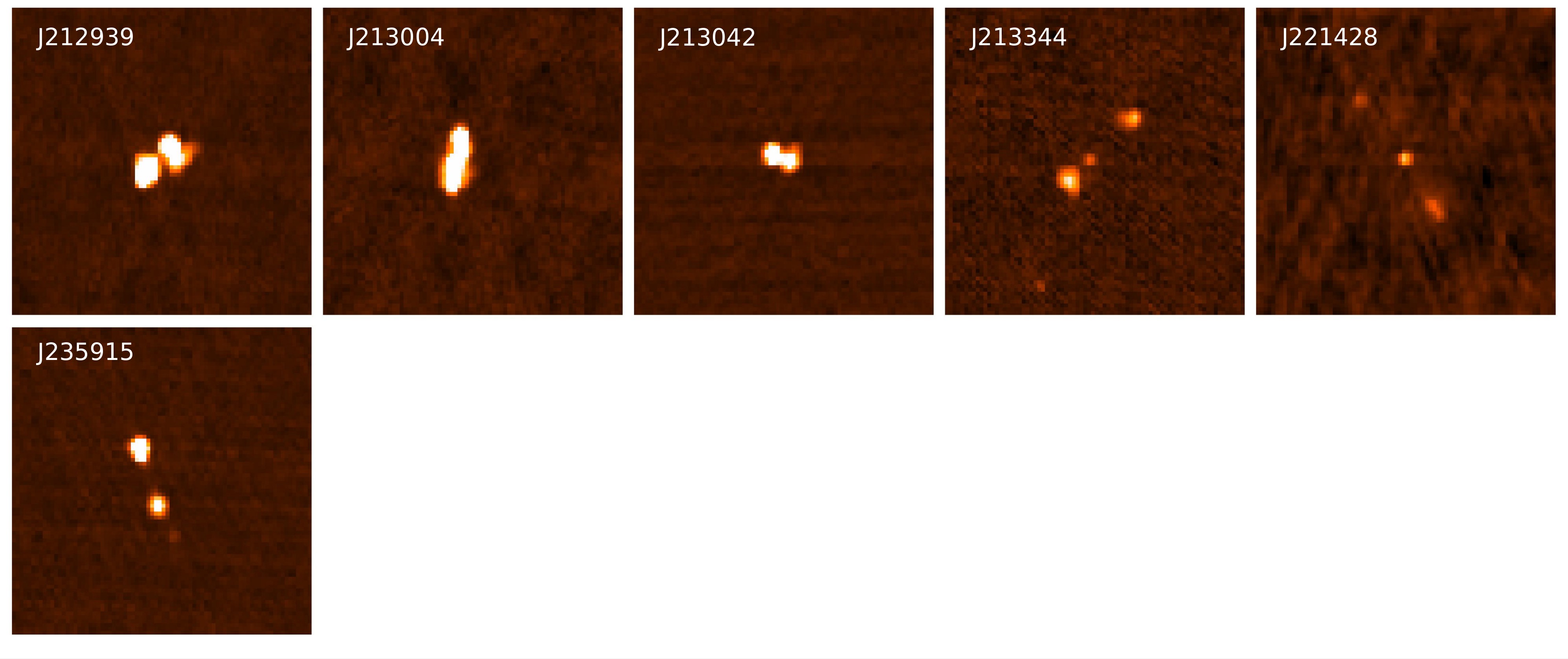}
\caption{Radio maps at 1.4 GHz for the FRII sample from the FIRST catalog with size of field 4.5' x 4.5' and a Maximum Intensity for Scaling equals to 10 mJy. The images follow the SDSS ID order of table \ref{tab:FRII} from left to right and from top to bottom.}
\label{fig:atla2}
\end{center}
\end{figure*}

\end{appendix}


\end{document}